\documentclass{pasa2}%

\usepackage{graphicx}
\usepackage{xcolor}
\definecolor{orcidlogocol}{HTML}{A6CE39}
\definecolor{dodgerblue}{RGB}{30, 144, 255}

\usepackage{aas_macros}
\usepackage{hyperref} 
\hypersetup{colorlinks,citecolor=dodgerblue,linkcolor=dodgerblue,urlcolor=dodgerblue}

\usepackage{amsmath}

\usepackage{booktabs}
\usepackage{threeparttable}
\usepackage{etoolbox}
\usepackage{sansmath}
\AtBeginEnvironment{tablenotes}{\sffamily\sansmath}
\usepackage{subcaption,cleveref}
\usepackage{cprotect}
\usepackage{multirow}

\usepackage{times}
\usepackage{mathptmx}  % times-like math fonts
\usepackage{tgheros}   % sans-serif
\usepackage{beramono}  % mono

\usepackage{pdflscape}  % Landscape pages

\usepackage{amsmath}	% Advanced maths commands
\usepackage{amssymb}	% Extra maths symbols
\usepackage{amsfonts}   % For a checkmark
\usepackage{siunitx}    % For units.
\usepackage{multirow}   % For row contents that span multiple rows in tables.
\usepackage{booktabs}
\usepackage{longtable}

\newcommand{\tabft}[1]{(#1)}
\newcounter{ft}
\newcommand{\ft}[1]{\noindent%
  ~\refstepcounter{ft}\tabft{\alph{ft}\label{#1}}}
   
\DeclareSIUnit\parsec{pc}
\DeclareSIUnit\lightyear{ly}
\DeclareSIUnit\year{yr}
\DeclareSIUnit\jansky{Jy}
\DeclareSIUnit\beam{beam}
\DeclareSIUnit\ev{eV}
\DeclareSIUnit\gauss{G}
\DeclareSIUnit\erg{erg}

\def\mjybeam{mJy\,beam$^{-1}$}

\def\degr{\ensuremath{^\circ}}
\def\arcmin{\ensuremath{'}}
\def\arcsec{\ensuremath{''}}

\def\tgsscontour{blue}
\def\nvsscontour{red}
\def\sumsscontour{red}

\newcommand\fs{\mbox{$.\!\!^{\mathrm s}$}}%

\newcommand\corrs[1]{{ #1}}

\title[Diffuse galaxy cluster emission at 168 MHz]{Diffuse galaxy cluster emission at 168 MHz within the Murchison Widefield Array Epoch of Reionization 0-hour field}

\author[S.~W. Duchesne et al.]{S.~W. Duchesne$^{1,2}$\thanks{email: \url{stefan.duchesne.astro@gmail.com}}, M. Johnston-Hollitt$^{2,3}$, A.~R. Offringa$^{4}$, G.~W. Pratt$^{5}$, Q. Zheng$^{6,7}$ and S. Dehghan$^{1}$
\affil{$^1$School of Chemical and Physical Sciences, Victoria University of Wellington, P.~O. Box 600, Wellington 6140, New Zealand}%
\affil{$^2$International Centre for Radio Astronomy Research (ICRAR), Curtin University, Bentley, WA 6102, Australia}
\affil{$^3$Curtin Institute for Computation, Curtin University,
GPO Box U1987, Perth, WA 6845, Australia}
\affil{$^4$Netherlands Institute for Radio Astronomy (ASTRON), PO Box 2, NL-7990 AA Dwingeloo, the Netherlands}
\affil{$^5$AIM, CEA, CNRS, Université Paris-Saclay, Université Paris Diderot, Sorbonne Paris Cité, F-91191 Gif-sur-Yvette, France}
\affil{$^6$School of Engineering and Computer Science, Victoria University of Wellington, PO Box 600, Wellington 6140, New Zealand}
\affil{$^7$Shanghai Astronomical Observatory, Chinese Academy of Sciences, 80 Nandan Road, Shanghai 200030, China}
}%

\jid{PASA}
\doi{10.1017/pas.\the\year.xxx}
\jyear{\the\year}
\documentstatus{\textcolor{green}{accepted} for publication in PASA}

\begin{document}

\begin{frontmatter}
\maketitle
\rule{\linewidth}{0.75pt}\vspace{11.5pt}
\begin{abstract}
We detect and characterise extended, diffuse radio emission from galaxy clusters at 168 MHz within the Epoch of Reionization 0-hour field: a $45\degr \times 45\degr$ region of the southern sky centred on R.~A.${}= 0\degr$, decl.${}=-27\degr$. We detect 29 sources of interest; a newly detected halo in Abell~0141; a newly detected relic in Abell~2751; 4 new halo candidates and a further 4 new relic candidates; and a new phoenix candidate in Abell 2556. Additionally, we find 9 clusters with unclassifiable, diffuse steep-spectrum emission as well as a candidate double relic system associated with RXC~J2351.0-1934. We present measured source properties such as their integrated flux densities, spectral indices {($\alpha$, where $S_\nu \propto \nu^\alpha$)}, and sizes where possible. We find several of the diffuse sources to {have ultra-steep spectra} including the halo in Abell 0141, if confirmed, showing $\alpha \leq -2.1 \pm 0.1$ with the present data making it one of the {steepest-spectrum} haloes known. Finally, we compare our sample of haloes with previously detected haloes and revisit established scaling relations of the radio halo power ($P_{1.4}$) with the cluster X-ray luminosity ($L_{\mathrm{X}}$) and mass ($M_{500}$). {We find that the newly detected haloes and candidate haloes are consistent with the $P_{1.4}$--$L_{\mathrm{X}}$ and $P_{1.4}$--$M_{500}$ relations, and see an increase in scatter in the previously found relations with increasing sample size likely caused by inhomogeneous determination of $P_{1.4}$ across the full halo sample. We show that the MWA is capable of detecting haloes and relics within most of the galaxy clusters within the \emph{Planck} catalogue of Sunyaev--Zel'dovich sources depending on exact halo or relic properties.}
\end{abstract}

\rule{\linewidth}{0.75pt}\vspace{11.5pt}
\end{frontmatter}

\section{Introduction}
\label{sec:intro}

Clusters of galaxies are among the largest structures in the Universe. Understanding how clusters form and their dynamics is key to understanding how the Universe behaves on some of the largest scales. Galaxy clusters are thought to form in the hierarchical model, where galaxies eventually clump together during sometimes intense merger events \citep{pee80}. The clusters themselves are primarily dark matter, diffuse gas that makes up the intra-cluster medium (ICM), and the galaxies for which they are named. Galaxy clusters are found to host magnetic fields on the order of 0.1--1~$\mu$G \citep{ckb01, mj-h, bfm+10}. The magnetic fields in clusters give rise to radio synchrotron emission; relativistic electrons accelerated by the magnetic fields with Lorentz factors of $\gamma > 1000$, where the spectral energy distribution (SED) of the emission gives insight into the ages of electron populations and the possible shock-driven re-acceleration from merger events \citep[see][for reviews]{fggm12,bj14,vda+19}. The steep spectral indices \footnote{The spectral index $\alpha$ is defined through $S_{\nu} \propto \nu^{\alpha}$ for flux density $S_\nu$ at frequency $\nu$.} of such synchrotron emission means that to detect the faintest non-thermal diffuse cluster emission low-frequency radio telescopes are required, \corrs{such as the Giant Metrewave Radio Telescope \citep[GMRT;][]{Ananthakrishnan1995}, the Murchison Widefield Array \citep[MWA;][]{tgb+13}, and the LOw Frequency ARray \citep[LOFAR;][]{lofar}.} As radio telescopes become more sensitive, more of this steep-spectrum diffuse emission is expected to be found \citep{cbn+12,joh17}.\par

Diffuse synchrotron emission comes in two main classes: cluster haloes and relics. Cluster relics can be broken down further into two types:~kpc-scale \emph{phoenices} and \emph{megaparsec-scale relics}. \corrs{The kpc-scale radio phoenices are thought to be emission from revived fossil plasma left over from long dormant radio galaxies \citep[see e.g.][]{Ensslin01,eb02} and are usually found near the cluster centre \citep[e.g.][]{srm+01}}. Megaparsec-scale relics (hereafter relics) are thought to trace shocks through the ICM during and after massive merger events. These are found on the periphery of clusters, usually aligned with the major merger axis and can come in adjacent pairs of so-called \emph{double relics} (e.g. Abell 3667; \citealt{mj-h}, Abell 3376; \citealt{bdnp06}, PSZ1~G108.18$-$11.53; \citealt{div+15}). For both types of relics the electrons must go through some re-acceleration process albeit on vastly different scales. These processes are thought to be through shocks typically resulting in an elongated or arc-like morphology \corrs{in the case of relics}. The key \corrs{observed} distinction between the two types of emission are their size and spectral properties. \corrs{Phoenices, thought to form through adiabatic shock compression, can show curved spectra \citep{Ensslin01} whereas relic spectra typically resemble power laws \citep[e.g.][]{hjh+14,George2017,Rajpurohit2020a}.}

Haloes also come in two main types: \emph{mini-haloes} and \emph{cluster haloes}. Mini-haloes are associated with strong active galactic nuclei (AGN), often the central dominant (cD) galaxy within the core of the cluster, and are smaller in extent though are otherwise morphologically similar to cluster haloes \citep[for a recent review see][]{bgb16}. Cluster haloes are centrally located within the cluster, morphologically regular, and are often found to coincide with the X-ray emitting plasma of the ICM. Haloes do not normally show any significant fractional polarisation however this is likely a limitation of the resolution of current-generation radio interferometers \citep{gmx+013}. The mechanism that generates these radio haloes is still under investigation. The primary, re-acceleration model of halo generation suggests the synchrotron emission occurs after electrons are re-accelerated through merger-driven turbulence in the magnetised ICM \citep[see e.g.][]{bsfg01, buo01, p01, pe08, cbn+12}. An alternate model is that of hadronic origin \citep[see e.g.][]{den80,de00}. In this secondary model, electrons are generated as secondary products of collisions between cosmic ray protons and ICM protons. Pions, a product in these proton-proton collisions, produce the electrons that will be accelerated by magnetic fields, as well as $\gamma$-rays. This model not only requires $\gamma$-ray emission from clusters, but also that all galaxy clusters host radio haloes \corrs{at some level}. The synchrotron emission from electrons produced through these proton-proton collisions will be significantly weaker than that seen through re-acceleration via turbulence \citep{bc99}. So far only upper limits for $\gamma$-ray emission have been presented \corrs{\citep[e.g.][]{Ackermann2014,pc14,Liang2016}}, and with current generation radio telescopes, the necessary sensitivity to detect haloes generated through the secondary model \corrs{alone} has not been reached. The primary and secondary models are not mutually exclusive, and there has been work to combine the two models \citep[e.g.][]{bb05, bl11, bl16}. The primary model is observationally supported by the fact that \corrs{predominantly unrelaxed, X-ray luminous clusters} are known to \corrs{host} radio haloes. However radio halo detection had been biased toward those clusters hosting highly X-ray luminous plasma as these are the clusters often targeted \citep[e.g.][]{gtf99, vgb+07, vgd+08, kvg+13, kvg+15}. Only recently have surveys been conducted to search for diffuse cluster emission without preselecting clusters based solely on their X-ray luminosities. For example, \citet{bvc+16} select clusters based on mass, and \citet{sjp16} survey clusters over a wide range of X-ray luminosities. \par%
Given the comparative rarity of diffuse cluster emission detection, we wish to perform larger surveys to properly ascertain the incidence and nature of these types of radio emission. In this paper we present the results of one such survey using a deep $45\degr \times 45\degr$ image produced by the MWA as part of the MWA Epoch of Reionization (EoR) project \citep{bck+13, oth+16}. This study forms the pilot for a larger search for diffuse cluster emission (Johnston-Hollitt et al. in prep.) using the recently released GaLactic and Extragalactic All-sky MWA survey \citep[GLEAM;][]{wlb+15}, which covers the entire southern sky below a declination of $+25\degr$ and covers the frequency range 72--231~MHz. In the following sections we discuss the \corrs{various images used} and the process involved in searching for diffuse cluster emission.\par%
This paper unless otherwise stated assumes a flat $\Lambda$CDM cosmology with $H_0 = 70$~km\,s$^{-1}$\,Mpc$^{-1}$, $\Omega_\mathrm{M} = 0.3$, and $\Omega_\Lambda = 1-\Omega_\mathrm{M}$.

\section{The search for diffuse cluster emission}
\subsection{The Epoch of Reionization 0-hour field}

As part of the MWA EoR project \citet{oth+16} present a $45\degr \times 45 \degr$ image centred on {$(\alpha_{\mathrm{J2000}},\,\delta_{\mathrm{J2000}}) =  (00^{\mathrm{h}}00^{\mathrm{m}}00^{\mathrm{s}}, -27\degr00\arcmin00\arcsec)$}, at a frequency of 168~MHz called the EoR0 field. This image is obtained from 45 hours of integration and has a resolution of 2.3~arcmin. The EoR0 field is the deepest, confusion limited image made with the 128-tile Phase I MWA \footnote{Since this work was undertaken the MWA has been upgraded to the so-called Phase II MWA (see \citealt{2018PASA...35...33W} for details).}. In addition to the overall sensitivity, the low surface brightness imaging capability provided by the number of short ($\leq60$~m) baselines makes the MWA a powerful tool to investigate extended, diffuse emission. Data collection, reduction, and imaging for the field used here is explained in detail in \citet{oth+16}. Whilst the primary purpose of the EoR0 field is the study of EoR, the image itself is incredibly sensitive \corrs{for an MWA image at this frequency}, reaching down to $\sim$2.3~\mjybeam\ near the centre of the image and increasing up to $\sim$100~\mjybeam out towards the image edges. This surface brightness sensitivity makes the EoR0 field useful in the search for steep spectrum cluster haloes and relics. The R.A. and decl. range used here is as follows: {$\left(22^{\mathrm{h}}29^{\mathrm{m}}55\fs2 \leq \alpha_{\mathrm{J2000}} \leq 01^{\mathrm{h}}29^{\mathrm{m}}57\fs6\right)$ and $\left(-44\degr41\arcmin24\arcsec \leq \delta_{\mathrm{J2000}} \leq -08\degr36\arcmin36\arcsec\right)$}, which is chosen to cut out the most significant noise at the edge of the image.

\subsection{Catalogues of galaxy clusters}\label{priors}

\defcitealias{planck15}{PSZ1}
\defcitealias{pap+11}{MCXC}
\defcitealias{aco89}{ACO}

\begin{figure}[!t]
\centering
\includegraphics[width=1\linewidth]{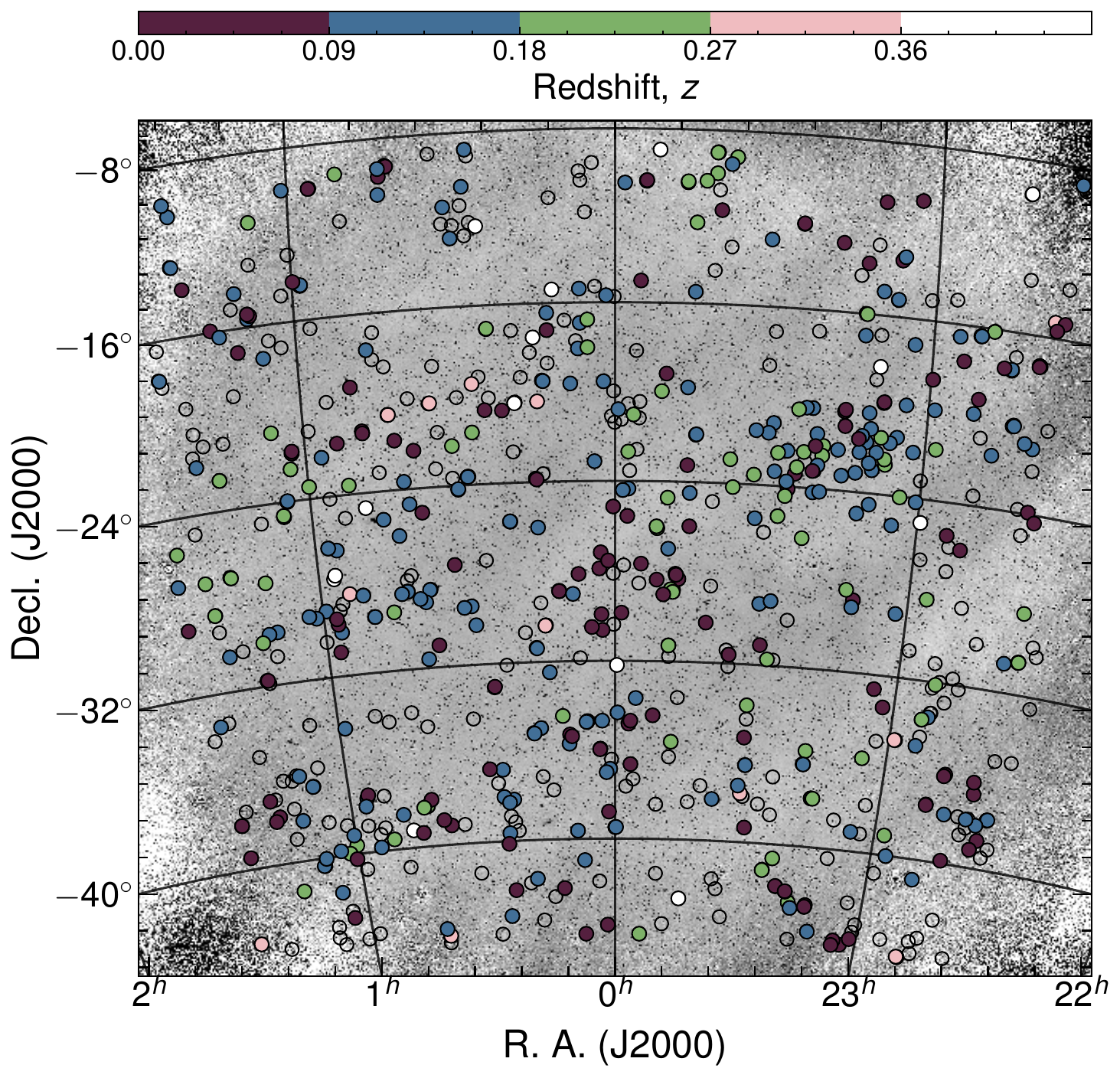}
\caption{ The central $\sim$42\degr~of the Epoch of Reoinization 0-hour field. Overlaid are the positions of galaxy clusters from the \citep{aco89} catalogues, \citetalias{pap+11}, and the \citetalias{planck15}. We cut the sample of clusters in an attempt to avoid the edges of the image where the noise is highest. The filled circles are coloured according to their redshift. Unfilled circles are those without a measured redshift. Note the side-lobe structure of the primary beam appearing in the corners of the image. The colourmap of the redshift distribution is an implementation of \texttt{cubehelix} \citep{cubehelix}.}
\label{fig:eor0_field}
\end{figure}

Within the EoR0 field we searched for diffuse emission within a $\sim$2~Mpc radius around clusters within the following catalogues: Abell revised North, South, and Supplementary catalogues \citep[][hereafter ACO, but see also \citealt{abell}]{aco89}; the Meta-Catalogue of X-ray detected Clusters of galaxies \cite[][hereafter MCXC]{pap+11}; the \emph{Planck} catalogue of Sunyaev--Zel'dovich sources \citep[][hereafter PSZ1]{planck15} \footnote{\corrs{Note that PSZ2 was not available when this work was started.}}. Within the region encompassed by the EoR0 field, and excluding those clusters that lie too far to the edge of the image, this constitutes 668 unique clusters, 505 unique to ACO, 70 unique to MCXC, and 19 unique to PSZ1, with 24 clusters present in all three catalogues. Fig.~\ref{fig:eor0_field} shows the distribution of ACO, PSZ1, and MCXC clusters within the EoR0 field, coloured by redshift where available. \par
All clusters are checked systematically for diffuse cluster emission except 217 clusters in the ACO catalogue without a redshift. For clusters without a redshift we are unable to determine the projected linear distance from the cluster centre, which makes determining if emission is part of the cluster difficult if not at the centre. Whilst this does not pose much problem for haloes, we also consider that ACO clusters without a redshift are unlikely to have auxiliary data in the form of cluster mass, X-ray luminosities, or information on cluster members. Further, cluster emission serendipitously found in clusters not part of the aforementioned catalogues is investigated when noticed.

\subsection{Source detection and measurement}

\subsubsection{Manual source-finding: eyeballing galaxy clusters}

\begin{table*}[!t]
\centering
\begin{threeparttable}
\caption{Existing sky surveys used as auxiliary data to the EoR0 field. \label{tab:surveys}}
\begin{tabular}[width=\textwidth]{l c c c c} \toprule
Survey & Frequency & Declination & Resolution \tnote{a} & $\sigma_{\rm{rms}}$  \\
{} &~MHz & (J2000, \degr) & (arcsec $\times$ arcsec) & (\mjybeam) \\\midrule
EoR0 field & 168 & {$-44\degr41\arcmin24\arcsec \leq \delta \leq -08\degr36\arcmin36\arcsec$}  & 138 $\times$ 138 & $\gtrsim 2.3$  \\
NVSS & 1400 & $\geq -40$ & $45 \times 45$ & $\gtrsim 0.45$ \\
SUMSS & 843 & $\leq -30$ & $\sim 2.2 \left(45 \times 45 \right)$ & $\gtrsim 2$ \\
TGSS & 147.5 & $\gtrsim -53$ &  $\sim 1.5 \left(25 \times 25 \right)$ & $\sim 3.5$ \\
VLSSr & 74 & $\geq -30$ & 75 $\times$ 75 & $\gtrsim 100$ \\
\bottomrule
\end{tabular}
\begin{tablenotes}[flushleft]
\footnotesize \item[a] At $\delta_{\mathrm{J2000}} = -27\degr$.
\end{tablenotes}
\end{threeparttable}
\end{table*}

While source-finding algorithms exist and are put to good use to produce point-source catalogues, automated source-finding can miss the extended, low surface brightness haloes and relics within clusters \citep{2012PASA...29..309H}. Therefore the EoR0 field is searched by eye for diffuse emission. Auxiliary radio data exists in the form of the following sky surveys: the NRAO VLA Sky Survey \footnote{National Radio Astronomy Observatory Very Large Array Sky Survey} \citep[NVSS;][]{ccg+98}, the Sydney University Molonglo Sky Survey \citep[SUMSS;][]{bls99, mmb+03}, the TFIR \footnote{\corrs{Tata Institute of Fundamental Research.}} GMRT Sky Survey  \cite[alternate data release, TGSS;][]{ijmf16}, and the VLA \corrs{Low-frequency} Sky Survey redux \citep[VLSSr;][]{lcv+14}. These surveys and their salient properties are summarised in Table~\ref{tab:surveys}. Beyond radio surveys, we use the \emph{R\"{O}entgen SATellite} \citep[\emph{ROSAT};][]{tru84} All-Sky Survey \citep[RASS;][]{vab+99}, the Digitized Sky Survey (DSS2), \corrs{the first Pan-STARRS \footnote{\corrs{Panoramic Survey Telescope And Rapid Response System; \citealt{kpc+10}.}} survey--- PS1 \citep{tsl+12,cmm+16}, the Dark Energy Survey Data Release 1 \citep[DES DR1;][]{des1,des2,decam},} as well as archival \emph{Chandra} data with the Advanced CCD Imaging Spectrometer (ACIS) instrument and XMM-\emph{Newton} data with the European Photon Imaging Camera (EPIC) instrument, where available. For a small selection of clusters, we utilise deep ($> 30$~ks exposure) X-ray images from the Representative XMM-\emph{Newton} Cluster Structure Survey \citep[{\sffamily REXCESS};][]{rexcess, pcab09}.  \par
To determine the nature of detected emission, we look for the following:
\begin{enumerate}
\item[i.] high-frequency counterparts (1.4~GHz and 843~MHz),
\item[ii.] low-frequency counterparts (147.5 and 74~MHz),
\item[iii.] optical identifications, and
\item[iv.] X-ray emission coincident with centrally located radio emission.
\end{enumerate}
(i) and (ii) are used as an easy method of checking if we are looking at blended point sources. (i) gives a quick insight into the spectral index of the source, with significant high frequency emission, at least comparably to 168~MHz, a flat spectral index is present which is uncharacteristic of diffuse cluster emission. (iii) is important as cluster haloes and relics are not associated with an optically visible galaxy, though in the case of cluster haloes there is expected to be a concentration of optically visible galaxies \corrs{due to the central location in the cluster}. If an optically visible galaxy is found at the peak of the diffuse emission or between two lobes, then the likelihood is that of extended, disturbed, or otherwise normal lobes of a radio galaxy. (iv) allows us to confidently classify centrally located diffuse emission as a cluster halo or relic. In particular, \emph{Chandra} or XMM-\emph{Newton} observations are detailed enough to provide the position and any elongation of the X-ray emission relative to any centrally located diffuse radio emission. With these points forming the foundations of our search, we eyeballed the subset of clusters described in Section~\ref{priors}, followed by measurements of relevant physical properties.

\subsubsection{Local noise properties}

The EoR0 field is a large image that has greatly varying rms noise throughout. However, corners of the image feature \corrs{significant} noise due to the primary beam null. \citet{oth+16} use \texttt{bane} \footnote{\url{https://github.com/PaulHancock/Aegean/wiki/BANE}} \citep{hmg+12} to estimate noise throughout the EoR0 field. The mean noise level is calculated to be $3.2 \pm 0.6$~\mjybeam\ for the central $10^\circ$ of the image. Large-scale diffuse structure of Galactic origin is seen streaking the image which leads to non-constant background signal affecting rms noise calculations. In regions with no Galactic emission the rms can be as low as $\sim$2~\mjybeam. \corrs{Additionally, typical local rms noise values for the various survey data are provided in Table \ref{tab:surveys}.}

\subsubsection{Integrated flux densities}

% \begin{figure}[!t]
% \includegraphics[width=0.99\linewidth]{flux_difference.pdf}
% \caption{Flux density differences for sources between 0.1 and 10~Jy in a the central 5 degrees of the EoR0 field, comparing with the 162--170~MHz band of GLEAM. Note that while the cross-matching had a 10~Jy limit, all sources were below 3.5~Jy. The dashed line is a weighted least-squares fit to the data, and the solid line is the expected 1:1 fit. Note that uncertainties in the y direction are the quadrature sum of all applicable uncertainties in the data.}
% \label{fig:flux_difference}
% \end{figure}

The software that generated the EoR0 field at the time did not calculate a correct synthesized beam for the restored, stacked image. As a result, the integrated flux densities \corrs{measured directly from the image are incorrect}. We find that the integrated flux density measurements of the EoR0 field differed by a systematic factor of approximately 30\% when compared to the nearly equivalent 162--170~MHz band in the GLEAM survey which is tied to the Baars flux scale \citep{bgpw77}. \par

To scale the integrated flux densities in the EoR0 field we choose six reasonably bright ($>2$~Jy) unresolved sources, exhibiting no side-lobe structure and no blending with nearby sources. From a linear fit between the GLEAM 162--170~MHz and EoR0 168~MHz flux densities we find a factor $0.69 \pm 0.05$ to be used for calibration of measured integrated flux densities (i.e., $S_{168,\text{corrected}} = 0.69 \times S_{168}$), and this calibration is used through the remainder of this paper. \par

Flux densities of extended sources are either calculated by purpose-built \texttt{fluxtools.py} \footnote{\url{https://gist.github.com/Sunmish/198ef88e1815d9ba66c0f3ef3b18f74c}} code or by using \texttt{aegean} if sources are blended, as the aforementioned \texttt{python} code does not fit sources, and assumes each source is discrete. Both methods measure the source flux densities down to the 2.6$\sigma_{\mathrm{rms}}$ level so as to include as much real contribution from the faint sources as possible \citep[e.g.][but see also \citealt{blobcat}]{ksc+17}. We do not use \texttt{aegean} for all sources as \texttt{aegean} is intended as a point-source--finder, and will give the best results measuring such sources. Each flux measurement has an uncertainty, $\sigma_{S_\nu}$, calculated as\corrs{
\begin{equation}\label{eq:flux_unc}
\sigma_{S_\nu} = \sqrt{ {S_\nu}^2 \left({\sigma_\text{scale}}^2  + {\sigma_\text{rescale}}^2 \right) + \left( \sigma_{\mathrm{rms}}  \sqrt{N_{\mathrm{beam}}} \right)^2} \quad [\mathrm{Jy}] \, ,
\end{equation}}

where $N_{\mathrm{beam}}$ is the number of beams crossing the extended source, \corrs{$\sigma_\text{scale} = 5\%$---the flux scale error as described in Section 4.1 of \citet{oth+16}, and $\sigma_\text{rescale} = 5\%$ for the additional uncertainty in rescaling the integrated flux density measurements. The last term is the standard error given to flux density measurements of extended sources.}

\subsubsection{Spectral indices and source sizes}

Where possible a spectral index is calculated for each source \corrs{assuming the SED follows a standard power law in the relevant frequency range}. This is either done as a two-point spectral index ($\alpha = \ln\left(S_1 / S_2 \right) / \ln\left(\nu_1 / \nu_2 \right)$) or by fitting a first-order polynomial to the flux density measurements in log-log space, hence fitting a power law to the data. Over the frequency range here (74--1400~MHz) haloes and relics tend not to show any turnovers or breaks and typically do not deviate from the assumed power law except in rare instances \citep[e.g. the relic in Abell 2443 which has a break near 325~MHz reported by][]{cc11}. \corrs{Phoenices, however, can show curved spectra \citep[e.g.][]{srm+01,kd12}. For the purpose of this work (and often in the absence of more than two flux density measurements) we assume power laws model these SEDs sufficiently in the relevant frequency regime as is often the case \citep[see e.g. Abell~0013;][]{George2017}.}\par

Where appropriate, we estimate limits to flux densities. In particular, we use this for estimating 1.4~GHz and 147.5~MHz limits when 168~MHz emission has no counterpart in the NVSS or TGSS survey images, respectively. These are used then to impose limits on the spectral indices. For such sources, we estimate the source area at 168~MHz, which is a function of the MWA beam at $B_{\mathrm{maj}}\approx 2.3$ arcmin, and attempt to correct for the difference in beam sizes between the VLA (NVSS), GMRT (TGSS), and MWA (EoR0) by naively taking the ratios of $B_{\mathrm{maj}}$ and correcting the area based on this ratio. The limit is then \begin{equation}
S_{\mathrm{limit}} = \sigma_{\mathrm{rms}} f A_{168} \times \dfrac{4 \ln 2}{\pi B_{\mathrm{maj}} B_{\mathrm{min}}} \quad [\si{\jansky}] \, ,
\label{eq:limit}
\end{equation}
where $f = B_{\mathrm{maj}}/B_{\mathrm{maj,168}}$ and $A_{168}$ is the source area measured at 168~MHz. \par

A largest-angular size/scale (LAS) is provided where possible. For extended sources that are confused and blend with nearby sources, we estimate an angular size by making an assumption on how far the diffuse source has blended into any nearby point sources. The size characterisation is important to determine if the detection is truly extended. For a non-blended source to be considered extended in this work it must have an LAS that is greater than $1.5 B_{\mathrm{maj}}$, where $B_{\mathrm{maj}} \approx 2.3$ arcmin, which is approximately the expected $B_{\mathrm{maj}}$ of the EoR0 field. Finally, any measured angular scale is deconvolved from the beam size before a linear project size is calculated, and we report on the deconvolved sizes only.

\subsection{Additional Australia Telescope Compact Array observations}\label{sec:atca}

\begin{table}[t!]
\centering
\begin{threeparttable}
\caption{Details of the 2.1-GHz ATCA observations of Abell S1063.\label{tab:observations}}
\centering
\begin{tabular}{c c c c}
\toprule
Array  & Date & $t_{\mathrm{scan}}$ &  $\theta_\text{max}$ \tnote{a} \\
 {}            &   {} & (min)            & (arcmin) \\
\midrule
EW352 & 2013 Jun 18,20,22 & $224$ & 19.6 \\
6A    & 2012 Feb 3--5 & $543$ & 1.8 \\
\bottomrule
\end{tabular}
\begin{tablenotes}[flushleft]
\footnotesize
\item[a] Maximum angular scale sensitivity.
\end{tablenotes}
\end{threeparttable}
\end{table}

\begin{table}[t!]
\centering
\begin{threeparttable}
\caption{Sub-band image properties for the ATCA observations of Abell S1063. \label{tab:subbands}}
\begin{tabular}{@{~}c@{~}c@{~}c@{~}c}
\toprule
Image & $\nu_\mathrm{c}$ \tnote{a} & Restoring beam & $\sigma_{\rm{rms}}$ \\
{} & (\si{\mega\hertz}) & ($\arcsec \times \arcsec, \degr$) & (\si{\micro\jansky\per\beam}) \\
\midrule
1332 & 1384 & $9.95 \times 4.45,\ -3.1$ & 50 \\
1844 & 1873 & $7.57 \times 3.89,\ 1.5$  & 21 \\
2356 & 2349 & $6.10 \times 3.15,\ 0.0$  & 22 \\
2868 & 2811 & $5.13 \times 2.74,\ -4.0$ & 26 \\
Stacked & 2034  & $9.95 \times 4.45,\ -3.1$ &  18 \\
Stacked \tnote{b} & 2251 & $ 122.9 \times 40.1,\ -17.7$ & 360 \\
\bottomrule
\end{tabular}
\begin{tablenotes}[flushleft]
\footnotesize 
\item[a] Effective central frequency of image. 
\item[b] Stacked after tapering sub-bands with a 60 arcsec Gaussian.
\end{tablenotes}
\end{threeparttable}
\end{table}
\setcounter{ft}{0}

One cluster, Abell~S1063, had unpublished archival Australia Telescope Compact Array \citep[ATCA;][]{fbw92} observations made with the Compact Array Broadband Backend \cite[CABB;][]{cabb}. The cluster was observed in two array configurations: EW352 (Project code C2837, PI: M. Johnston-Hollitt) and 6A (Project code C2585, PI: R. Kale), and the data were retrieved from the Australia Telescope Online Archive. Table~\ref{tab:observations} summarises the properties of the observations. \par
The reduction of ATCA data follows standard procedures of continuum data reduction with \texttt{miriad}. After radio frequency inteference flagging, flux and bandpass calibration with PKS~B1934$-$638, gain and phase calibration is performed with PKS~B2326$-$477 and MRC~2117$-$614 for the EW352 and 6A configurations, respectively. Imaging is performed with the multi-frequency CLEAN task \texttt{mfclean} with a `Briggs' \citep{bri95} \verb|robust = 0.0| image weighting after splitting the data into 512~MHz subbands. Two rounds of phase-only self-calibration are performed on each subband independently. An additional stacked image is made for the full 2-GHz bandwidth, and one final tapered, stacked image is made. Table~\ref{tab:subbands} summarises the image properties.

\section{Results}\label{sec:results}

\subsection{Diffuse cluster emission at 168 MHz}

\begin{table*}
\centering
\begin{threeparttable}
\caption{Select physical properties of clusters found to host diffuse emission.\label{table:clusters}}
\begin{tabular}{l c c c c c c}
\toprule
Cluster name & $\alpha_{\mathrm{J2000}}$ & $\delta_{\mathrm{J2000}}$ & $z$ & $M_{500}$ & $L_{\mathrm{X},500}$ & References \\
             &                   &                   &  & ($\times 10^{14}$~M$_\odot$) & ($\times 10^{44}$~\si{\erg\per\s}) & \\
\midrule
\hyperlink{link:a13}{Abell 0013} & 00$^{\mathrm{h}}$13$^{\mathrm{m}}$38\fs3 & $-$19\degr30\arcmin07\arcsec & 0.0940 & $2.79^{+0.36}_{-0.38}$ & 1.236 & \tabft{\ref{aco_table}}/\tabft{\ref{mcxc_table}}/\tabft{\ref{psz1_table}}/\tabft{\ref{mcxc_table}} \\
\hyperlink{link:a22}{Abell 0022} & 00$^{\mathrm{h}}$20$^{\mathrm{m}}$42\fs8 & $-$25\degr42\arcmin37\arcsec & 0.1424  & $4.56^{+0.42}_{-0.44}$ & 2.872 & \tabft{\ref{aco_table}}/\tabft{\ref{pse+06_table}}/\tabft{\ref{psz1_table}}/\tabft{\ref{mcxc_table}} \\
\hyperlink{link:a33}{Abell~0033} & 00$^{\mathrm{h}}$27$^{\mathrm{m}}$07\fs0 & $-$19\degr30\arcmin24\arcsec & 0.2395 & -                      & -     & \tabft{\ref{aco_table}}/\tabft{\ref{wh13_table}}/-/- \\
\hyperlink{link:a85}{Abell 0085} & 00$^{\mathrm{h}}$41$^{\mathrm{m}}$50\fs1 & $-$09\degr18\arcmin06\arcsec & 0.0551 & $4.90^{+0.21}_{-0.22}$ & 5.100 & \tabft{\ref{aco_table}}/\tabft{\ref{oh01_table}}/\tabft{\ref{psz1_table}}/\tabft{\ref{mcxc_table}} \\
\hyperlink{link:a22}{Abell 0122} & 00$^{\mathrm{h}}$57$^{\mathrm{m}}$24\fs7 & $-$26\degr16\arcmin50\arcsec & 0.1135 & 1.727                  & 0.861 & \tabft{\ref{aco_table}}/\tabft{\ref{zgz06_table}}/\tabft{\ref{mcxc_table}}/\tabft{\ref{mcxc_table}} \\
\hyperlink{link:a133}{Abell 0133} & 01$^{\mathrm{h}}$02$^{\mathrm{m}}$42\fs1 & $-$21\degr52\arcmin25\arcsec & 0.0562 & $3.08^{+0.23}_{-0.24}$ & 1.460 & \tabft{\ref{aco_table}}/\tabft{\ref{wqi97_table}}/\tabft{\ref{psz1_table}}/\tabft{\ref{mcxc_table}} \\
\hyperlink{link:a141}{Abell 0141} & 01$^{\mathrm{h}}$05$^{\mathrm{m}}$34\fs8 & $-$24\degr39\arcmin16\arcsec & 0.230  & $4.48^{+0.66}_{-0.73}$ & 5.161 & \tabft{\ref{aco_table}}/\tabft{\ref{sr99_table}}/\tabft{\ref{psz1_table}}/\tabft{\ref{mcxc_table}} \\
\hyperlink{link:a2496}{Abell 2496} & 22$^{\mathrm{h}}$51$^{\mathrm{m}}$00\fs6 & $-$16\degr24\arcmin24\arcsec & 0.1221 & $2.98^{+0.41}_{-0.44}$ & 2.031 & \tabft{\ref{aco_table}}/\tabft{\ref{mcxc_table}}/\tabft{\ref{psz1_table}}/\tabft{\ref{mcxc_table}} \\
\hyperlink{link:a2556}{Abell 2554} & 23$^{\mathrm{h}}$12$^{\mathrm{m}}$20\fs7 & $-$21\degr30\arcmin02\arcsec & 0.1108 & $3.05^{+0.37}_{-0.39}$ & 1.431 & \tabft{\ref{aco_table}}/\tabft{\ref{cmkw02_table}}/\tabft{\ref{psz1_table}}/\tabft{\ref{mcxc_table}} \\
\hyperlink{link:a2556}{Abell 2556} & 23$^{\mathrm{h}}$13$^{\mathrm{m}}$00\fs9 & $-$21\degr37\arcmin54\arcsec & 0.0871 & 2.476                  & 1.509 & \tabft{\ref{aco_table}}/\tabft{\ref{cmkw02_table}}/\tabft{\ref{mcxc_table}}/\tabft{\ref{mcxc_table}} \\
\hyperlink{link:a2680}{Abell 2680} & 23$^{\mathrm{h}}$56$^{\mathrm{m}}$28\fs3 & $-$21\degr02\arcmin17\arcsec & 0.1771 & - & -                          & \tabft{\ref{aco_table}}/\tabft{\ref{wh13_table}}/-/- \\
\hyperlink{link:a2693}{Abell 2693} & 00$^{\mathrm{h}}$02$^{\mathrm{m}}$09\fs6 & $-$19\degr33\arcmin17\arcsec & 0.173  & - & -                          & \tabft{\ref{aco_table}}/\tabft{\ref{cac+09_table}}/-/- \\
\hyperlink{link:a2721}{Abell 2721} & 00$^{\mathrm{h}}$06$^{\mathrm{m}}$03\fs0 & $-$34\degr43\arcmin27\arcsec & 0.1144 & $3.77^{+0.35}_{-0.37}$ & 1.810 & \tabft{\ref{aco_table}}/\tabft{\ref{zgz06_table}}/\tabft{\ref{psz1_table}}/\tabft{\ref{mcxc_table}} \\
\hyperlink{link:a2744}{Abell 2744} & 00$^{\mathrm{h}}$14$^{\mathrm{m}}$18\fs9 & $-$30\degr23\arcmin21\arcsec & 0.3066 & $9.56^{+0.49}_{-0.51}$ & 11.818 & \tabft{\ref{aco_table}}/\tabft{\ref{mcxc_table}}/\tabft{\ref{psz1_table}}/\tabft{\ref{mcxc_table}} \\
\hyperlink{sec:a2751}{Abell 2751} & 00$^{\mathrm{h}}$16$^{\mathrm{m}}$19\fs8 & $-$31\degr21\arcmin55\arcsec & 0.107  &  1.261                  & 0.495 & \tabft{\ref{aco_table}}/\tabft{\ref{sr99_table}}/\tabft{\ref{mcxc_table}}/\tabft{\ref{mcxc_table}}  \\
\hyperlink{link:a2751}{APMCC 039}  & 00$^{\mathrm{h}}$17$^{\mathrm{m}}$37\fs6 & $-$31\degr28\arcmin14\arcsec & 0.082  & -                  & -     & \tabft{\ref{dmse97_table}}/\tabft{\ref{dmse97_table}}/-/- \\
\hyperlink{link:a2798}{Abell 2798} & 00$^{\mathrm{h}}$37$^{\mathrm{m}}$27\fs0 & $-$28\degr31\arcmin52\arcsec & 0.105  & 1.315                  & 0.546 & \tabft{\ref{aco_table}}/\tabft{\ref{sr99_table}}/\tabft{\ref{mcxc_table}}/\tabft{\ref{mcxc_table}} \\
\hyperlink{link:a2811}{Abell 2811} & 00$^{\mathrm{h}}$42$^{\mathrm{m}}$08\fs7 & $-$28\degr32\arcmin08\arcsec & 0.1079 & $3.67^{+0.35}_{-0.37}$ & 2.734 & \tabft{\ref{aco_table}}/\tabft{\ref{zgz06_table}}/\tabft{\ref{psz1_table}}/\tabft{\ref{mcxc_table}} \\
\hyperlink{link:a4038}{Abell 4038} & 23$^{\mathrm{h}}$47$^{\mathrm{m}}$43\fs2 & $-$28\degr08\arcmin29\arcsec & 0.0282 & 2.038                  & 1.030 & \tabft{\ref{aco_table}}/\tabft{\ref{sfs11_table}}/\tabft{\ref{mcxc_table}}/\tabft{\ref{mcxc_table}} \\
\hyperlink{link:as84}{Abell S0084} & 00$^{\mathrm{h}}$49$^{\mathrm{m}}$24\fs0 & $-$29\degr31\arcmin27\arcsec & 0.1080 & 2.368                 & 1.438 & \tabft{\ref{aco_table}}/\tabft{\ref{zgz06_table}}/\tabft{\ref{mcxc_table}}/\tabft{\ref{mcxc_table}} \\
\hyperlink{link:as1099}{Abell S1099} & 23$^{\mathrm{h}}$13$^{\mathrm{m}}$15\fs7 & $-$23\degr08\arcmin39\arcsec & 0.1104  & -                 & -     & \tabft{\ref{aco_table}}/\tabft{\ref{cmkw02_table}}/-/- \\
\hyperlink{link:as1121}{Abell S1121} & 23$^{\mathrm{h}}$25$^{\mathrm{m}}$13\fs0 & $-$41\degr12\arcmin29\arcsec & 0.3580 & $7.05^{+0.61}_{-0.60}$ & -    & \tabft{\ref{aco_table}}/\tabft{\ref{lhd+15_table}}/\tabft{\ref{psz1_table}}/- \\
\hyperlink{link:as1136}{Abell S1136} & 23$^{\mathrm{h}}$36$^{\mathrm{m}}$17\fs0 & $-$31\degr36\arcmin37\arcsec & 0.0625 & 1.289                 & 0.504 & \tabft{\ref{aco_table}}/\tabft{\ref{shl+00_table}}/\tabft{\ref{mcxc_table}}/\tabft{\ref{mcxc_table}} \\
\hyperlink{link:rxc}{RXC~J2351.0$-$1954} & 23$^{\mathrm{h}}$51$^{\mathrm{m}}$01\fs4 & $-$19\degr56\arcmin42\arcsec & 0.2477 & $5.60^{+0.59}_{-0.62}$ & $4.33 \pm  0.84$ & \tabft{\ref{cb12_table}}/\tabft{\ref{psz1_table}}/\tabft{\ref{psz1_table}}/\tabft{\ref{cb12_table}} \\
\hyperlink{link:macs}{MACS~J2243.3$-$0935} & 22$^{\mathrm{h}}$43$^{\mathrm{m}}$21\fs5 & $-$09\degr35\arcmin44\arcsec & 0.447 & $10.07^{+0.58}_{-0.60}$ & 15.200 & \tabft{\ref{eeh01_table}}/\tabft{\ref{eem+10_table}}/\tabft{\ref{psz1_table}}/\tabft{\ref{mcxc_table}} \\
\hyperlink{link:whl}{PSZ1~G082.31$-$67.01} & 23$^{\mathrm{h}}$51$^{\mathrm{m}}$47\fs8 & $-$08\degr58\arcmin35\arcsec & 0.3939  & $5.90^{+0.78}_{-0.84}$  &  - & \tabft{\ref{psz1_table}}/\tabft{\ref{planck14_table}}/\tabft{\ref{psz1_table}}/- \\
\hyperlink{link:as1063}{Abell S1063} & 22$^{\mathrm{h}}$48$^{\mathrm{m}}$43\fs5 & $-$44\degr31\arcmin44\arcsec & 0.3475  & $11.41^{+0.43}_{-0.44}$ & 27.167 & \tabft{\ref{aco_table}}/\tabft{\ref{bsg+04_table}}/\tabft{\ref{psz1_table}}/\tabft{\ref{mcxc_table}} \\
\bottomrule
\end{tabular}
\begin{tablenotes}[flushleft]
\footnotesize \item References (catalogue/redshift/mass/X-ray luminosity):
\ft{aco_table}~\citetalias{aco89};
\ft{mcxc_table}~\citetalias{pap+11};
\ft{psz1_table}~\citetalias{planck15};
\ft{pse+06_table}~\citet{pse+06};
\ft{wh13_table}~\citet{wh13};
\ft{oh01_table}~\citet{oh01};
\ft{zgz06_table}~\citet{zgz06};
\ft{wqi97_table}~\citet{wqi97};
\ft{sr99_table}~\citet{sr99};
\ft{cmkw02_table}~\citet{cmkw02};
\ft{cac+09_table}~\citet{cac+09};
\ft{dmse97_table}~\citet{dmse97};
\ft{sfs11_table}~\citet{sfs11};
\ft{lhd+15_table}~\citet{lhd+15};
\ft{shl+00_table}~\citet{shl+00};
\ft{planck14_table}~\citet{planck14};
\ft{cb12_table}~\citet{cb12};
\ft{eeh01_table}~\citet{eeh01};
\ft{eem+10_table}~\citet{eem+10};
\ft{bsg+04_table}~\citet{bsg+04}.
\end{tablenotes}
\end{threeparttable}
\end{table*}
\setcounter{ft}{0}

{Here we present the cluster emission detected in the EoR0 field from the ACO, PSZ1, and MCXC catalogues. We detect 30 objects of interest, of which 29 are candidate relics, phoenices, or haloes associated with 25 clusters. The clusters found to host candidate diffuse emission are presented in Table~\ref{table:clusters} along with their physical properties}. The detection rate for such emission within the EoR0 field is $\sim$6.4\%, which on average is lower than previous surveys, (e.g. $\sim$32\%: \citealt{vgb+07,vgd+08}, $\sim$17\%: \citealt{bvc+16}, $\sim$12\%: \citealt{sjp16}), however as mentioned, previous surveys target the most massive, and X-ray luminous clusters. Included are previously detected relics in Abell 0013, Abell 0085, and Abell 2744 \citep{sr84, srm+01, gfg+01}, phoenices in Abell 0133 and Abell 4038 \citep{sr84, sr98, srm+01}, haloes in Abell 2744 and MACS~J2243.3$-$0935 \citep{gfg+01, cso+16}, as well as the large, ambiguous emission seen in Abell 0133 \citep{rcn+10}. For the purpose of distinguishing between relics and phoenices, we place a limit of 400~kpc as a maximum size of a phoenix. Where emission scale approaches this size we look at the spectral index and location, where a linear size approaching 400~kpc with ultra-steep spectral indices ($\alpha < -1.5$, \citealt{kbc+04}) and a location closer to the cluster's centre would be suggestive of phoenices rather than relics. Table~\ref{table:diffuse_sources} summarises the results of the diffuse emission search. Following this, Section~\ref{sec:individual} describes each cluster along with the diffuse emission detected within it. Images featuring optical DSS2 backgrounds are three-colour images with red, green, and blue \corrs{(RGB)} corresponding to infrared, red, and blue respectively unless otherwise stated. \corrs{Insets on figures are RGB images made with the PS1 ($z$, $i$, $r$ bands) or DES~DR1 ($i$, $r$, $g$ bands) images.} Radio contours in images increase with factors of 2 unless otherwise noted.

% \clearpage
% \pagebreak

\begin{landscape}

\begin{table}
\centering
\begin{threeparttable}
\caption{List of diffuse emission presented in this paper, in the order presented in Section~\ref{sec:individual}.\label{table:diffuse_sources}}
% \resizebox{\linewidth}{!}{
\begin{tabular}{l c c c c c c c c}
\toprule
Cluster &  Type \tnote{a} & New & $\alpha_{\mathrm{J2000}}$ \tnote{b} & $\delta_{\mathrm{J2000}}$ \tnote{b} & $S_{168}$ & $\alpha$ & LAS \tnote{l} & LLS \tnote{l} \\
{} & {} &  {} & {} & {} & (mJy) & {} & (arcmin) & (kpc)\\
\midrule
\hyperlink{link:a13}{Abell 0013} \tnote{c}  & R & $\times$ & 00$^{\mathrm{h}}$13$^{\mathrm{m}}$28\fs8 & $-$19\degr30\arcmin00\arcsec & $1850 \pm 130$ & $\alpha_{168}^{1400} = -1.96 \pm 0.08$ & 5.9 & 610 \\
\hyperlink{link:a22}{Abell 0022} &  cR or cH & $\checkmark$ & 00$^{\mathrm{h}}$20$^{\mathrm{m}}$38\fs4 & $-$25\degr39\arcmin36\arcsec & - & - & - & - \\
\hyperlink{link:a33}{Abell~0033} &  cR of RG & $\checkmark$ & 00$^{\mathrm{h}}$27$^{\mathrm{m}}$33\fs6 & $-$19\degr32\arcmin24\arcsec & $26 \pm 5$ & $\alpha_{168}^{1400} \leq -0.4(\pm 0.1)$ & 6.3 & 1400 \tnote{d} \\
\hyperlink{link:a85}{Abell 0085} \tnote{c}  & P & $\times$ & 00$^{\mathrm{h}}$41$^{\mathrm{m}}$31\fs2 & $-$09\degr22\arcmin12\arcsec & $9390 \pm 960$ & $\alpha_{147.5}^{300} = -1.85 \pm 0.03$ & 6.7 & 430 \\
\hyperlink{link:a122}{Abell 0122}  & cmH or RG & $\checkmark$ & 00$^{\mathrm{h}}$57$^{\mathrm{m}}$24\fs0 & $-$26\degr17\arcmin24\arcsec & $329 \pm 25$ & $\alpha_{168}^{1400} \leq -1.52 \pm 0.04$ & 4.3 & 530 \\
\multirow{2}*{\hyperlink{link:a133}{Abell 0133}}  & P & $\times$ & 01$^{\mathrm{h}}$02$^{\mathrm{m}}$40\fs8 & $-$21\degr52\arcmin12\arcsec & - & - & - & -\\
 & R or RG & $\times$ & - & - & - & - & - & - \\
\hyperlink{link:a141}{Abell 0141} & H & $\checkmark$ & 01$^{\mathrm{h}}$05$^{\mathrm{m}}$33\fs6 & $-$24\degr38\arcmin24\arcsec & $110 \pm 11$ & $\alpha_{168}^{610} \leq -2.1 \pm 0.1$ & 5.0 & 1100 \\
\hyperlink{link:a2496}{Abell 2496} & cR or cH & $\checkmark$ & 22$^{\mathrm{h}}$50$^{\mathrm{m}}$52\fs8 & $-$16\degr26\arcmin60\arcsec & $561 \pm 42$ & $\alpha_{74}^{1400}= -1.26 \pm 0.02$ & $\sim$4.2 & $\sim$560 \\
\hyperlink{link:a2556}{Abell 2556} & cP & $\checkmark$ & 23$^{\mathrm{h}}$13$^{\mathrm{m}}$12\fs0 & $-$21\degr28\arcmin12\arcsec & $29.3 \pm 5.5$ & $\alpha_{168}^{1400}= -1.22 \pm 0.14$ & 2.4 & 240 \\
\hyperlink{link:a2680}{Abell 2680}  & cH & $\checkmark$ & 23$^{\mathrm{h}}$56$^{\mathrm{m}}$31\fs2 & $-$21\degr02\arcmin24\arcsec & $23 \pm 8$ & $\alpha_{168}^{1400} \leq -1.2 \pm 0.2$ & $\sim$2.2 & $\sim$400\\
\hyperlink{link:a2693}{Abell 2693} & cH & $\checkmark$ & 00$^{\mathrm{h}}$02$^{\mathrm{m}}$14\fs4 & $-$19\degr33\arcmin00\arcsec & $50 \pm 6$ & $\alpha_{168}^{1400} \leq -0.88 \pm 0.06$ & 3.0 & 530 \\
\hyperlink{link:a2721}{Abell 2721} & cR or cH & $\checkmark$ & 00$^{\mathrm{h}}$06$^{\mathrm{m}}$14\fs4 & $-$34\degr43\arcmin48\arcsec & $54 \pm 14$ & $\alpha_{168}^{1400} \leq -0.96 \pm 0.12$ & 4.0 & 500 \\
\multirow{2}*{\hyperlink{link:a2744}{Abell 2744} \tnote{e}} & H & $\times$ & 00$^{\mathrm{h}}$14$^{\mathrm{m}}$19\fs2 & $-$30\degr23\arcmin24\arcsec & $550 \pm 51$ & $\alpha_{168}^{1400} = -1.11 \pm 0.04$ & $\sim$6.9 & $\sim$1900\\
& R & $\times$ & 00$^{\mathrm{h}}$14$^{\mathrm{m}}$38\fs4 & $-$30\degr19\arcmin48\arcsec & $237 \pm 24$ & $\alpha_{168}^{1400} = -1.19 \pm 0.05$ & $\sim$5.2 & $\sim$1400 \\
\hyperlink{link:a2751}{Abell 2751}  & R & $\checkmark$ & 00$^{\mathrm{h}}$16$^{\mathrm{m}}$55\fs2 & $-$31\degr23\arcmin24\arcsec & $323 \pm 62$ & $\alpha_{168}^{1400} = -1.27 \pm 0.11$ & $\sim$8.7 & $\sim$1000 \\
\hyperlink{link:a2751}{APMCC 039}  & cR or RG & $\checkmark$ &  00$^{\mathrm{h}}$17$^{\mathrm{m}}$50\fs4 & $-$31\degr18\arcmin36\arcsec & $60 \pm 8$ & $-1.3(\pm0.1) \leq \alpha_{168}^{1400} \leq -0.4(\pm0.1)$ & 8.5 & 1000 \tnote{f}\\
\hyperlink{link:a2798}{Abell 2798}  & cR & $\checkmark$ & 00$^{\mathrm{h}}$37$^{\mathrm{m}}$38\fs4 & $-$28\degr30\arcmin36\arcsec & $110 \pm 9$ & $\alpha_{168}^{1400} = -1.2 \pm 0.1$ & 4.2 & 490\\
\hyperlink{link:a2811}{Abell 2811} & cH or mH & $\checkmark$ & 00$^{\mathrm{h}}$42$^{\mathrm{m}}$09\fs6 & $-$28\degr31\arcmin48\arcsec & $81 \pm 17$ & $\alpha_{168}^{1400} \leq -1.5 \pm 0.1$ & $\sim$3.4 & $\sim$400\\
\hyperlink{link:a4038}{Abell 4038} \tnote{c} & P & $\times$ & 23$^{\mathrm{h}}$47$^{\mathrm{m}}$40\fs8 & $-$28\degr09\arcmin00\arcsec & $4790 \pm 250$ & - & - & - \\
\hyperlink{link:as84}{Abell S0084} & cH & $\checkmark$ & 00$^{\mathrm{h}}$49$^{\mathrm{m}}$19\fs2 & $-$29\degr30\arcmin36\arcsec & $32 \pm 5$ & $\alpha_{168}^{1400} \leq -1.3 \pm 0.1$ & 3.5 & 420 \\
\hyperlink{link:as1099}{Abell S1099}  & U & $\checkmark$ & 23$^{\mathrm{h}}$13$^{\mathrm{m}}$04\fs8 & $-$23\degr08\arcmin24\arcsec & $180 \pm 20$ & $\alpha_{168}^{1400} = -1.0 \pm 0.2$ & $\sim$9.5 & $\sim$1100 \\
\hyperlink{link:as1121}{Abell S1121} & H or R & $\checkmark$ & 23$^{\mathrm{h}}$25$^{\mathrm{m}}$14\fs4 & $-$41\degr12\arcmin36\arcsec & $80 \pm 13$ & $\alpha_{168}^{843} = -1.2 \pm 0.2$ & $\sim$3.6 & $\sim$1100 \tnote{g}\\
\hyperlink{link:as1136}{Abell S1136} & U & $\checkmark$ & 23$^{\mathrm{h}}$36$^{\mathrm{m}}$19\fs2 & $-$31\degr36\arcmin36\arcsec & $586 \pm 46$ & - & $\sim$6.8 & $\sim$490\\
\multirow{3}{*}{\hyperlink{link:rxc}{RXC~J2351.0$-$1954}} & cH & $\checkmark$ & 23$^{\mathrm{h}}$51$^{\mathrm{m}}$02\fs4 & $-$19\degr56\arcmin24\arcsec & $ 87 \pm 17$ & $\alpha_{168}^{1400} \leq -1.4 \pm 0.1$ & $\sim$1.6 & $\sim$370\\
& cR (A) & $\checkmark$ & 23$^{\mathrm{h}}$51$^{\mathrm{m}}$28\fs8 & $-$19\degr59\arcmin24\arcsec & $57 \pm 9$ & $ \alpha_{168}^{1400} \leq  -1.2 \pm 0.1$ & 5.8 & 1400 \\
 & cR (B) & $\checkmark$ & 23$^{\mathrm{h}}$50$^{\mathrm{m}}$21\fs6 & $-$19\degr48\arcmin36\arcsec & $147 \pm 13$ & $ \alpha_{168}^{1400} \leq  -1.68 \pm 0.04$ & 5.4 & 1300 \\
\hyperlink{link:macs}{MACS~J2243.3$-$0935}  \tnote{h} & H & $\times$ &  22$^{\mathrm{h}}$43$^{\mathrm{m}}$26\fs4 & $-$09\degr35\arcmin24\arcsec & $80 \pm 40$  & $\alpha_{168}^{610} = -1.6 \pm 0.4$ & - & - \\
\hyperlink{link:whl}{GMBCG~J357.91841$-$08.97978} \tnote{j}& cH & $\checkmark$ & 23$^{\mathrm{h}}$51$^{\mathrm{m}}$38\fs4 & $-$08\degr58\arcmin48\arcsec & $128 \pm 20$ & $\alpha_{168}^{1400} = -1.62 \pm 0.10$ & 3.2 & 1000 \\
\hyperlink{link:as1063}{Abell S1063} & H and RG \tnote{k} & $\checkmark$ & 22$^{\mathrm{h}}$48$^{\mathrm{m}}$45\fs6 & $-$44\degr30\arcmin36\arcsec & $265 \pm 38$ & $\alpha_{168}^{843} = -1.36 \pm 0.11$ & - & - \\
\bottomrule
\end{tabular}
\begin{tablenotes}[flushleft]
\footnotesize \item[a] Classification (H: radio halo; R: radio relic; P: radio phoenix; mH: mini-halo, RG: individual or blended [remnant] radio galaxy; U: undecided---requires further information; c: candidate object). 
\item[b] Flux-weighted average right ascension and declination of the emission, or peak flux density position if using \texttt{aegean}, or estimated central coordinates based on morphology.
\item[c] Reported by \citet{sr84}.
\item[d] Assuming a redshift of $z=0.2395$.
\item[e] Reported by \citet{gfg+01}.
\item[f] Assuming a redshift of $z=0.107$.
\item[g] Assuming a redshift of $z=0.3580$.
\item[h] Reported by \citet{cso+16}.
\item[j] We consider GMBCG~J357.91841$-$08.97978 and WHL J235151.0-0.085929 to be the same cluster.
\item[k] The emission is comprised of blended radio sources in data presented here, but \citet{Xie2020} report a radio halo.
\item[l] Size deconvolved from the 2.3~arcmin beam.
\end{tablenotes}
\end{threeparttable}
\end{table}
\setcounter{ft}{0}
\end{landscape}

\subsection{Individual galaxy clusters}\label{sec:individual}

\hypertarget{link:a13}{}\subsubsection{Abell 0013}

\begin{figure}[!t]
\centering
\includegraphics[width=1\linewidth]{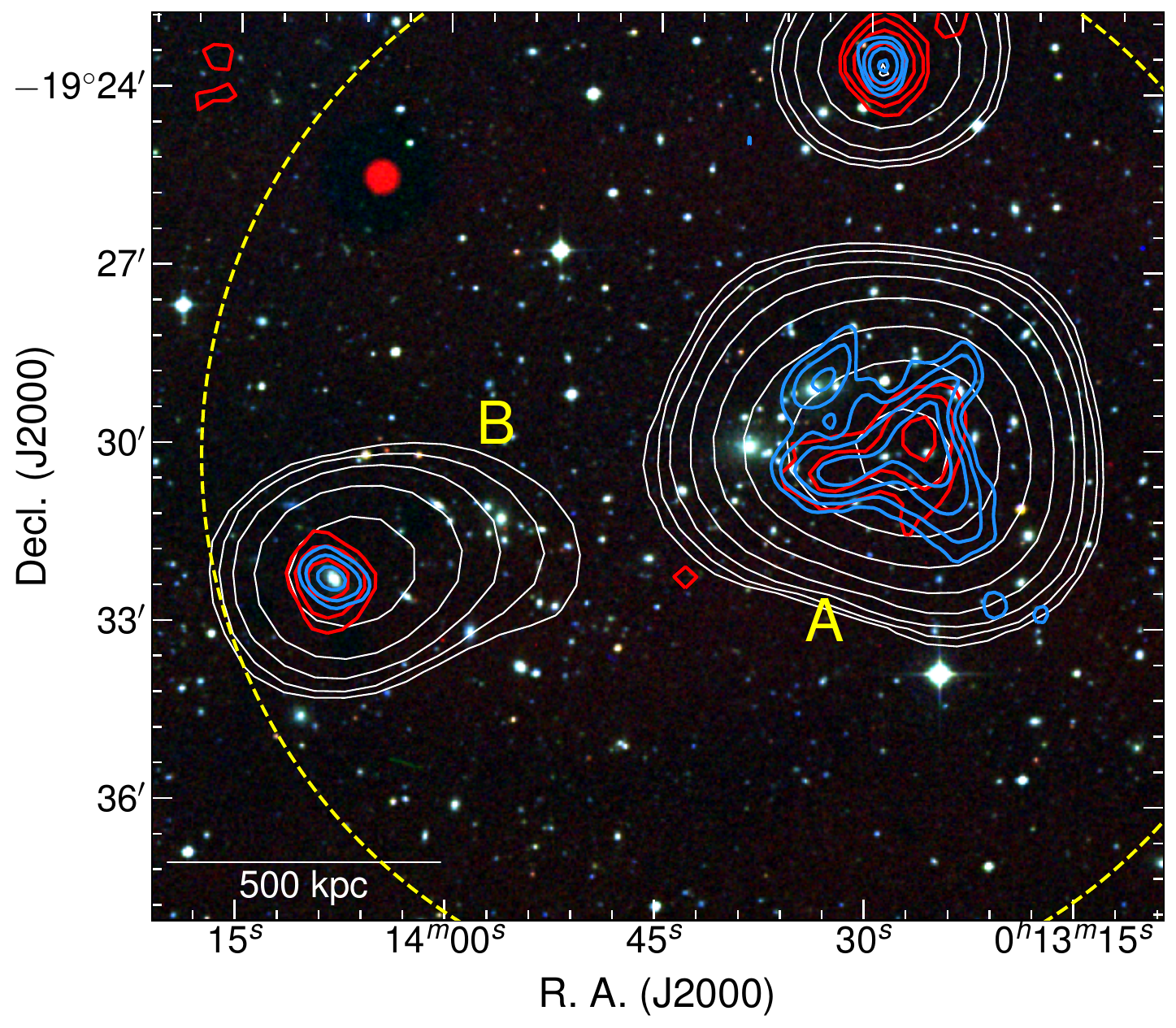}
\caption{Abell 0013. \corrs{DSS2} RGB image with contours overlaid as follows: EoR0 field, white, starting at 7~\mjybeam; NVSS, \nvsscontour, beginning at 1.5~\mjybeam; TGSS, \tgsscontour, beginning at 13.2~\mjybeam, all increasing with factors of 2. `A' marks the relic. The dashed circle has a 1~Mpc radius centered on the cluster, and the linear scale is set at the cluster's redshift.}
\label{fig:a13}
\end{figure}

{\citet{sr84} report the detection of a steep-spectrum radio phoenix in Abell~0013 at GHz frequencies with filamentary structure \citep[see also][]{srm+01,George2017}. We detect the same emission, shown as contours in Fig. \ref{fig:a13} (labelled `A'), also detected in the NVSS and TGSS surveys. We measure $S_{168} = 1.85 \pm 0.13$~Jy {with an LAS of 5.9~arcmin and largest linear scale (LLS) at the cluster's redshift of 610~kpc.} From the NVSS image we measure $S_{1400} =  28.7 \pm 4.1$~mJy, resulting in a spectral index between 168--1400~MHz of $\alpha_{168}^{1400} = -1.96 \pm 0.08$. \corrs{While classified as a radio phoenix, the SED is well-modelled by a power law in the frequency range here \citep{George2017}, though we note that the two-point index here is steeper than that reported by \citet{George2017} largely due to the NVSS measurement.} We also note the detection of additional diffuse emission in the optical subcluster to the East of the phoenix, labelled `B', connected to the nearby point source, though we cannot comment on its nature. \par

\hypertarget{link:a22}{}\subsubsection{Abell 0022}

\begin{figure*}[!t]
\includegraphics[width=0.478\linewidth]{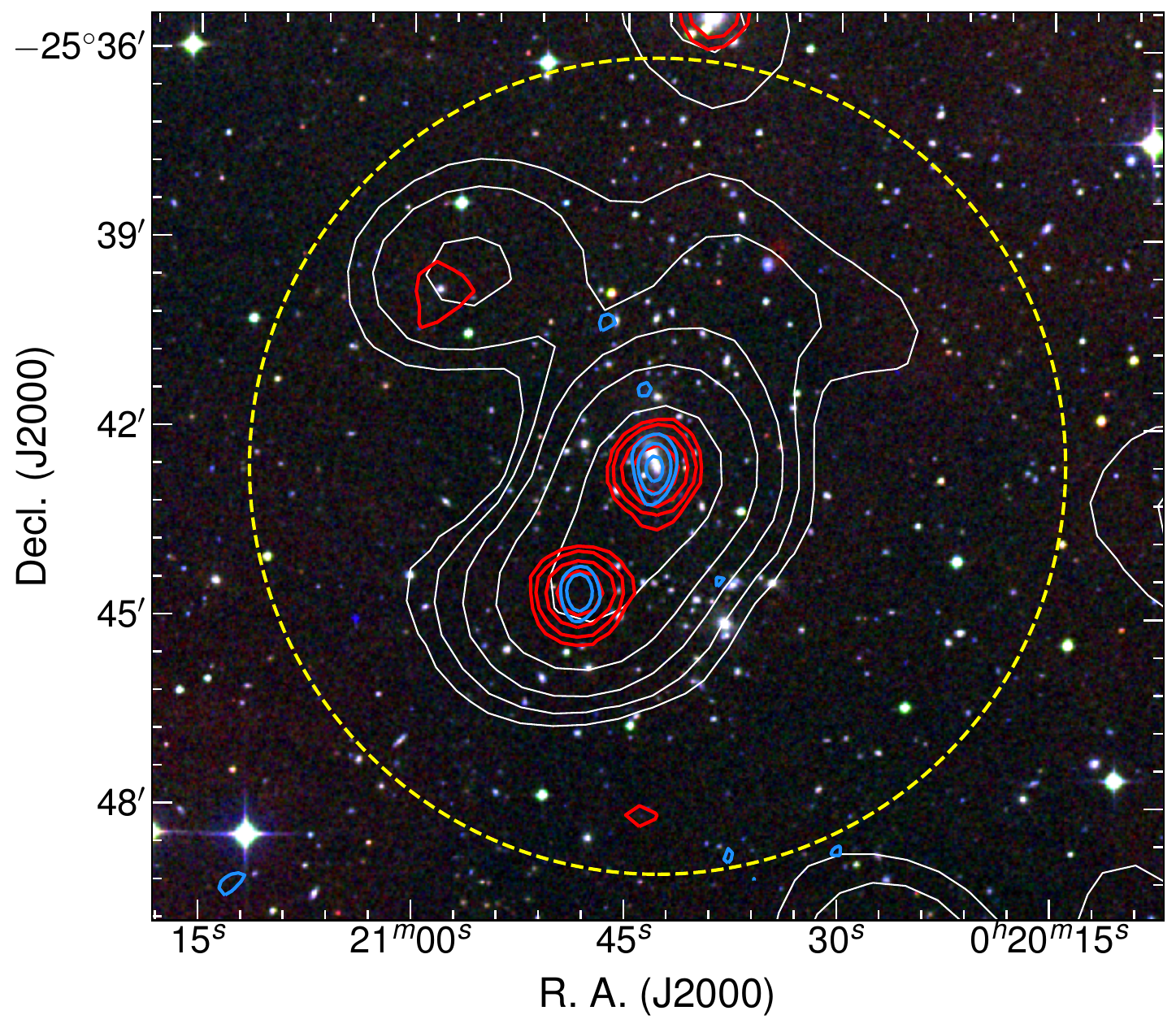}\hfill
\includegraphics[width=0.478\linewidth]{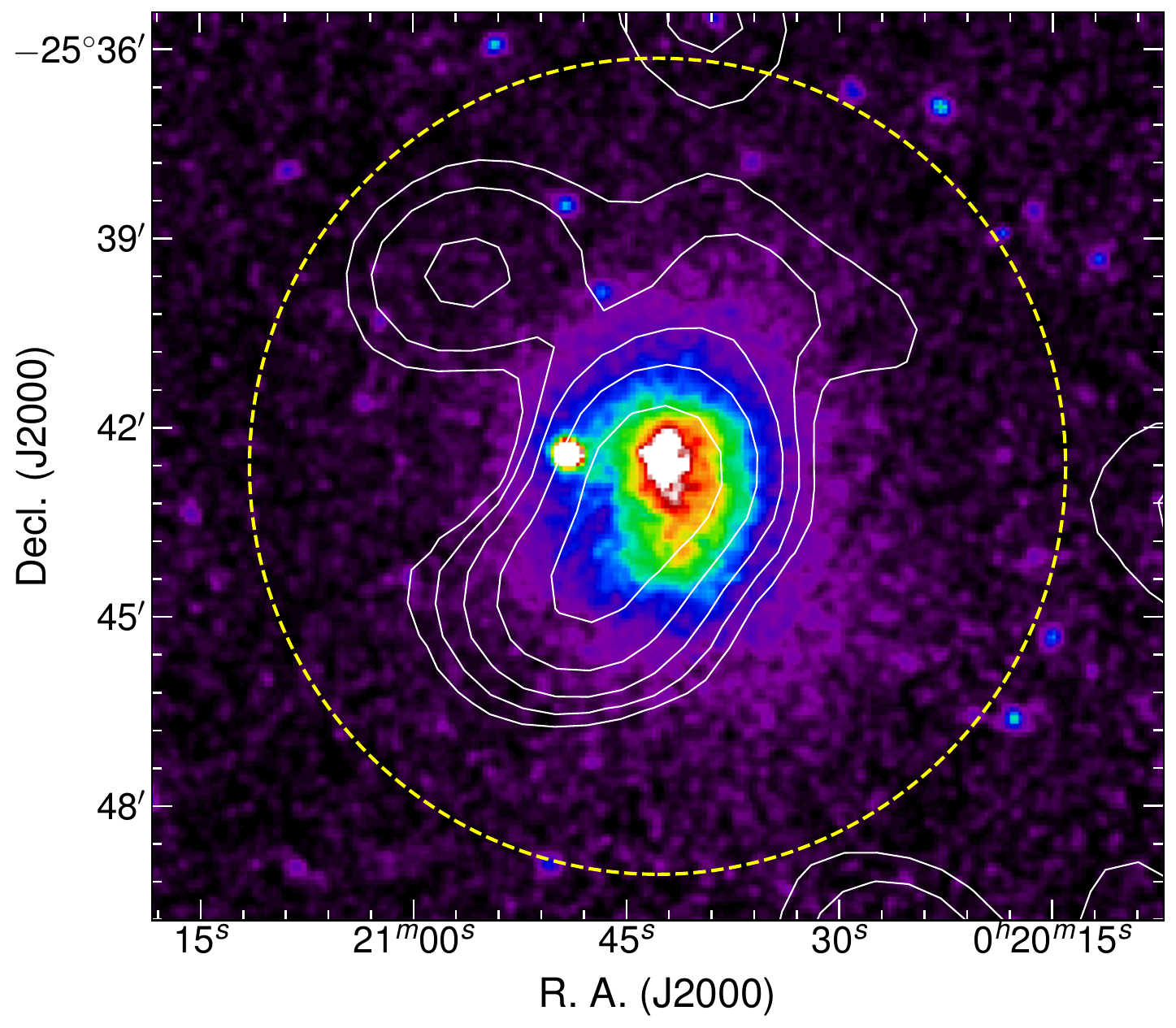}
\caption{Diffuse emission within Abell 0022. \emph{Left:} \corrs{DSS2} RGB image with contours overlaid as follows: EoR0 field, white, beginning at 7~\mjybeam; NVSS, \nvsscontour, beginning at 1.5~\mjybeam; TGSS, \tgsscontour, beginning at 21~\mjybeam. Image features as in Fig. \ref{fig:a13}. \emph{Right:} {Exposure corrected, smoothed XMM-\emph{Newton} image from the {\sffamily REXCESS} survey with EoR0 contours overlaid as in the left panel.}}
\label{fig:a22}
\end{figure*}

{Abell 0022 features extremely diffuse, faint emission that appears to permeate the cluster.} Fig.~\ref{fig:a22} shows the emission extending from the centre of the cluster northward. We see from the NVSS and TGSS data that the MWA emission is coincident with three point sources: NVSS~J002042$-$254239, associated with a member of the intervening galaxy triple DUKST 473$-$042; NVSS~J002048$-$254437; and NVSS~J002058$-$253957, emission associated with the cluster member 2MASX~J00205811$-$2539516. The MWA data extends considerably further north reminiscent of the cluster halo in Abell 3888 \citep{sjp16}, though also appears connected to a steep-spectrum source that appears point-like at the MWA resolution to the East. We do not obtain a flux density measurement for the extended emission due to complex blending of sources. \par
XMM-\emph{Newton} data is shown in the right panel of Fig.~\ref{fig:a22} (Obs. ID 0201900301, PI B\"ohringer), which were taken and reduced as part of the {\sffamily {\sffamily REXCESS}} survey \citep{rexcess,pcab09}. The 168~MHz radio emission extends far beyond the X-ray emission, however some of the emission may coincide with the discrete point sources at the cluster center. We cannot unabiguously classify this emission but suggest a higher-resolution follow-up may reveal its nature. 

\hypertarget{link:a33}{}\subsubsection{Abell~0033}

\begin{figure}[t!]
\includegraphics[width=0.99\linewidth]{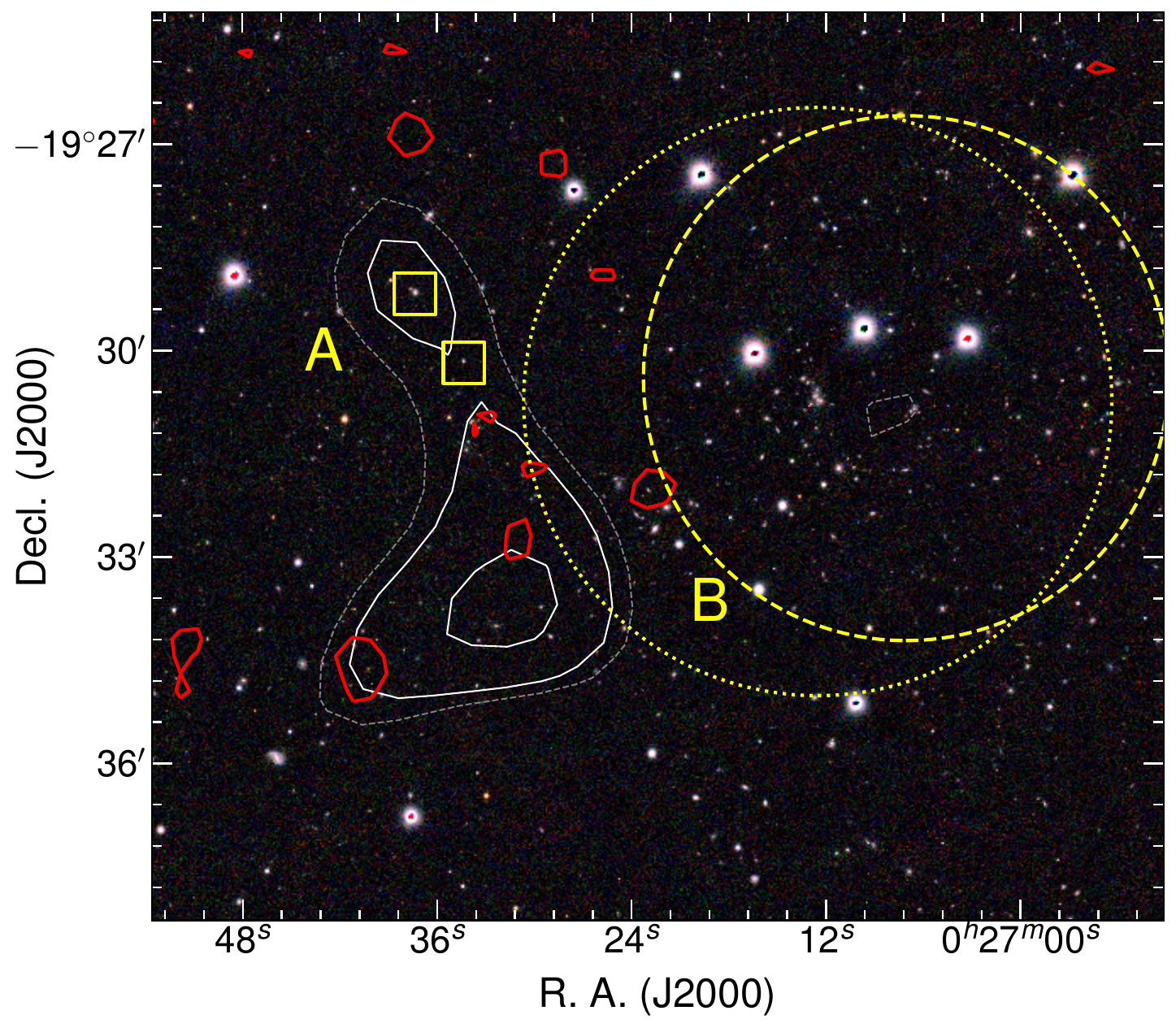}
\caption{Candidate relic on the periphery of Abell~0033. The background is \corrs{a PS1} RGB image with contours overlaid as follows: EoR0, white, 3$\sigma_{\rm{rms}}$ beginning at 6.9~\mjybeam and a single grey, dashed, 2$\sigma_{\rm{rms}}$ contour at 4.6~\mjybeam; NVSS, \nvsscontour, beginning at 1.5~\mjybeam. No TGSS emission is seen above the 3$\sigma_{\rm{rms}}$ level of 25.8~\mjybeam. The dashed circle is centre on the position of Abell~0033 and the dotted circled is centred on WHL J002712.5-193045, both with 1~Mpc radii at the reported redshifts. They are suspected to be the same cluster (see main text). The boxes indicate possible optical IDs for the diffuse emission.}
\label{fig:a33}
\end{figure}

Fig.~\ref{fig:a33} shows emission on the periphery of both Abell~0033 ($z=0.28$, photometric; \citealt{lv77}) and WHL J002712.5-193045 ($z=0.2395$, spectroscopic; \citealt{wh13}). The white circles in Fig.~\ref{fig:a33} have 1~Mpc radii about the cluster centres. The two clusters are separated by an angular distance of $\sim$80~arcsec, and given the clear concentration of optical galaxies seen in the DSS2 images, they are likely the same cluster and we hereafter consider there to be only Abell~0033 at the redshift of $z=0.2395$. The grey, dashed contour in Fig.~\ref{fig:a33} is at the 2$\sigma_{\rm{rms}}$ level to indicate the possibility of the two objects, Obj.~A and B, being a single piece of extended emission on the cluster periphery. If this is the case, the entire structure has a flux density of {$S_{168} = 26 \pm 5$~mJy, and an LAS is 6.3~arcmin which translates to an LLS of 1.4~Mpc} at $z=0.2395$. {The NVSS does not show emission within the area of the 168~MHz emission. We provide an upper limit on the 1.4~GHz flux density of $S_{1400} \leq 10$~mJy resulting in $\alpha_{168}^{1400} \leq -0.4(\pm 0.1)$}, consistent with many radio sources and does not aid in classification. Potential optical IDs are highlight on Fig. \ref{fig:a33}, though neither provide further clarification on the classification of the source. While the source shares properties with radio relics and dead radio galaxies, we recommend sensitive follow-up observations of the source to confirm its nature. \par

\hypertarget{link:a85}{}\subsubsection{Abell 0085}

\begin{figure*}[t!]
\includegraphics[width=0.478\linewidth]{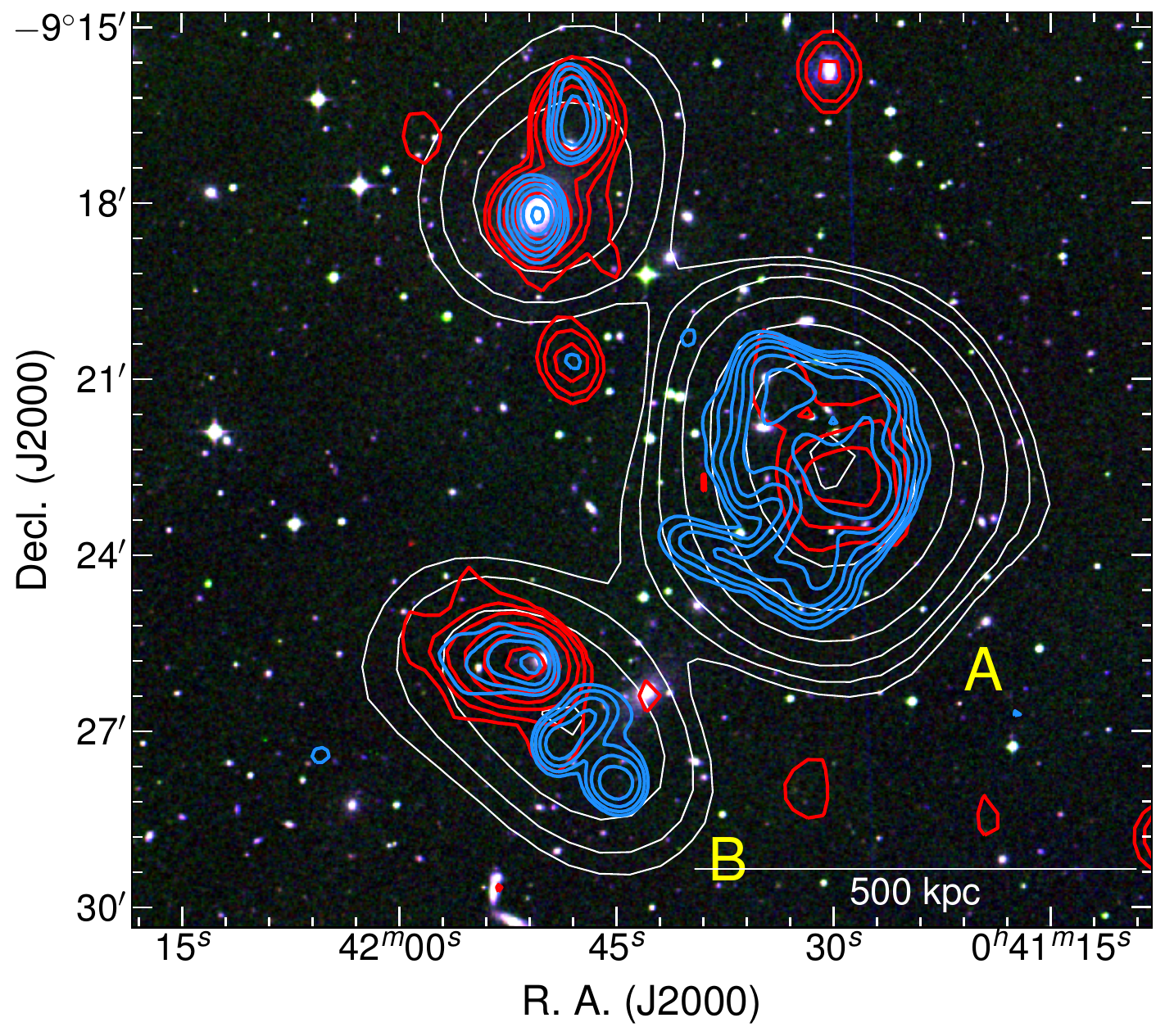}\hfill
\includegraphics[width=0.478\linewidth]{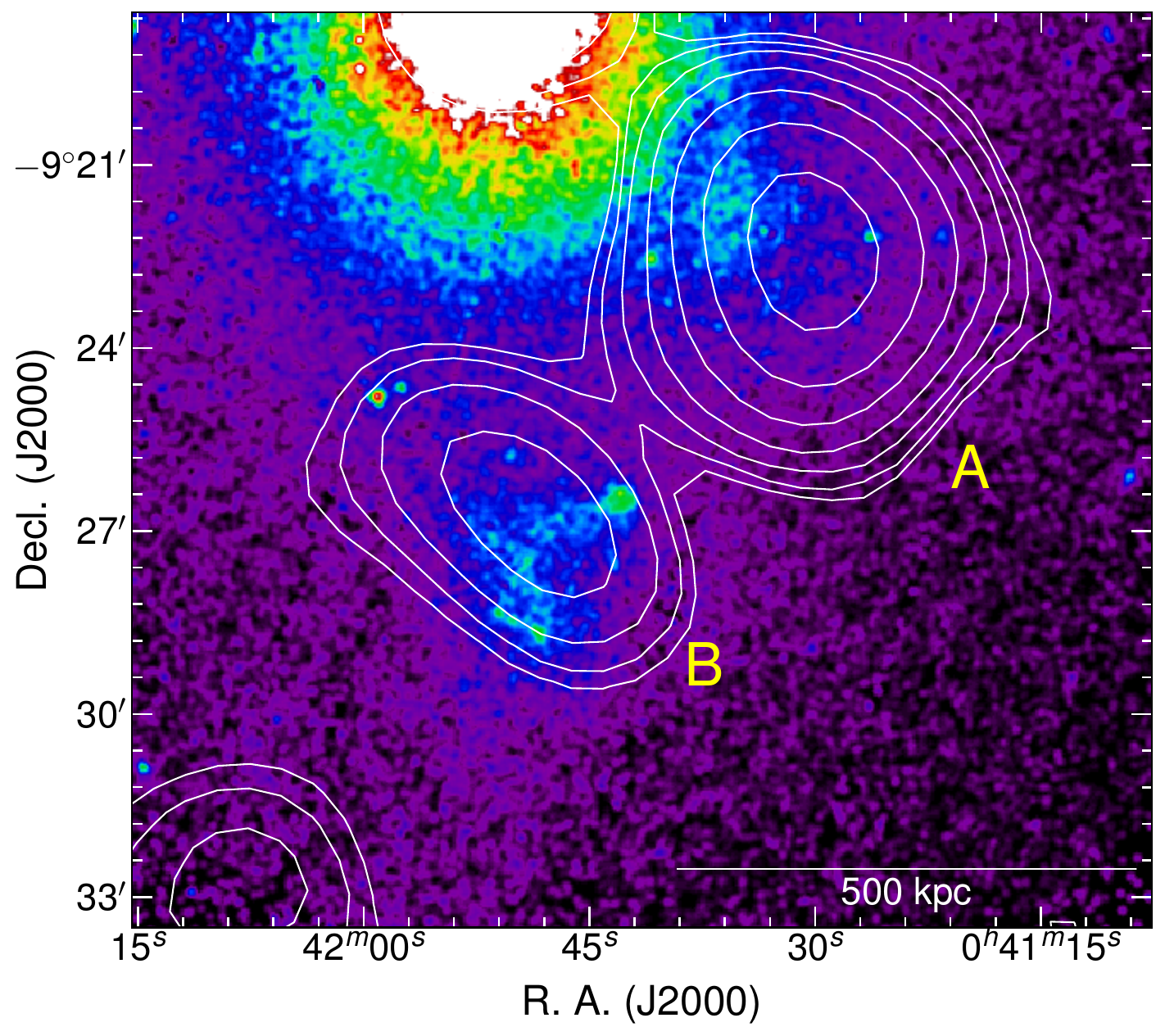}
\caption{Abell 0085. \emph{Left:} \corrs{DSS2} RGB image with contours overlaid as follows: EoR0 field, white, beginning at 49.7~\mjybeam; NVSS, \nvsscontour, beginning at 1.5~\mjybeam; TGSS, \tgsscontour, beginning at 9.6~\mjybeam. \emph{Right:} Exposure corrected, smoothed XMM-\emph{Newton} image with EoR0 contours overlaid as in the left panel. Note that the right panel has a smaller field of view and is centred to show the subcluster `A'.  Both panels show the linear scale at the cluster's redshift.}
\label{fig:a85}
\end{figure*}

{\citet{sr84} report the detection of a phoenix offset from the centre of Abell~0085}, and \citet{gf00} provide follow-up 300~MHz imaging with the VLA and ascertain an LLS for the source of 386~kpc (corrected for this cosmology). 168~MHz emission coincides with the previously detected phoenix (Obj.~A in Fig.~\ref{fig:a85}), with $S_{168} = 9.39 \pm 0.96$~Jy. The TGSS shows 147.5~MHz emission beyond that of the NVSS despite similar resolutions with an extended structure to the southeast, tracing the emission at 300~MHz shown by \citeauthor{gf00}. From the MWA, TGSS, and 300-MHz data we find $\alpha_{147.5}^{300} = -1.85 \pm 0.03$, though note that the TGSS image is likely missing flux due to resolution and missing \corrs{short} baselines, which suggests the relic may have an even steeper spectral index. \corrs{Given the small frequency range and the SED shown by \citet{srm+01}, we suggest a power law in this regime adequately models the observed data.} We measure an LAS of 6.7~arcmin (LLS of 430~kpc).
The radio source to the southeast of the relic (Obj.~B in Fig.~\ref{fig:a85}) has extended 168~MHz emission beyond the source seen in the NVSS which is likely associated with the galaxy SDSS~J004150.17-092547.4. The TGSS 147.5~MHz data shows two distinct sources within this extended, steep-spectrum emission. The right panel of Fig.~\ref{fig:a85} shows a zoomed-in view of Obj.~B, with {EoR0 field} contours overlaid on exposure corrected, smoothed XMM-\emph{Newton} data (Obs. ID 0723802201, PI de Plaa). Obj.~B features an extension to the bulk of the X-ray emitting plasma at the cluster's core. \citet{ksr02} suggest that this extension of X-ray emission, along with the complex of radio sources Obj.~B, is representative of subcluster asymmetrically merging with the main cluster of Abell 0085.

\hypertarget{link:a122}{}\subsubsection{Abell 0122}

\begin{figure*}[t!]
\includegraphics[width=0.478\linewidth]{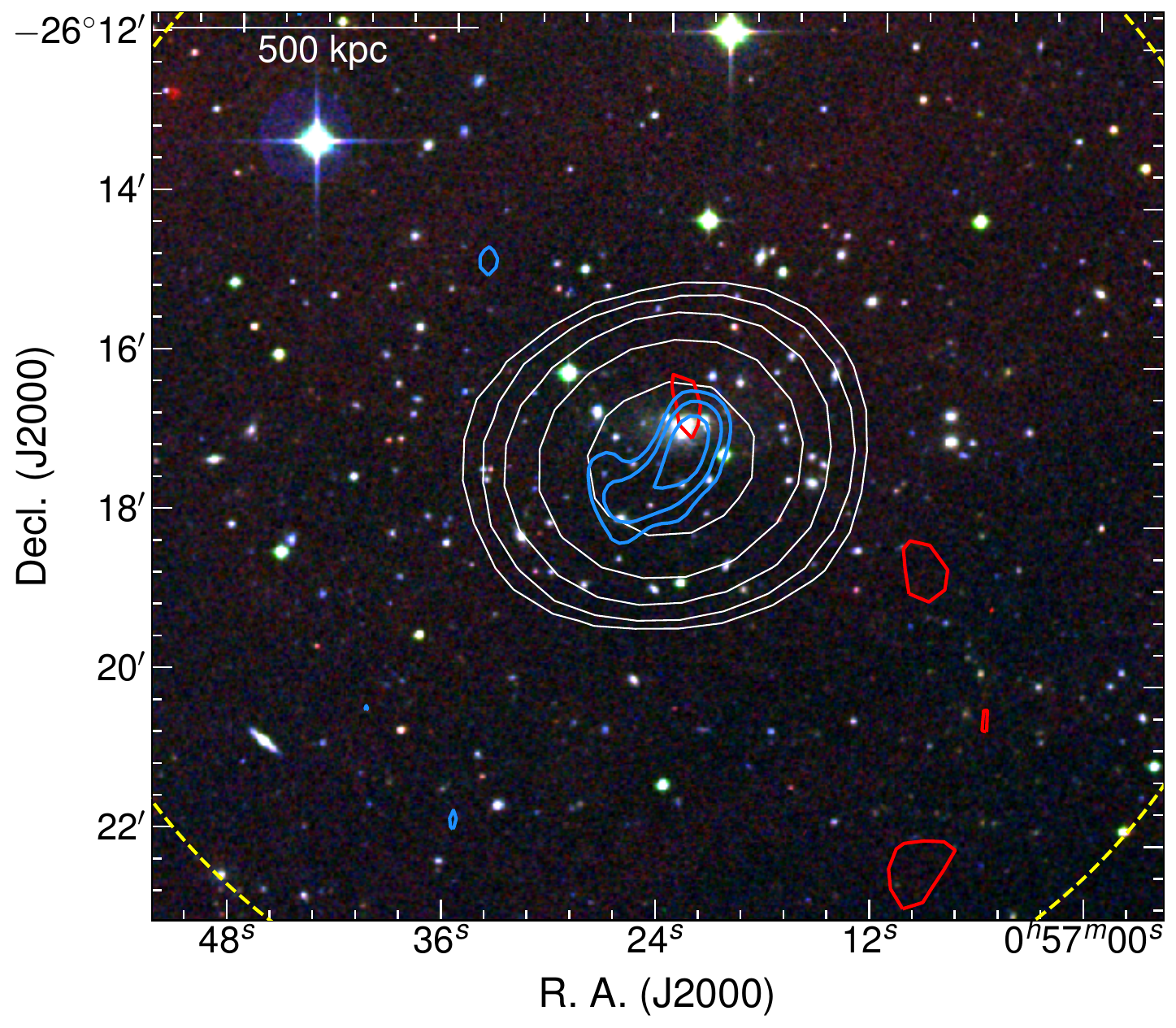}\hfill
\includegraphics[width=0.478\linewidth]{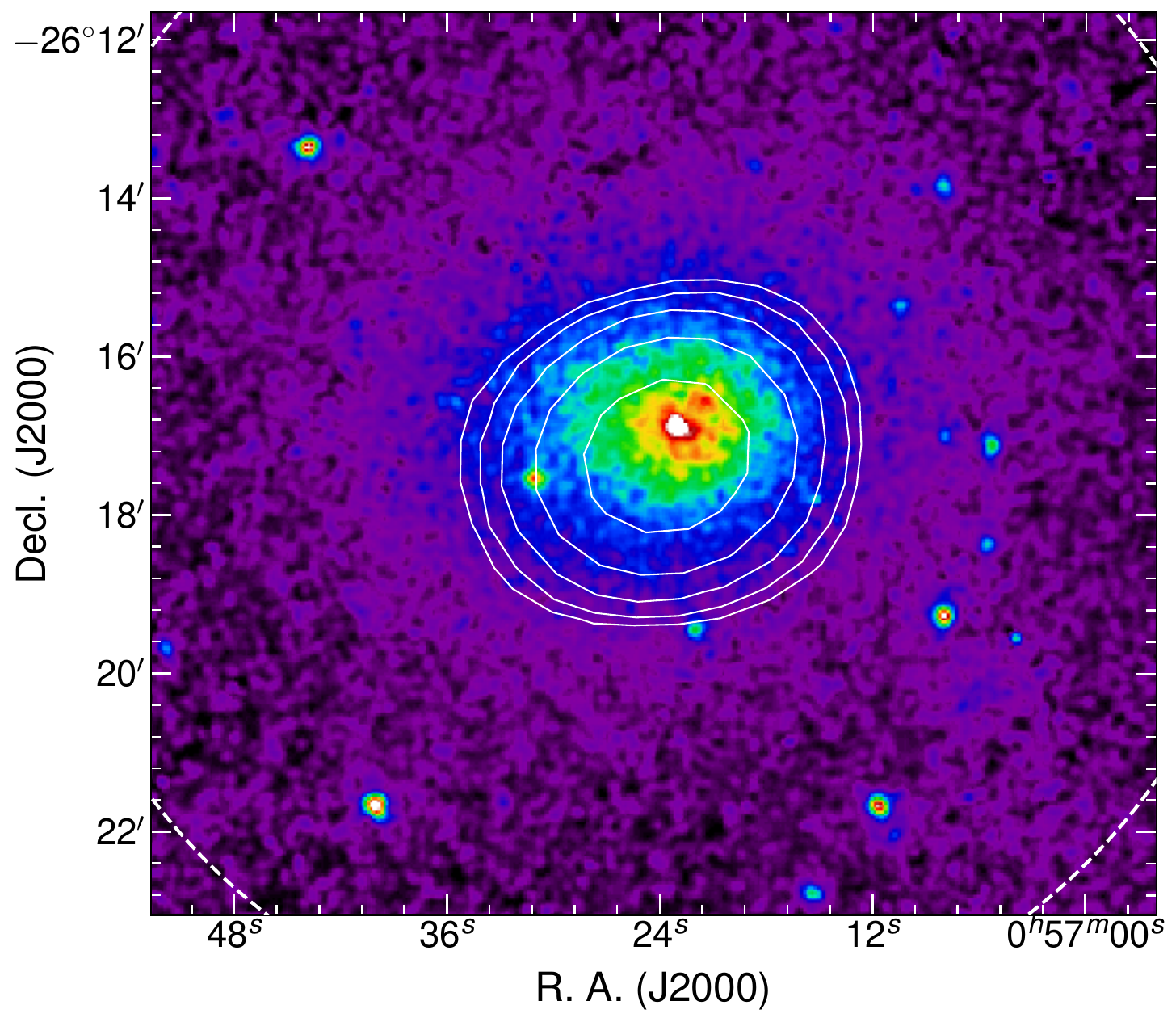}
\caption{Steep-spectrum emission at the centre of Abell 0122. \emph{Left:} \corrs{DSS2} RGB image with contours overlaid as follows: EoR0 field, white, beginning at 12~\mjybeam; NVSS, \nvsscontour, beginning at 1.5~\mjybeam; TGSS, \tgsscontour, beginning at 13.5~\mjybeam. The linear sale is at the cluster's redshift. \emph{Right:} Exposure corrected, smoothed XMM-\emph{Newton} image with EoR0 field contours overlaid as in the left panel.}
\label{fig:a122}
\end{figure*}

Abell 0122 features a strong extended source at its centre with a flux density of $S_{168} = 329 \pm 25$~mJy and LAS of 4.3~arcmin (with an LLS of 530~kpc)}. There is no significant 1.4~GHz emission seen with the NVSS {image}, though the 147.5~MHz TGSS data shows extended emission morphologically similar to a head-tail radio galaxy. We provide a 1.4~GHz flux limit of $S_{1400} \leq 13$~mJy and a corresponding spectral index of $\alpha_{168}^{1400} \leq -1.52 \pm 0.04$. \par
The right panel of Fig:~\ref{fig:a122} shows exposure corrected, smoothed XMM-\emph{Newton} data (Obs ID 0504160101, PI Sivanandam). The 168~MHz radio emission fills the X-ray plasma. Abell 0122 shows no evidence in the either X-ray emission or the optical density that would suggest the cluster is undergoing, or had undergone, a merger event and the source size points towards towards a mini-halo if not a radio galaxy, though we cannot confirm the classification here.

\hypertarget{link:a133}{}\subsubsection{Abell 0133}

\begin{figure}[t!]
\includegraphics[width=0.99\linewidth]{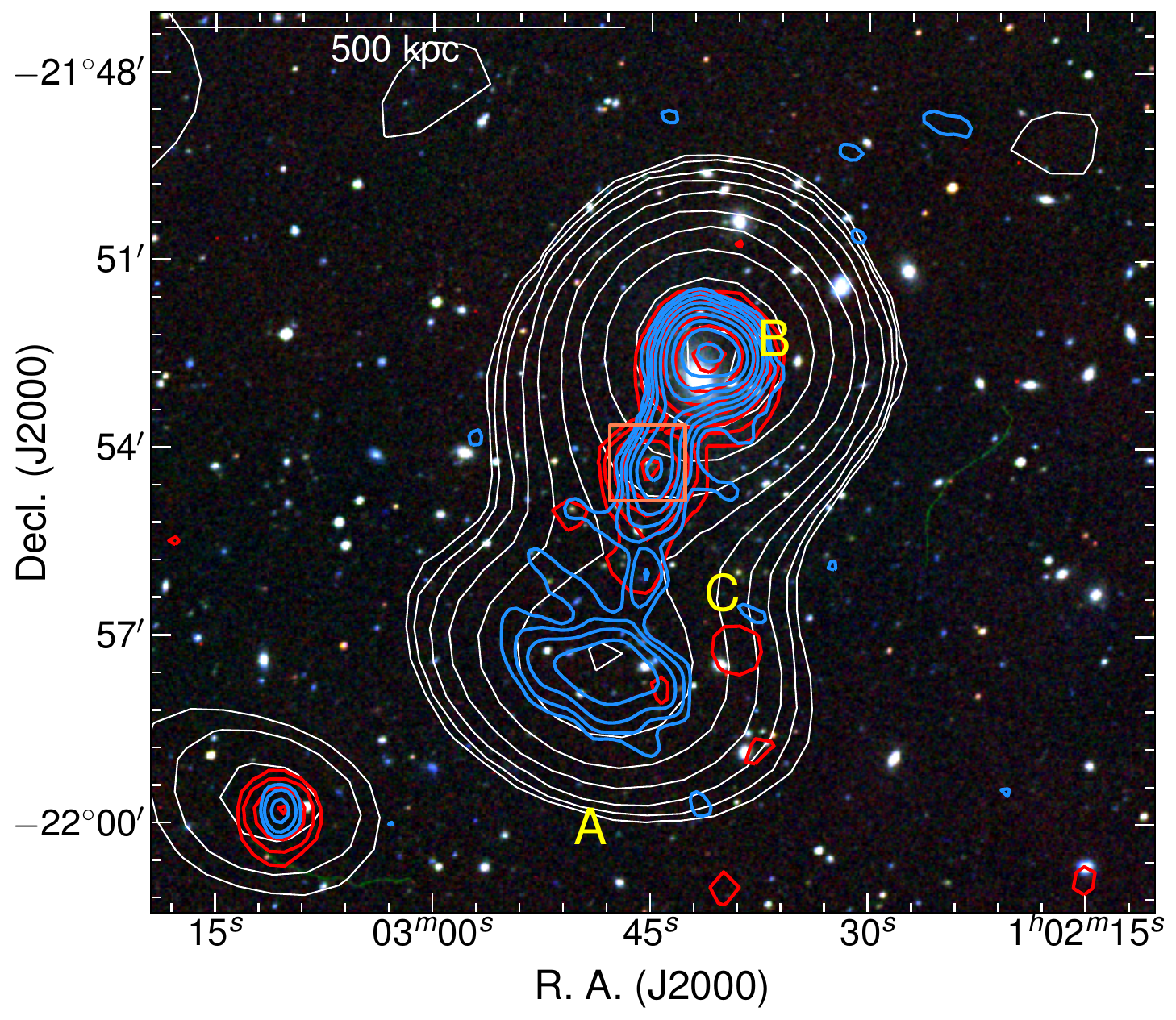}
\caption{The centre of Abell 0133. \corrs{DSS2} RGB image with contours overlaid as follows: EoR0 field, white, beginning at 15~\mjybeam; NVSS, \nvsscontour, beginning at 1.5~\mjybeam; TGSS, \tgsscontour, 13.2~\mjybeam. The linear scale is at the redshift of the cluster, and {Obj.~A, `B', and `C' are discussed in the text.}}
\label{fig:a133}
\end{figure}

{Abell 0133} has been studied extensively in X-ray \citep[e.g.][]{rmc+81, fsk+02, fsr+04} along with the multi-wavelength study by \citet{rcn+10} which all point towards the disturbed, dynamic nature of the cluster. 
A radio phoenix was detected by \citet[][but see also \citealt{srm+01}]{sr84}. A weak shock coincides with the phoenix source \citep{fsr+04}. \par
Fig.~\ref{fig:a133} shows the cluster centre with the emission of interest, with Obj.~A a large, possible lobe, Obj.~B the radio phoenix, and the orange square indicating the possible ID for double-lobe--like structure, along with Obj.~C, an interesting knot-like feature. \citet{rcn+10} discuss the possibility \corrs{that} the entire structure is a giant, background radio galaxy. As part of this interpretation, the phoenix is thought to be a separate entity. We consider an alternative explanation not covered by \citet{rcn+10} where the southern lobe `A' is a radio relic. This explanation is akin to the relic in 1E~0657-56 \citep[the Bullet Cluster;][]{lhba00, leh+01, sbf+14, smb+15, sri15}. \citet{lhba00} show low resolution radio imaging of the Bullet Cluster, and further X-ray observations provide high resolution imaging to show the directionality of the shock \citep{mar06} with clear diffuse emission located to the east of the west-ward X-ray shock. This piece of diffuse emission is considered a relic, created through back-shock of the massive, merging system \citep{sbf+14}. We consider the possibility of a similar relic in Obj.~A. Figure 6 of \citet{rcn+10} shows \emph{Chandra} data overlaid with 1400 and 330~MHz radio data, indicating the potential relic sitting beyond the X-ray emission towards the periphery of the cluster. Obj.~C in Fig.~\ref{fig:a133} marks a knot in the filament, clear in the \tgsscontour TGSS contours, and seen in Figure 5(d) of \citet{rcn+10}. This has no optical ID so is not necessarily an unassociated point source. The structure (`A'--`C'--part of `B') is considered to be a GRG \citep{rcn+10}. The supposed optical host (at `C') has a redshift of $z=0.2930$ \citep[2MASX~J01024529-2154137:][]{olk95,srm+01,rcn+10} placing it far behind Abell~0133. We find that the 147.5~MHz TGSS contours in Fig.~\ref{fig:a133} show that the peak of this emission near the core of the GRG does not align with the proposed optical ID, marked with an orange square, though the 1.4~GHz NVSS contours do align well with 2MASX~J01024529$-$2154137. \par
If the emission is that of a radio galaxy, we find {LAS to be 10.1 arcmin, which at $z=0.2930$ corresponds to an LLS of 2.7~Mpc and at $z=0.0562$ an LLS of 660~kpc.} In the relic scenario, we measure east-west dimensions: the LAS is found to be 5.5 arcmin (an LLS of 360~kpc at the cluster's redshift. We cannot confirm the nature of the emission with the available data. In Table~\ref{table:diffuse_sources} we list the phoenix, as well as the ambiguous emission as either a radio galaxy or radio relic.

\hypertarget{link:a141}{}\subsubsection{Abell 0141}

\begin{figure}[t!]
\includegraphics[width=0.99\linewidth]{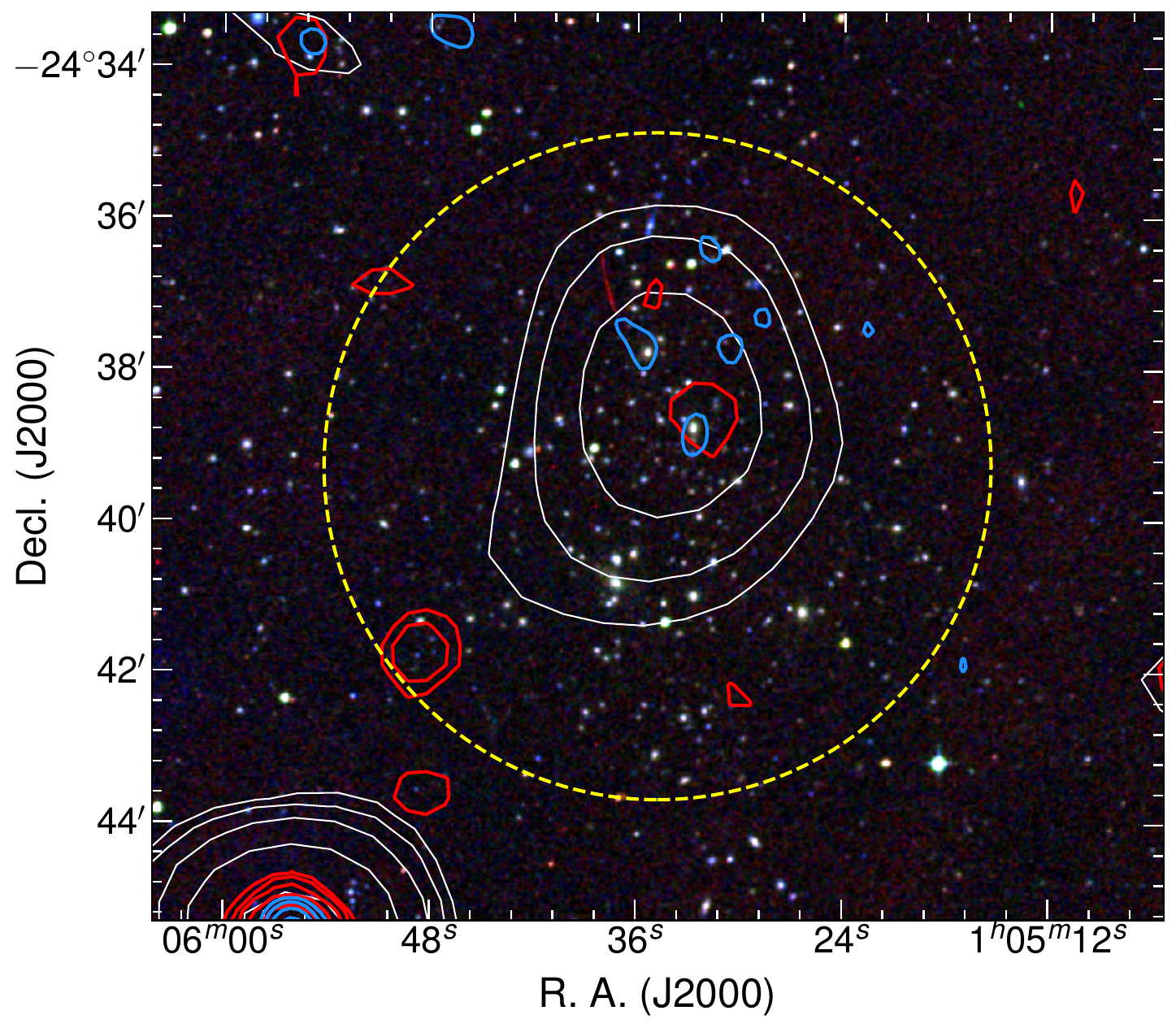}
\caption{Radio halo at the centre of Abell 0141. \corrs{DSS2} RGB image with contours overlaid as follows: EoR0 field, white, beginning at 10~\mjybeam; NVSS, \nvsscontour, beginning at 1.5~\mjybeam; TGSS, \tgsscontour, beginning at 13.8~\mjybeam. The dashed circle is centred on the cluster with a 1~Mpc radius.}
\label{fig:a141}
\end{figure}

{We present a hitherto undetected radio halo at the centre of Abell 0141 coinciding with the optical concentration of galaxies.} The left panel of Fig.~\ref{fig:a141} shows the cluster with an RGB image as a background with the EoR0 field contours overlaid to illustrate the radio halo's location relative to the cluster. Previously, \citet{vgb+07,vgd+08} reported a non-detection at 610~MHz with the GMRT with an upper limit of $S_{610} \leq 7.0$~mJy, assuming a standard spectral index of $-1.3$. From the EoR0 field image, the radio halo is measured to have a flux density of $S_{168} = 110 \pm 11$~mJy and an LAS of 5.0~arcmin (LLS of 1.1~Mpc). This suggest a spectral index of {$\alpha_{168}^{610} \leq -2.1 \pm 0.1$}. This places the halo within Abell 0141 at least equal in spectral steepness to the halo detected in Abell 0521, which has an average spectral index of {$\alpha \approx -2.1$} \citep{bgc+08}. \par
{\citet{cag18} have performed an X-ray analysis of the cluster, suggesting it may be in an early stage of the merger with both the northern and southern subclusters being described as ``moderately disturbed non-cool core structures''. Additionally, \citet{dki+02} comment on the ill-defined optical centre, noting that the two optical density peaks occur $\sim$2 arcmin apart, consistent with the X-ray emission. Radio halos have been found in pre-merging clusters \citep[Abell~0399 and Abell~0400;][and in MACS~J0416.1-2403; \citealt{2015ApJ...812..153O}]{2010A&A...509A..86M}, and in these known cases it is likely that each of the subclusters hosts its own halo. This may be the case here, where the source we detect is the convolution of a radio halo in each of the northern and southern subclusters. Future work with the Australian Square Kilometre Array Pathfinder (ASKAP) will provide insight into the nature of this radio halo (Duchesne et al., in prep.).

\hypertarget{link:a2496}{}\subsubsection{Abell 2496}

\begin{figure*}[t!]
\includegraphics[width=0.478\linewidth]{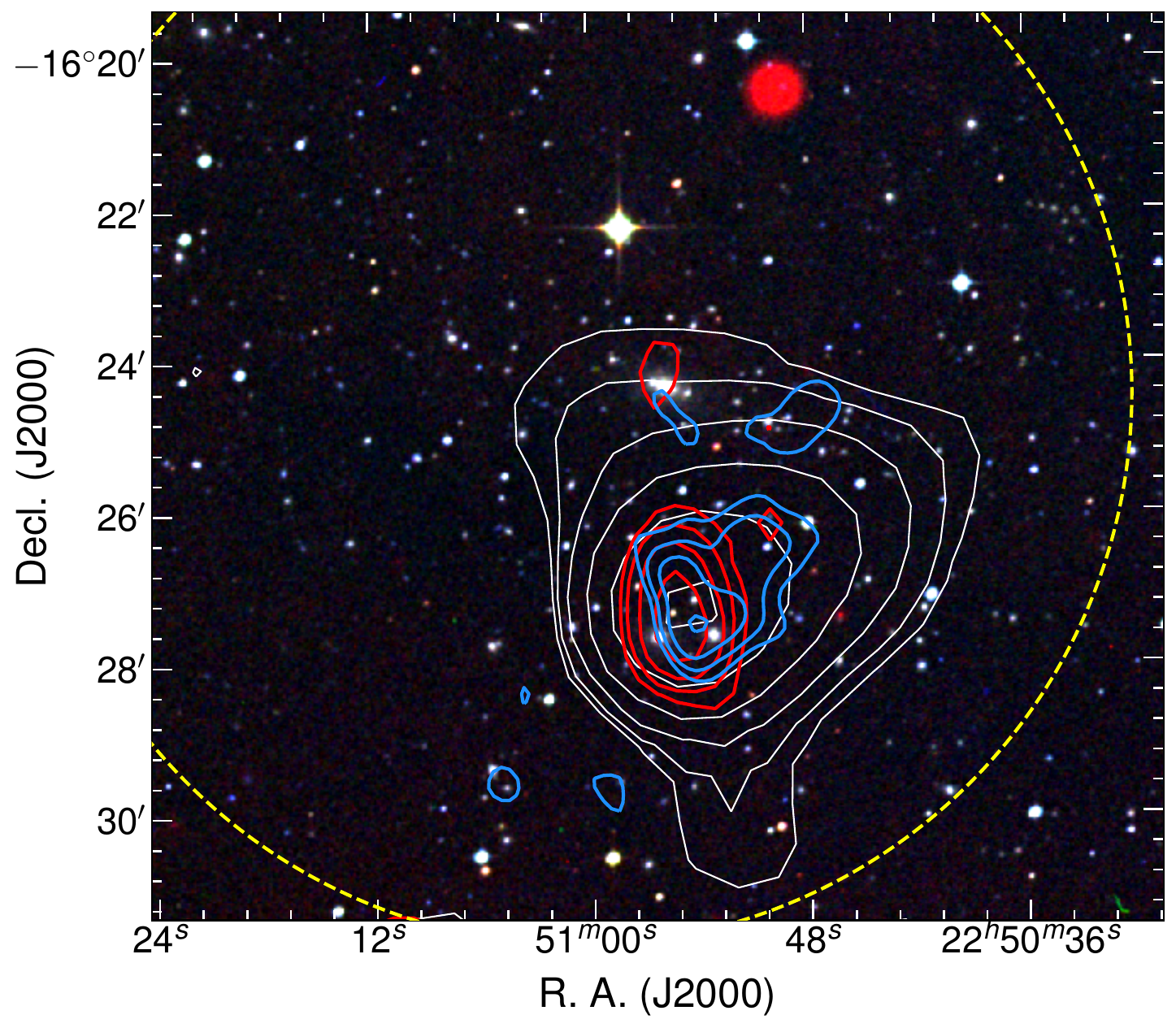} \hfill
\includegraphics[width=0.478\linewidth]{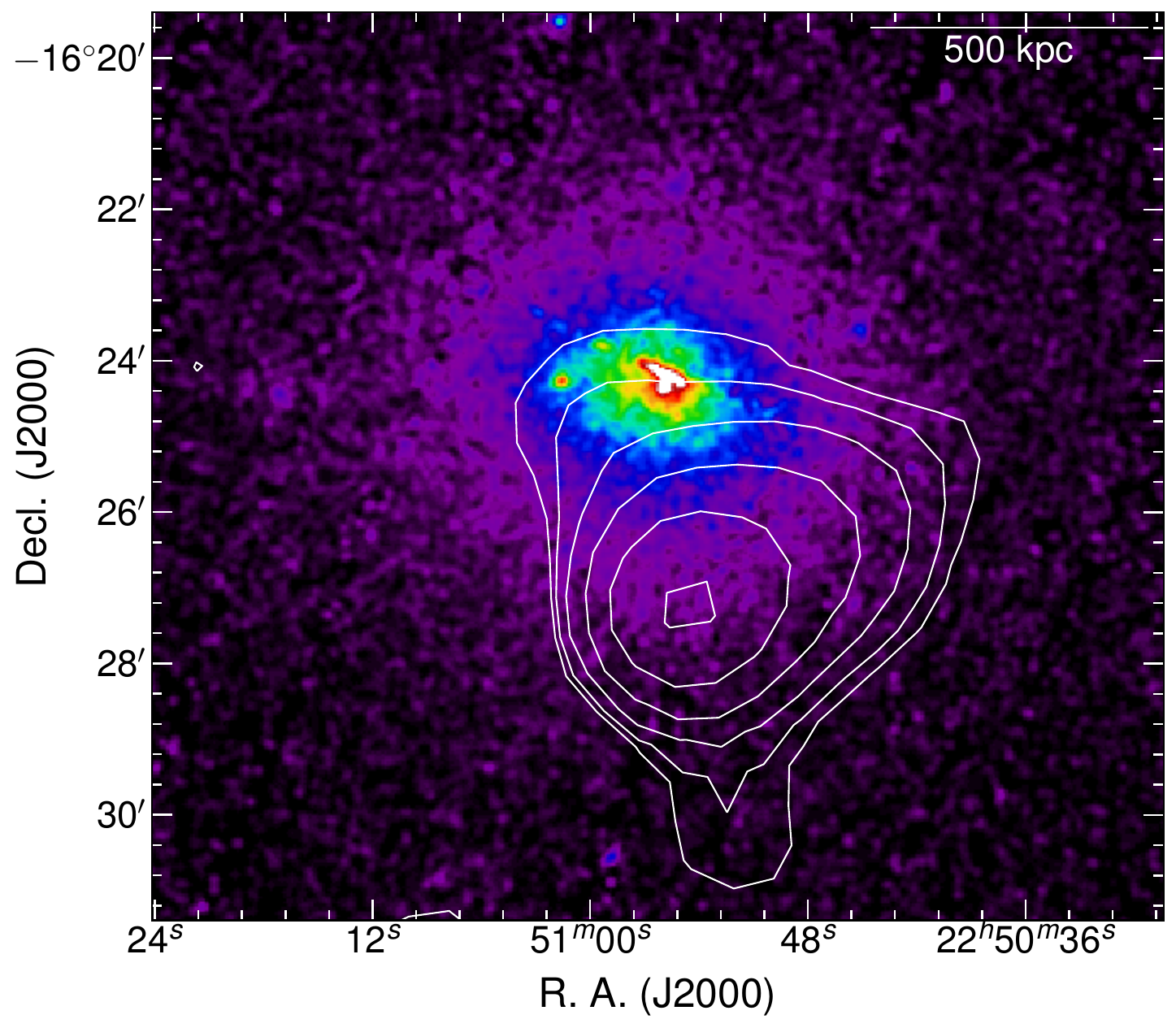}
\caption{Diffuse emission within Abell 2496. \emph{Left:} \corrs{DSS2} RGB image with contours overlaid as follows: EoR0 field, white, beginning at 15~\mjybeam; NVSS, \nvsscontour, beginning at 1.5~\mjybeam. TGSS, \tgsscontour, 12~\mjybeam. \emph{Right:} Exposure corrected, smoothed XMM-\emph{Newton} data with EoR0 field contours overlaid as in the left panel. The dashed circle is centred on the MCXC coordinates with radius of 1~Mpc.}
\label{fig:a2496}
\end{figure*}

{Fig.~\ref{fig:a2496} shows the centre of Abell 2496 with extended, diffuse emission with an irregular morphology.} We measure $S_{168} =  561 \pm 42$~mJy within the full EoR0 field contours, and obtain $S_{1400} = 37.7 \pm 2.0$~mJy \citep{ccg+98}, $S_{147.5} = 659.4 \pm 67.0$~mJy \citep{ijmf16}, and {$S_{74} = 1340 \pm 250 $~mJy} \citep{lcv+14}. From these measurement we obtain $\alpha_{74}^{1400} = -1.26 \pm 0.02$. We note that the NVSS and TGSS contours may represent a discrete cluster source, with the extended components in the EoR0 field and TGSS data separate emission such as a radio halo. Additionally, the TGSS data may be resolving out some of the emission if the full EoR0 field contours comprise a single source. \par
 The bulk of the radio emission is offset from the X-ray emission seen with the exposure corrected, smoothed XMM-\emph{Newton} data in the right panel of Fig.~\ref{fig:a2496} (Obs. ID 0765030801, PI Reiprich). The radio emission does extend towards the X-ray peak. If the total radio emission represents a single source, we measure an LAS of $\sim$4.2~arcmin (LLS of $\sim$560~kpc). In this case, this may be a ``face-on'' radio relic. If the NVSS contours represent a discrete cluster source, then extended lower frequency emission may represent a mini-halo, however, we cannot confirm the nature of the source with the present data.

\hypertarget{link:a2556}{}\subsubsection{Abell 2556 and Abell 2554}

\begin{figure}[t!]
\includegraphics[width=0.99\linewidth]{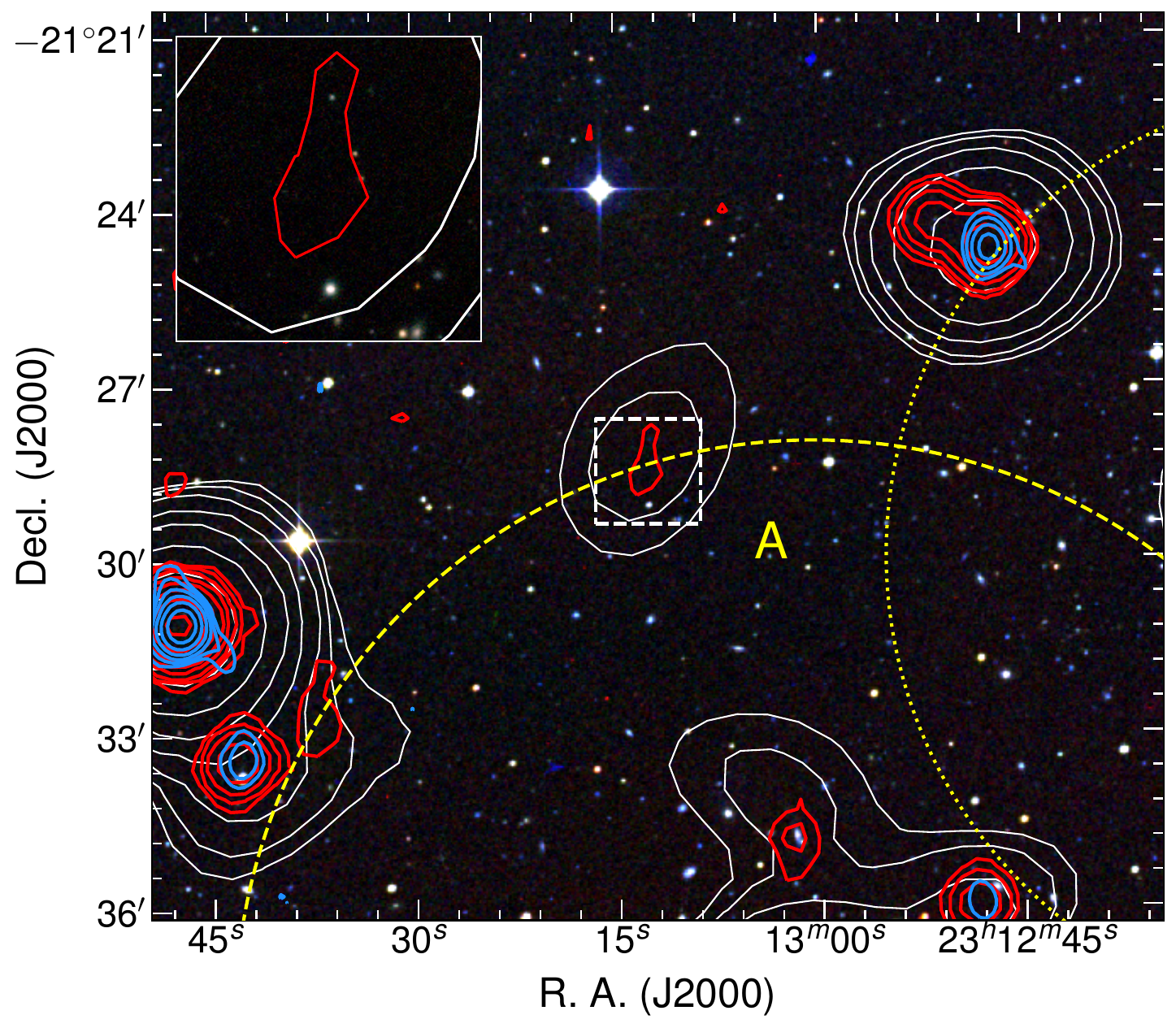}
\caption{Diffuse emission, Obj.~A, in Abell 2556. \corrs{DSS2} RGB image with contours overlaid as follows: EoR0 field, white, beginning at 10~\mjybeam; NVSS, \nvsscontour, beginning at 1.5~\mjybeam; TGSS, \tgsscontour, beginning at 13.4~\mjybeam. The dashed circle is centred on Abell 2556 and the dotted circle on Abell 2554, each with radii of 1~Mpc. \corrs{The inset is the PS1 data with its location indicated on the image as a dashed, white box. EoR0 field and NVSS contours are shown on the inset as in the main figure.}}
\label{fig:a2556}
\end{figure}

{Fig.~\ref{fig:a2556} shows the two clusters Abell 2556 and Abell 2554 which have centres within 13~arcmin of each other, but have redshifts of $z=0.0871$ and $z=0.1108$ \citep{cmkw02} respectively.} To the north of Abell 2556, 1~Mpc from its centre (east of Abell 2554, over 1~Mpc) an elongated diffuse source is seen, labelled `A' in Fig.~\ref{fig:a2556}, with flux densities $S_{168} = 29.3 \pm 5.5$~mJy and $S_{1400} = 2.2 \pm 0.5$~mJy \citep{ccg+98}, corresponding to $\alpha_{168}^{1400} = -1.22 \pm 0.14$. The LAS of the source is 2.4~arcmin (LLS of 240~kpc at $z=0.0871$). \corrs{No optical host is seen in the PS1 inset in Fig.~\ref{fig:a2556} at the centre of the emission.} Too small for a radio relic, we note that radio phoenices are more often found towards cluster centres but this would be consistent with the spectral index, where phoenices closer to the centre become much steeper. We consider this a candidate radio phoenix.

\hypertarget{link:a2680}{}\subsubsection{Abell 2680}

\begin{figure}[t!]
\includegraphics[width=0.99\linewidth]{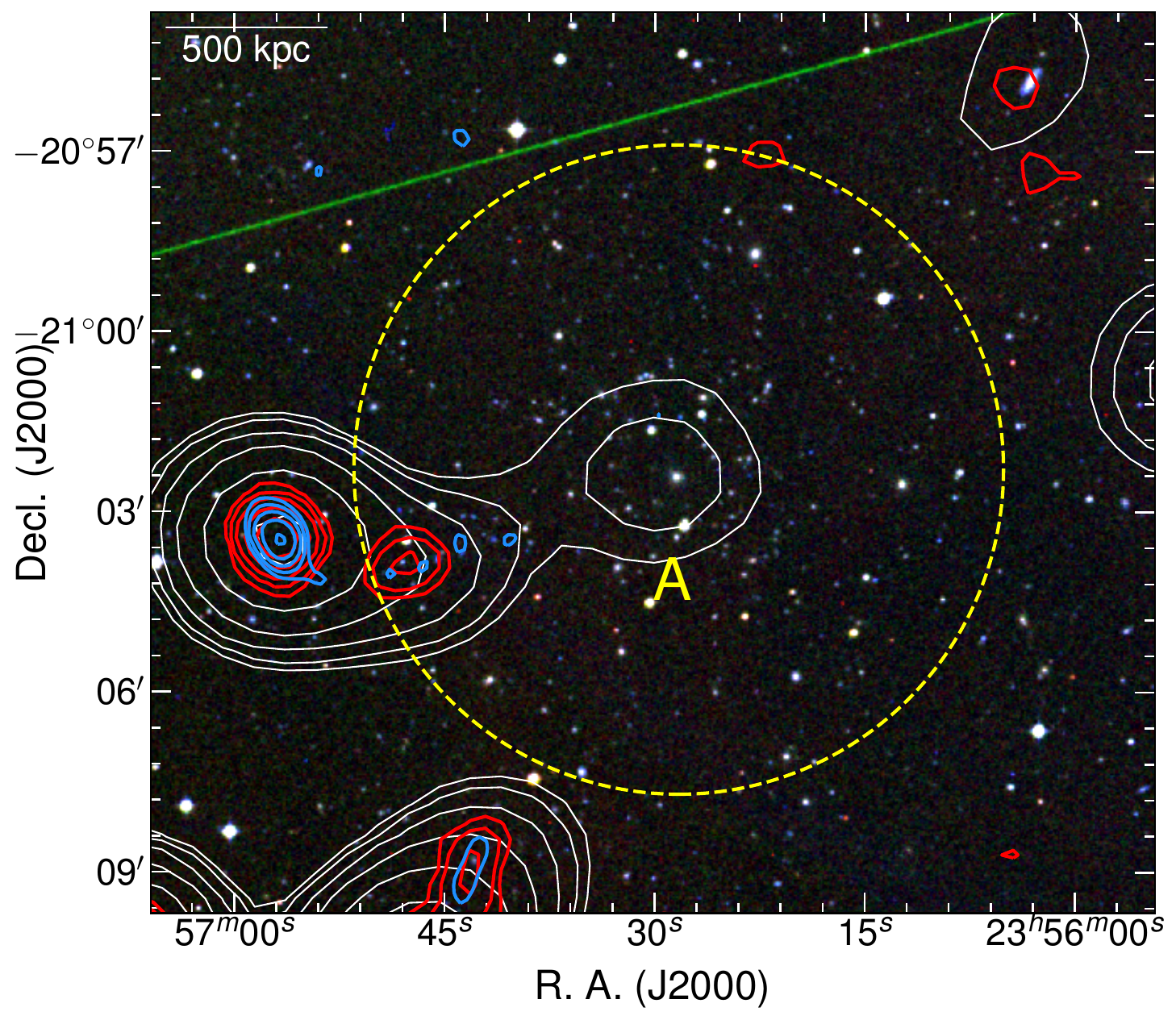}
\caption{Abell 2680 with a candidate halo marked with an `A'. \corrs{DSS2} RGB image with contours overlaid as follows: EoR0 field, white, beginning at 7~\mjybeam; NVSS, \nvsscontour, beginning at 1.5~\mjybeam; TGSS, \tgsscontour, beginning at 11.1~\mjybeam. The dashed circle has a 1~Mpc radius about the cluster centre.}
\label{fig:a2680}
\end{figure}

Fig.~\ref{fig:a2680} shows a patch of steep-spectrum emission at the centre of Abell 2680, with no counterparts in the NVSS or TGSS images (Obj.~A). The emission may be slightly elongated east-west, though this apparent elongation may just be the result of blending with the eastern sources. We make an approximate measurement of the flux density yielding $S_{168} = 22.8 \pm 8.0$~mJy, where the uncertainty is given by Eq.~\ref{eq:flux_unc} with an additional contribution to account for the slight blending to the east. We estimate a 1.4~GHz upper limit of $1.8$~mJy giving $\alpha_{168}^{1400} \leq -1.2 \pm 0.2$. The LAS is estimated to be $\sim$2.2~arcmin (LLS of $\sim$400~kpc). The physical extent of the source and coincidence with the cluster centre core suggests a cluster halo. This particular case requires observations at different resolutions to determine if the source is actually extended but we consider this a candidate radio halo or mini-halo.

\hypertarget{link:a2693}{}\subsubsection{Abell 2693}

\begin{figure}[t!]
\includegraphics[width=0.99\linewidth]{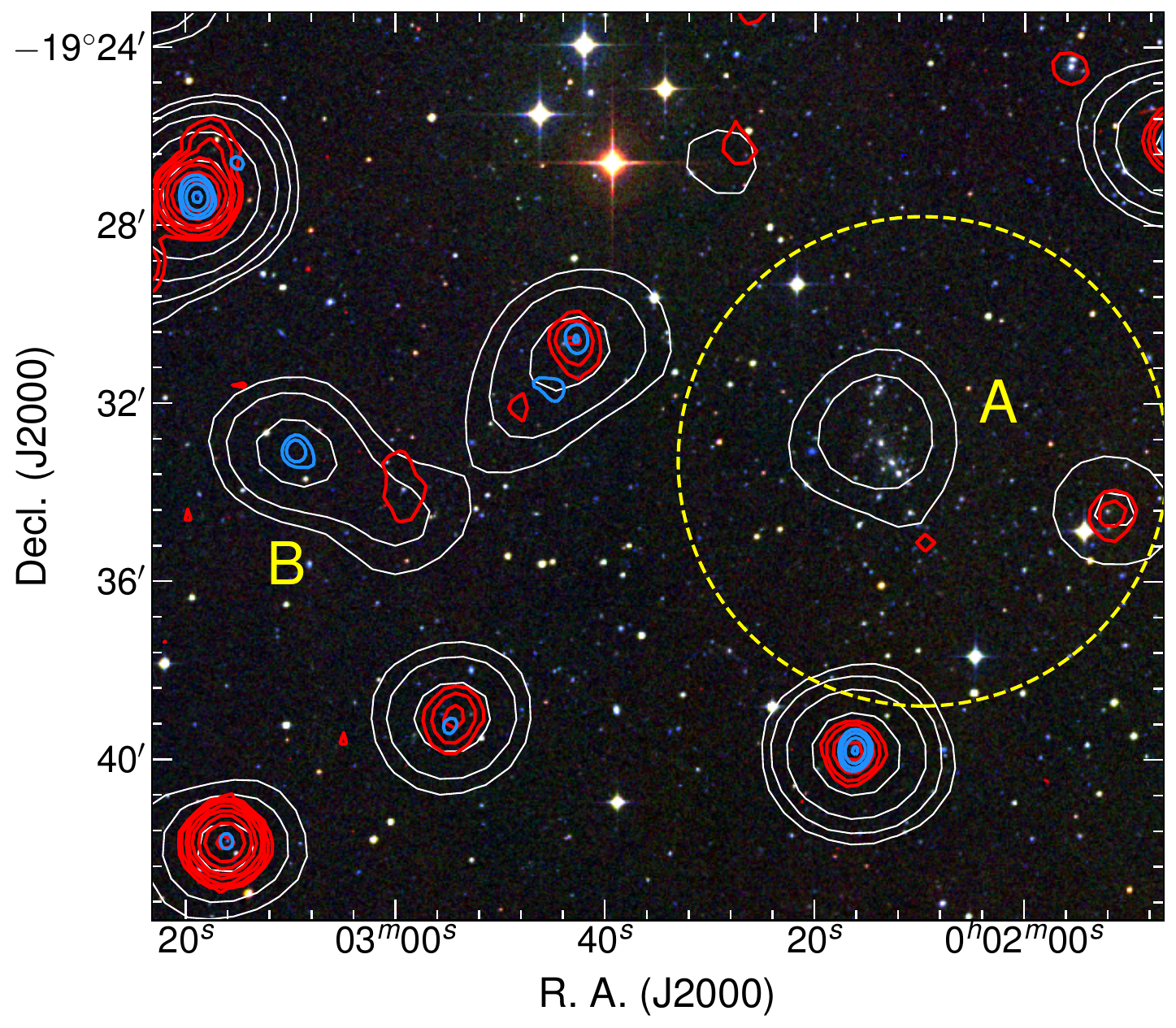}
\caption{Candidate radio halo A and steep-spectrum source B within and nearby Abell 2693. \corrs{DSS2} RGB image with contours overlaid as follows: EoR0 field, white, beginning at 10~\mjybeam; NVSS, \nvsscontour, beginning at 1.5~\mjybeam; TGSS, \tgsscontour, beginning at 12~\mjybeam. The dashed circle is centred on the cluster and has a 1~Mpc radius.}
\label{fig:a2693}
\end{figure}

{Abell 2693 is found to host an extended source at its centre.} We consider this a candidate halo, marked `A' in Fig.~\ref{fig:a2693}, has an LAS of {3.0~arcmin} (LLS of 530~kpc). We measure $S_{168} = 50 \pm 6$~mJy and $S_{1400} \leq 7.7$~mJy from the NVSS image, resulting in $\alpha_{168}^{1400} \leq -0.88 \pm 0.06$. The spectral index limit is inconclusive in the halo classification, however, the location and size suggest that it may be a halo, and we classify this emission as a candidate halo. \par
To the west of the cluster there is an elongated steep-spectrum source marked B in Fig.~\ref{fig:a2693}, though it appears as a point source in the TGSS image.

\hypertarget{link:a2721}{}\subsubsection{Abell 2721}

\begin{figure*}[t!]
\includegraphics[width=0.478\linewidth]{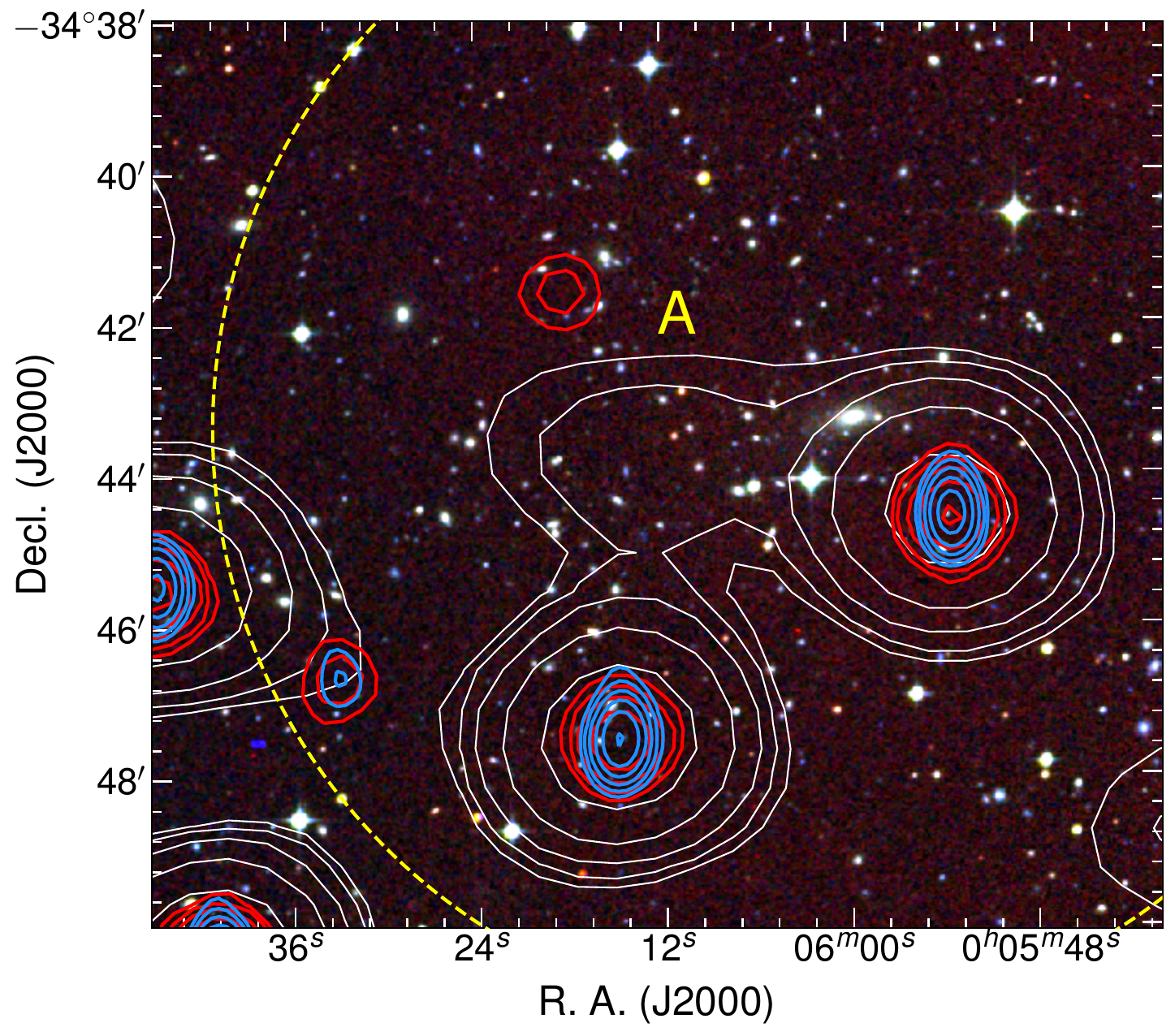}\hfill
\includegraphics[width=0.478\linewidth]{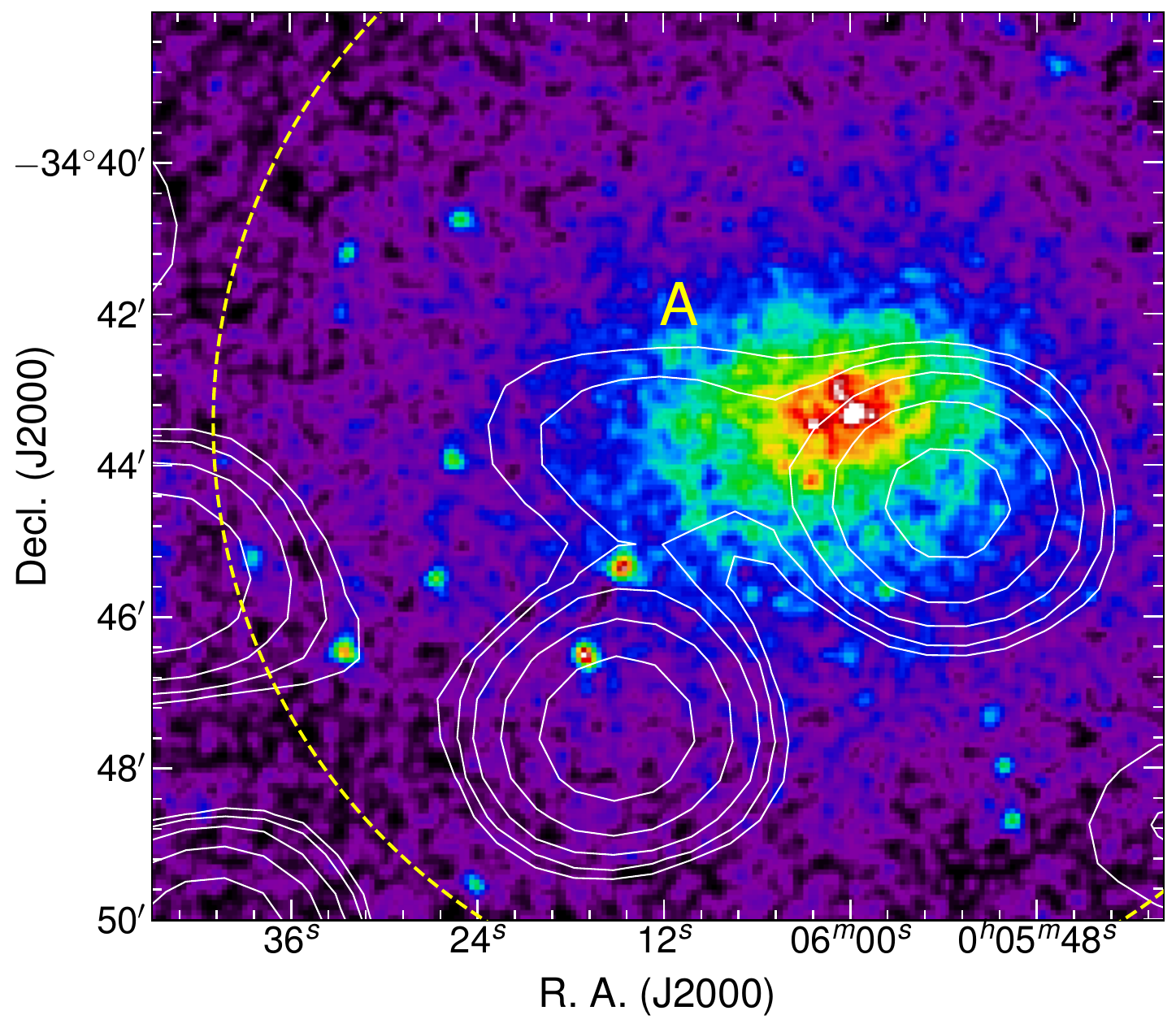}
\caption{Diffuse emission within Abell 2721, marked with an `A'. \emph{Left:} \corrs{DSS2 RGB image} with contours overlaid as follows: EoR0 field, white, beginning at 10~\mjybeam; NVSS, \nvsscontour, beginning at 1.5~\mjybeam; TGSS, \tgsscontour, beginning at 8.1~\mjybeam. \emph{Right:} {Exposure corrected, smoothed XMM-\emph{Newton} X-ray image from the {\sffamily REXCESS} survey with EoR0 contours overlaid as in the left panel.} The dashed circle is centred on the cluster with a radius of 1~Mpc.}
\label{fig:a2721}
\end{figure*}

{Fig.~\ref{fig:a2721} shows Abell 2721.} Diffuse radio emission is seen at 168~MHz offset to the east of the cluster centre (Obj.~A). The lack of emission seen in the NVSS or SUMSS suggests a steep spectral index, and a lack of emission in TGSS is likely due to lack of sensitivity. We estimate LAS of the emission to be $\sim4.0$~arcmin (LLS of $\sim$500~kpc).  \par
As part of the ATCA {\sffamily REXCESS} Diffuse Emission Survey (ARDES), deep 1.4 and 2.1~GHz imaging of the cluster was obtained with the ATCA, finding no evidence of a halo \citep{sjp16}. An upper limit to the emission is $S_{1400} \leq 7$~mJy (Shakouri, private comms.). After subtraction of the blended point sources, we find $S_{168} = 54 \pm 14$~mJy for the extended emission, resulting in $\alpha_{168}^{1400} \leq -0.96 \pm 0.12$. \par
\corrs{The} right panel of Fig.~\ref{fig:a2721} shows the {\sffamily {\sffamily REXCESS}} X-ray data overlaid with MWA contours. The extended radio emission is offset from the X-ray peak, ruling out a halo. A relic-type source is possible, but the spectral index likely rules out a phoenix. We suggest this source may be a relic viewed somewhat along the line of sight, and consider it a candidate radio relic. 

\hypertarget{link:a2744}{}\subsubsection{Abell 2744}
\label{sec:a2744}

\begin{figure}[t!]
\includegraphics[width=0.99\linewidth]{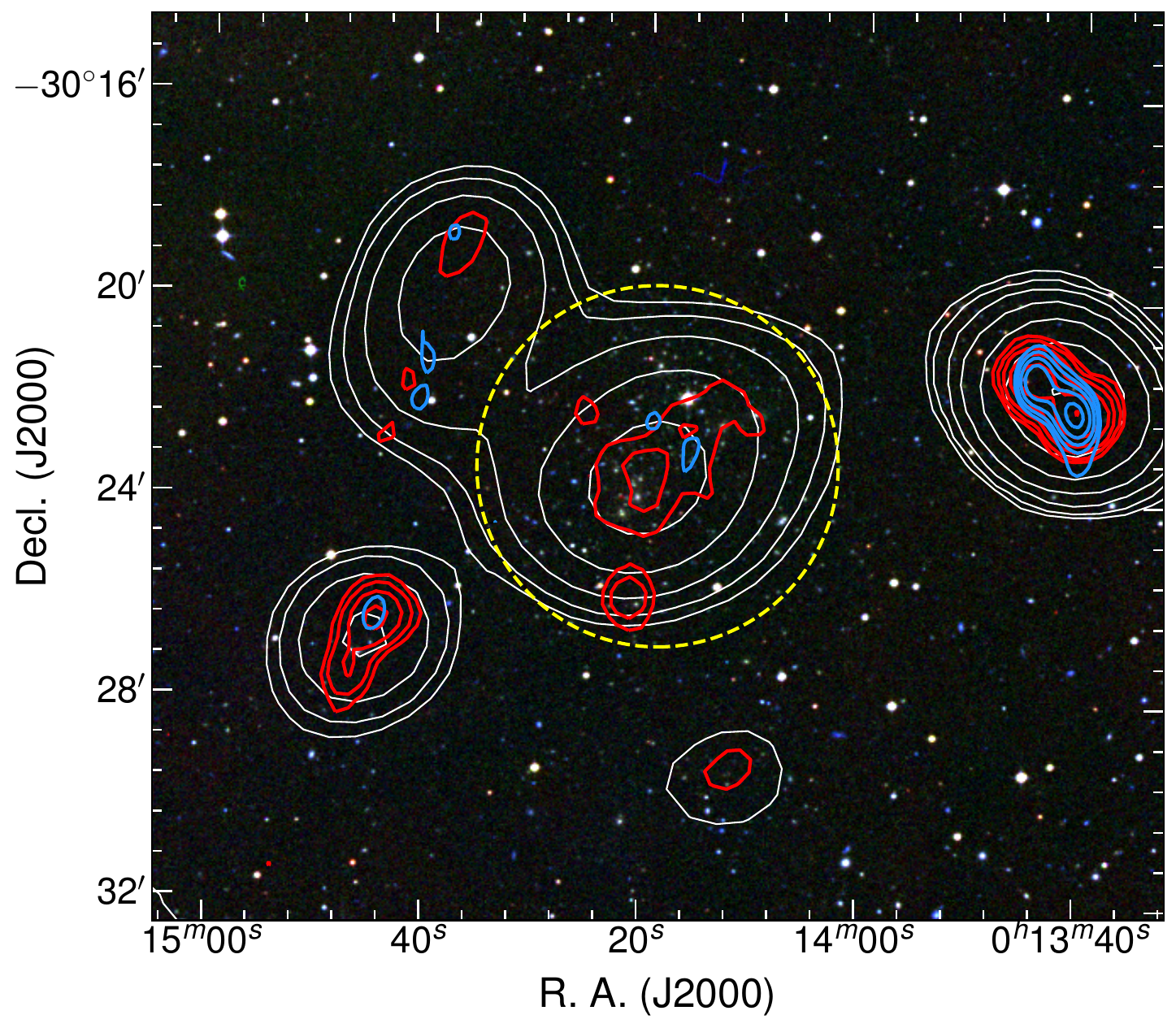}
\caption{Abell 2744 with giant radio halo and relic. \corrs{DSS2} RGB image with contours overlaid as follows: EoR0 field, white, beginning at 10~\mjybeam; NVSS, \nvsscontour, beginning at 1.5~\mjybeam; TGSS, \tgsscontour, beginning at 12.9~\mjybeam. The dashed circle is centred on the cluster with a radius of 1~Mpc.}
\label{fig:a2744}
\end{figure}

{Abell 2744 is a \emph{Hubble} Frontier Fields cluster} \citep{lkc+17} showing gravitational lensing of the high-redshift background galaxies \citep[see {e.g.}][]{cam+16}. Fig.~\ref{fig:a2744} shows Abell 2744 with both a centrally located giant radio halo (GRH, defined to have an LLS $>$ 1~Mpc) and a mega-parsec scale radio relic on its northeast periphery \citep[][]{gfg+01}. Both of these objects are seen in the EoR0 field at 168~MHz, blending together along the northeast edge of the cluster. The GRH fills the entire cluster out to 1~Mpc having an approximate LLS of 1.9~Mpc (LAS of $\sim$6.9~{arcmin}) and the relic with an LLS on the order 1.4~Mpc (LAS $\sim$5.2~arcmin). \par
With \texttt{aegean}, with fit Gaussians to the relic and GRH to decompose and measure the emission, finding $S_{168}^{\rm{halo}} = 550 \pm 51$~mJy and $S_{168}^{\rm{relic}} = 237 \pm 24$~mJy. From these and literature measurements \citep{vgd+13} we find $\alpha_{\rm{halo}} = -1.11 \pm 0.04$ and $\alpha_{\rm{relic}} = -1.19 \pm 0.05$, consistent with spectral indices reported by \citet{vgd+13}.

\subsubsection{Abell 2751 and APMCC 039\label{sec:a2751}}

\begin{figure*}[t!]
\includegraphics[width=0.99\linewidth]{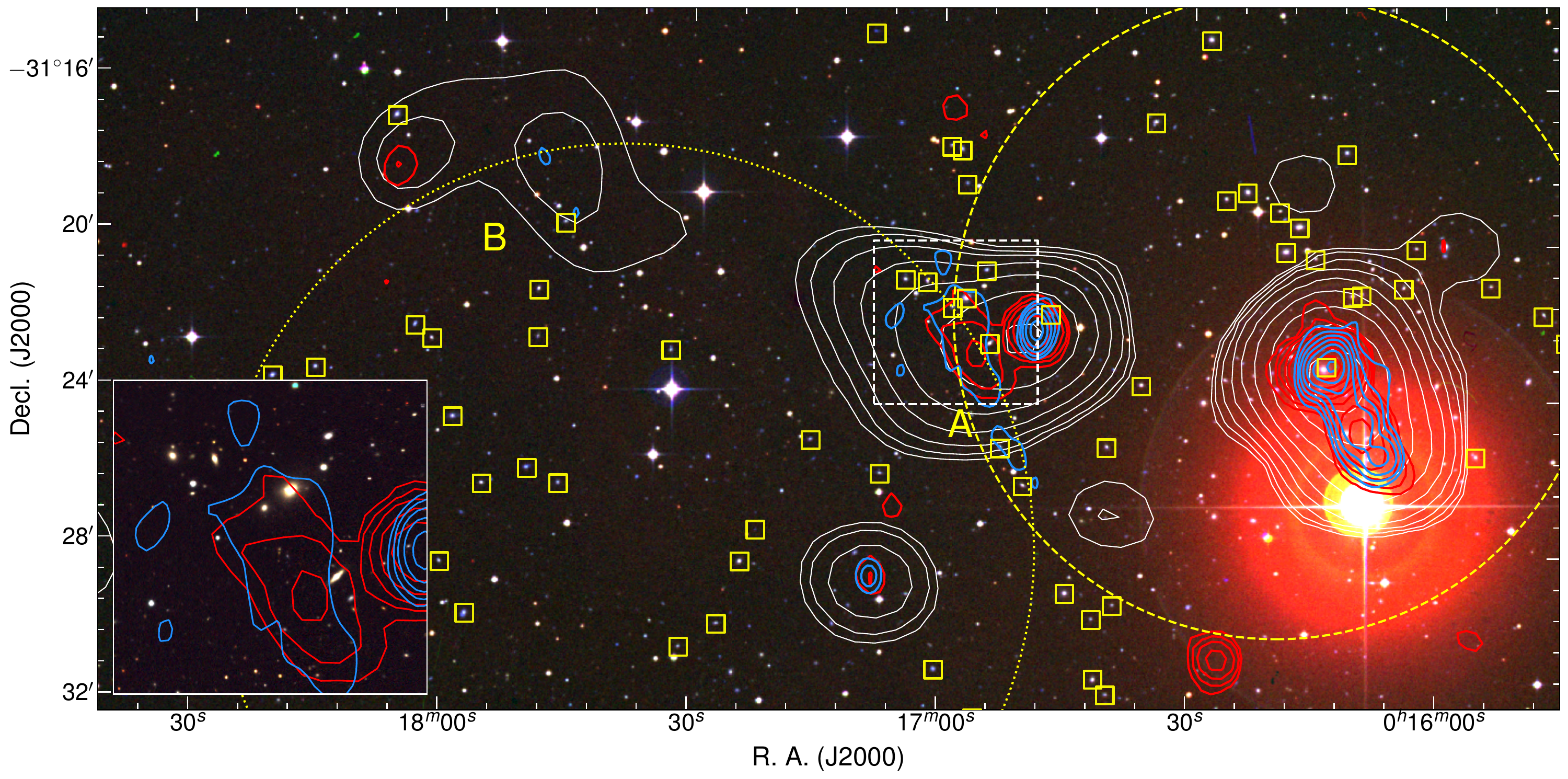}
\caption{A candidate relic and a faint radio galaxy, near Abell 2751 and APMCC 039, marked as A and B. \corrs{DSS2} RGB image with contours overlaid as follows: EoR0 field, white, beginning at 7~\mjybeam; NVSS, \nvsscontour, beginning at 1.5~\mjybeam; TGSS, \tgsscontour, beginning at 13.5~\mjybeam. The dashed circle is centred on Abell 2751 and the dotted on APMCC 039, each with radii of 1~Mpc. The squares indicate galaxies with redshifts in the range $0.1 \leq z \leq 0.114$. \corrs{The inset of Obj.~A is the DES~DR1 data with its location indicated by the dashed, white box. TGSS and NVSS contours are overlaid as in the main image.}}
\label{fig:a2751}
\end{figure*}

\hypertarget{link:a2751}{}~{Abell 2751 and APMCC 039} have an angular separation of 17.7~arcmin and redshifts of $z=0.107$ \citep{sr99} and $z=0.082$ \citep{dmse97}, respectively. Fig.~\ref{fig:a2751} shows the two clusters, with the dashed and dotted circles indicating 1~Mpc radii about the cluster centres at their reported redshifts. The small \corrs{yellow} squares indicate galaxies with redshifts in the range 0.1--0.114, which is $cz \approx 2000$~km\,s$^{-1}$ around the redshift of Abell 2751. There are no galaxies in the vicinity at the reported redshift of APMCC 039. From this \corrs{galaxy distribution} we suggest the clusters are likely interacting \corrs{or are otherwise a single system}. \par
We detect a new candidate relic to the east of Abell 2751 (Obj.~A in Fig.~\ref{fig:a2751}), blending with the point source NVSS~J001648-312223. The 168~MHz emission appears to simply be a radio tail extending from NVSS~J001648$-$312223, however the 147.5~MHz TGSS emission is resolved enough to show that the emission is not necessarily associated with the point source. After subtraction of the flux density contribution of the blended point source, $S^{\mathrm{relic}}_{168} = 323 \pm 62$~mJy. We consider this emission a candidate radio relic. We also measure $S^{\mathrm{relic}}_{1400} = 22 \pm 3$~mJy from the NVSS map and obtain $\alpha_{168}^{1400} = -1.27 \pm 0.11$, consistent with relic sources. We measure the LAS to be $\sim$8.7~arcmin (LLS of $\sim$1.0~Mpc). \par
Obj.~B north of APMCC~039 also may be a relic or HT radio galaxy. We measure $S_{168} = 60 \pm 8$~mJy and an LAS of 8.5 arcmin (LLS of 1.0~Mpc at $z=0.107$). With a limit from the NVSS we obtain {$-1.3(\pm0.1) \leq \alpha_{168}^{1400} \leq -0.4(\pm0.1)$} consistent with either scenario. While this may be a relic, we suggest the likeliest case is a HT radio galaxy. 

\hypertarget{link:a2798}{}\subsubsection{Abell 2798}

\begin{figure}[t!]
\includegraphics[width=0.99\linewidth]{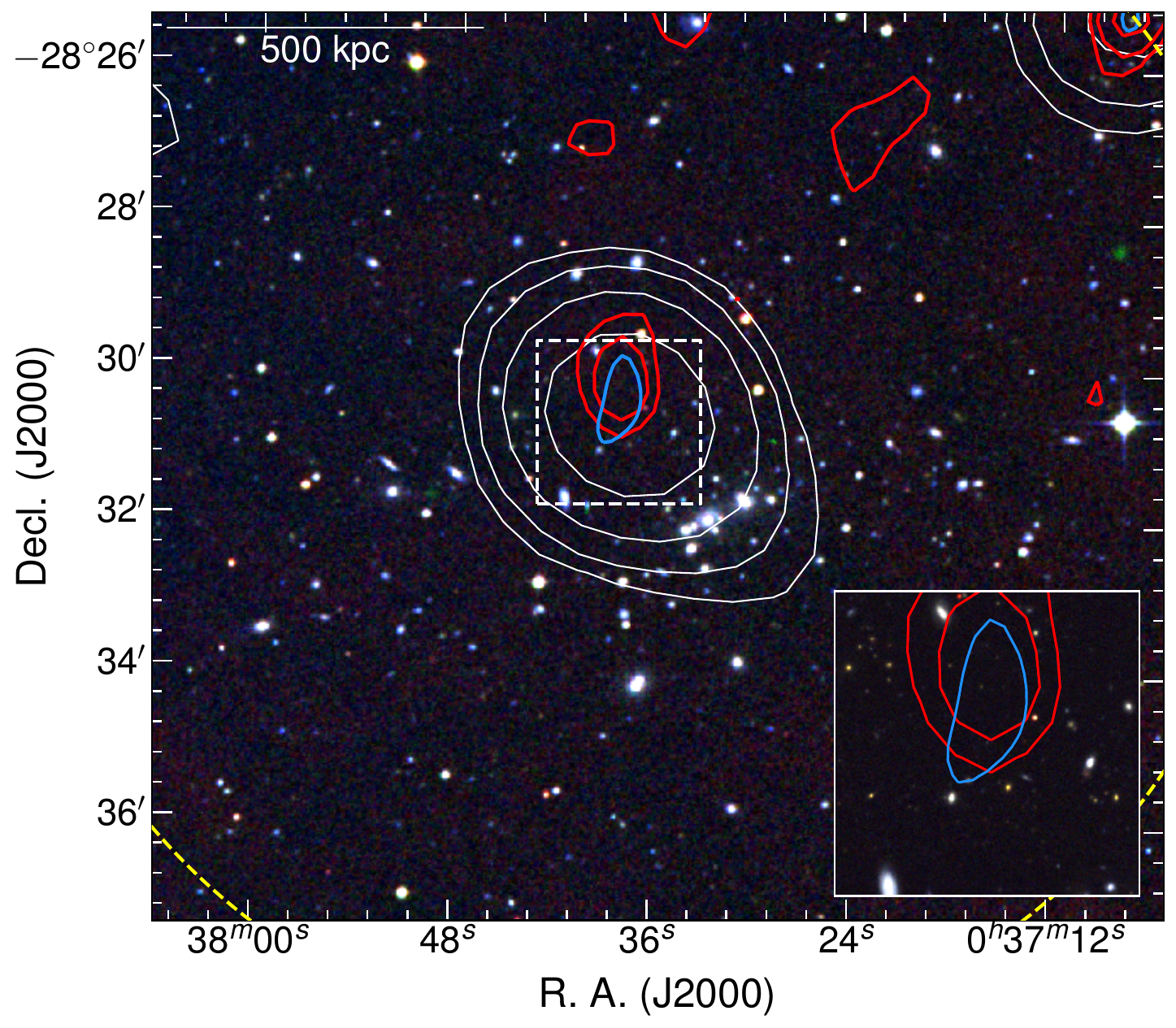}
\caption{Candidate radio relic within Abell 2798. \corrs{DSS2} RGB image with contours overlaid as follows: EoR0 field, white, beginning at 7~\mjybeam; NVSS, \nvsscontour, beginning at 1.5~\mjybeam; TGSS, \tgsscontour, beginning at 13.8~\mjybeam. \corrs{The inset is the DES~DR1 data with its location indicated by the dashed, white box. NVSS contours are overlaid as in the main image.}}
\label{fig:a2798}
\end{figure}

{Fig.~\ref{fig:a2798} shows the centre of Abell 2798 hosting a steep-spectrum radio source, slightly offset from centre.} We find {$S_{168} = 110 \pm 9$~mJy} and an LAS of {4.2~arcmin} (an LLS of {490~kpc}). Both the TGSS and NVSS surveys show extended emission offset slightly from the centroid of the 168~MHz emission. There are no \corrs{obvious} optical IDs for this emission. The NVSS source is NVSS~J003738$-$283008 and has a flux density of $S_{1400} = 9.0 \pm 1.3$~mJy \citep{ccg+98}. This yields a spectral index for the source of {$\alpha_{168}^{1400} = -1.2 \pm 0.1$}. The RASS broad band (0.1--2.4~keV) count image shows no significant X-ray emission within the cluster which is consistent with the low cluster mass. We classify this emission similarly to that in Abell 0013: a candidate radio phoenix either near the cluster centre or projected onto it. Further high-resolution imaging will be necessary to fully determine the nature of this emission.

\hypertarget{link:a2811}{}\subsubsection{Abell 2811}
\label{sec:a2811}

\begin{figure*}[t!]
\includegraphics[width=0.478\linewidth]{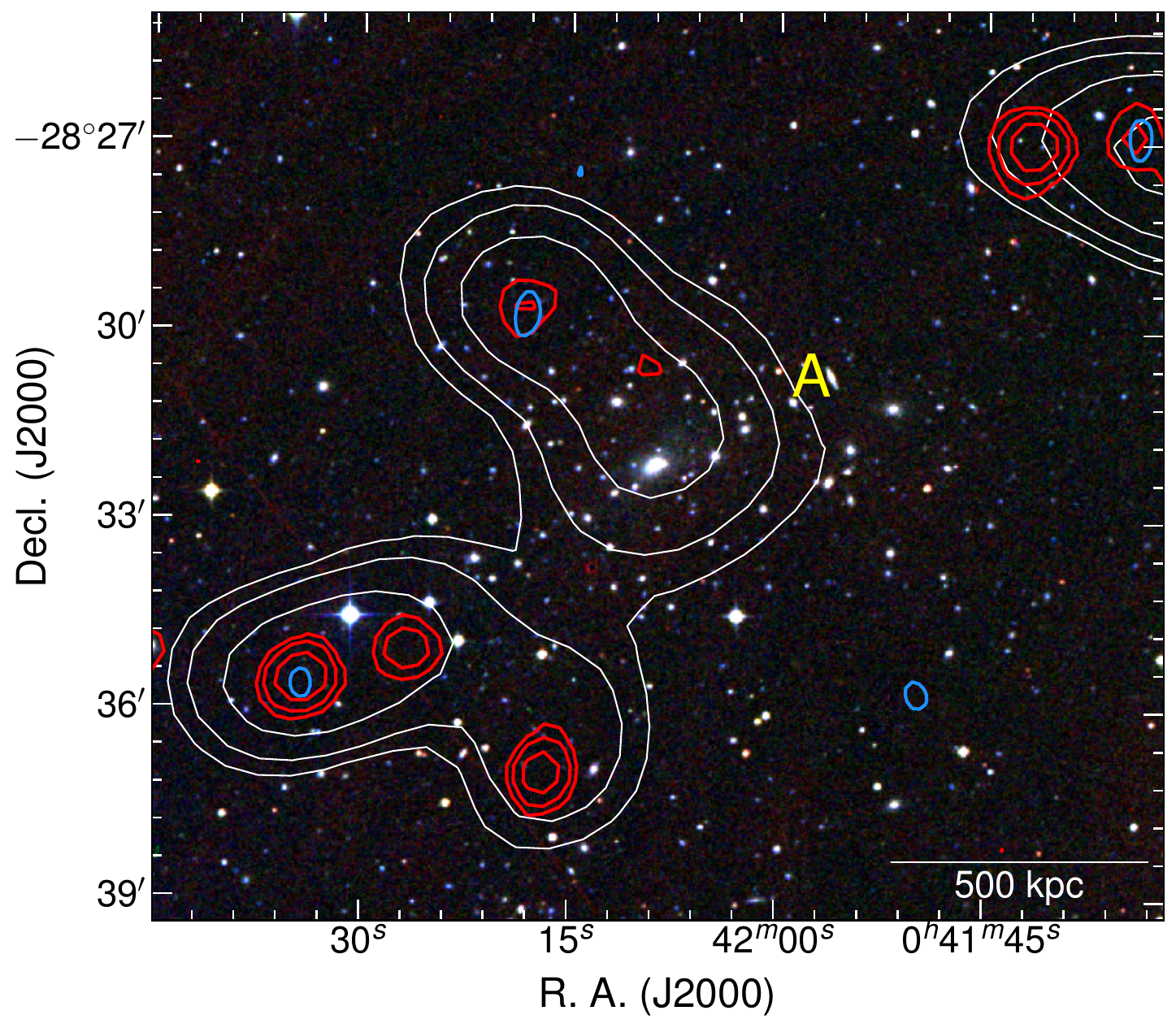} \hfill
\includegraphics[width=0.478\linewidth]{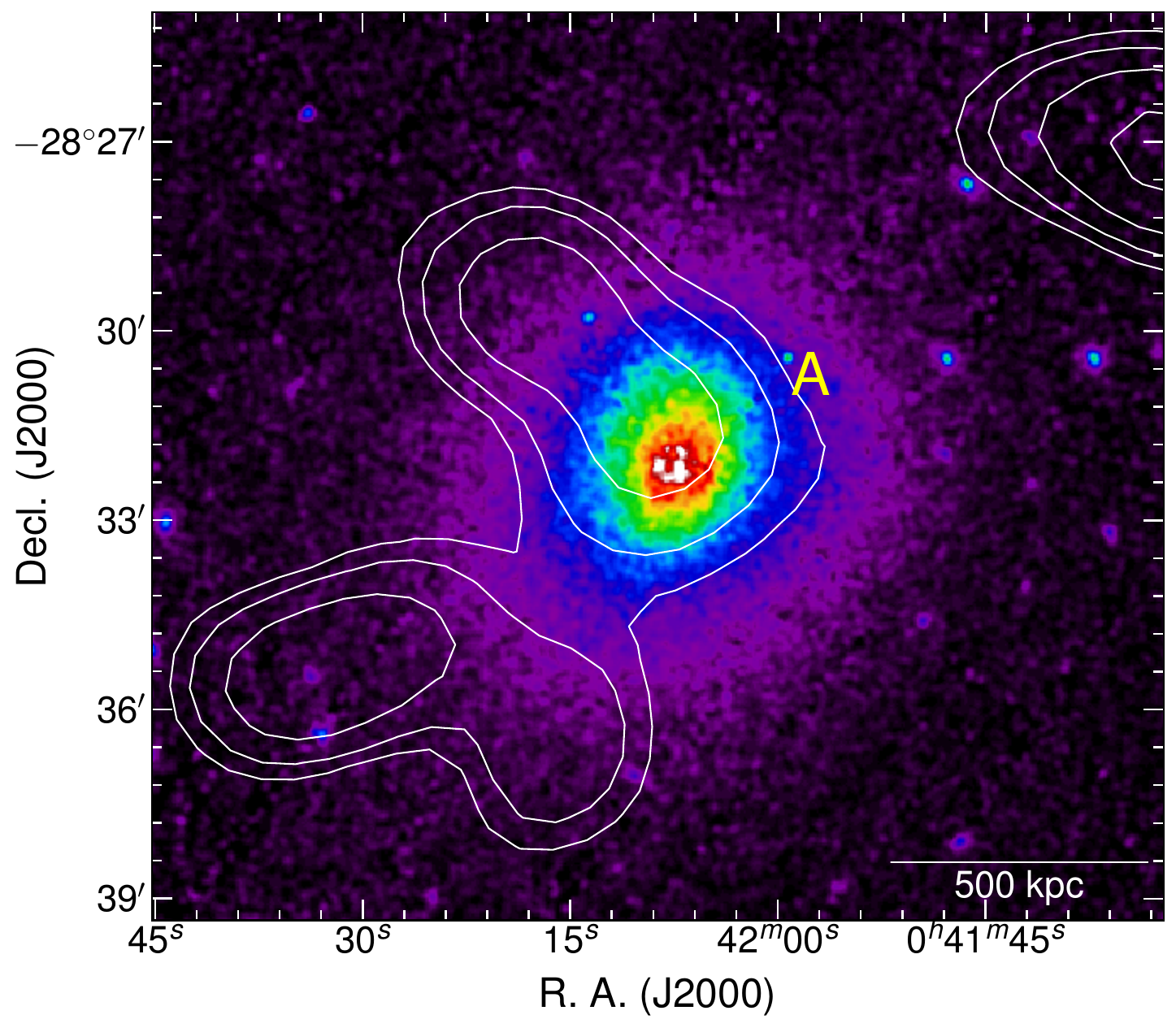}
\caption{Radio halo within Abell 2811, marked with an `A'. \emph{Left:} \corrs{DSS2} RGB image with contours overlaid as follows: EoR0 field, white, beginning at 7~\mjybeam; NVSS, \nvsscontour, beginning at 1.5~\mjybeam; TGSS, \tgsscontour, beginning at 12.6~\mjybeam. \emph{Right:} {Exposure corrected, smoothed XMM-\emph{Newton} X-ray image} of Abell 2811 with the EoR0 field contours overlaid as in the left panel. In both panels the linear scale is at the redshift of the cluster.}
\label{fig:a2811}
\end{figure*}

{The left panel of Fig.~\ref{fig:a2811} shows Abell 2811.} At the centre of the cluster we make a new detection of a faint radio halo (Obj.~A). As part of the XMM-\emph{Newton} survey of the soft X-ray background \citet{hs13} consider this emission a galactic halo, however, \citet{szz+09} note that the surrounding X-ray emission is offset from the BCG by 27~arcsec ($\sim$55~kpc), which suggests that the cluster is in a dynamic, merging state and that the 168~MHz radio emission seen in Fig.~\ref{fig:a2811} is a cluster halo. The XMM-\emph{Newton} data (Obs. ID 0404520101, PI Sivanandam) shown in Fig.~\ref{fig:a2811} reveals slight N-S elongation of the ICM, further hinting at the dynamical state. \par
This radio halo is on the order of {$\sim$3.4~arcmin} (LLS of {$\sim$400~kpc}). Measuring $S_{168} = 81 \pm 17$~mJy and  $S_{1400} < 3.1$~mJy we find $\alpha_{168}^{1400} \leq -1.5 \pm 0.1$. This would class this as an ultra-steep spectrum radio halo \citep[USSRH; e.g.][]{cbs06,bgc+08}. 

\hypertarget{link:a4038}{}\subsubsection{Abell 4038}

\begin{figure}[t!]
\includegraphics[width=0.99\linewidth]{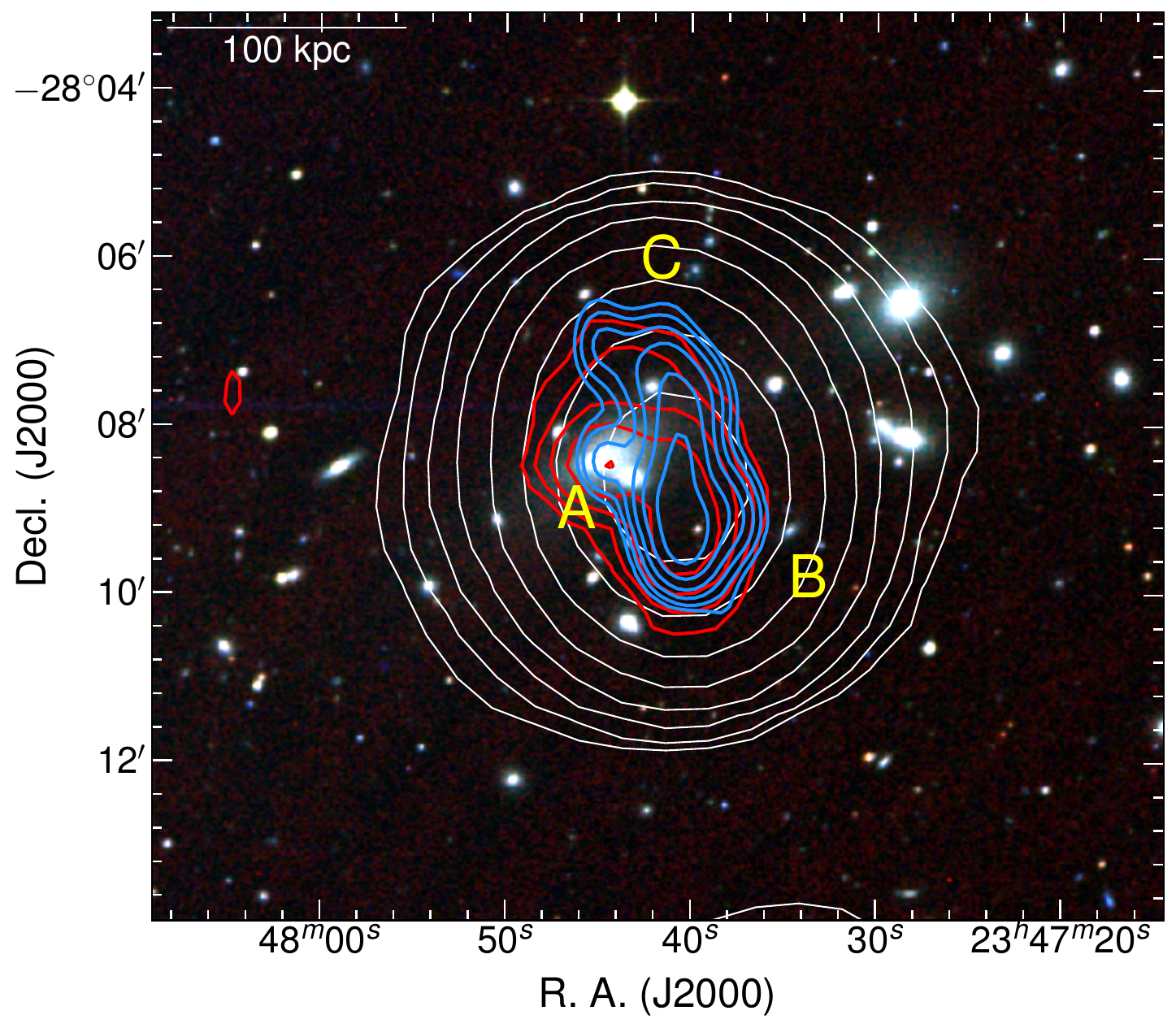}
\caption{The centre of Abell 4038.\corrs{DSS2} RGB image with contours overlaid as follows: EoR0 field, white, beginning at 20~\mjybeam; NVSS, \nvsscontour, beginning at 1.5~\mjybeam; TGSS, \tgsscontour, beginning at 20~\mjybeam. Marked objects are described in the text.}
\label{fig:a4038}
\end{figure}

{\citet{sr84} report a steep-spectrum source (Obj.~B and C in Fig.~\ref{fig:a4038}), and \citet{sr98,srm+01} follow-up and class it as a radio phoenix. The emission of the phoenix blends with the radio emission from IC~5358 and 2MASX~J23474209$-$2807335 (Obj.~A in Fig.~\ref{fig:a4038}). From measurements provided by \citet{kd12,srm+01,fj73} we subtract flux density contributions from discrete cluster sources and measure {$S_{168} = 4.79 \pm 0.25$~Jy} for the phoenix. 

\hypertarget{link:as84}{}\subsubsection{Abell S0084}

\begin{figure*}[t!]
\includegraphics[width=0.478\linewidth]{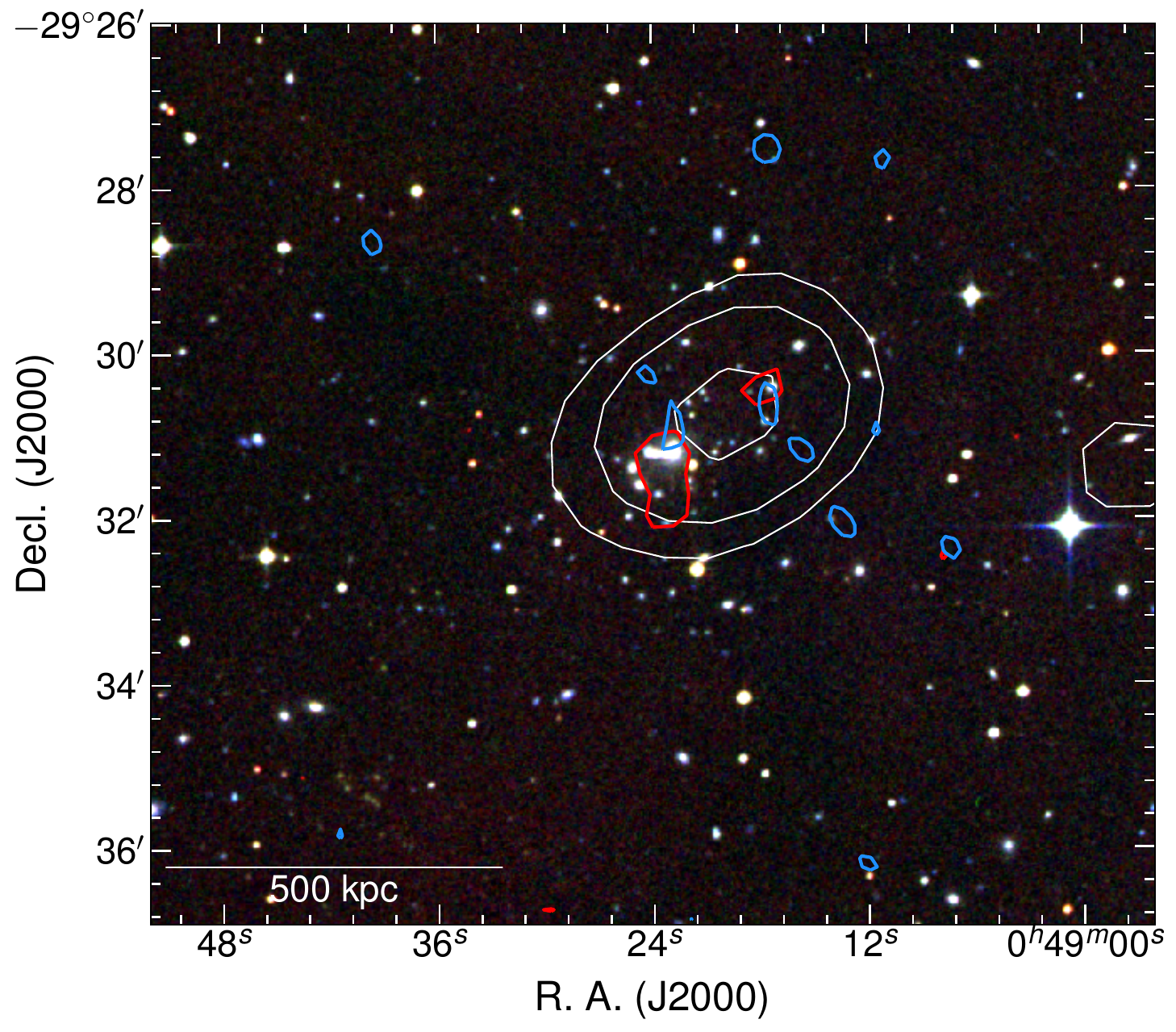}\hfill
\includegraphics[width=0.478\linewidth]{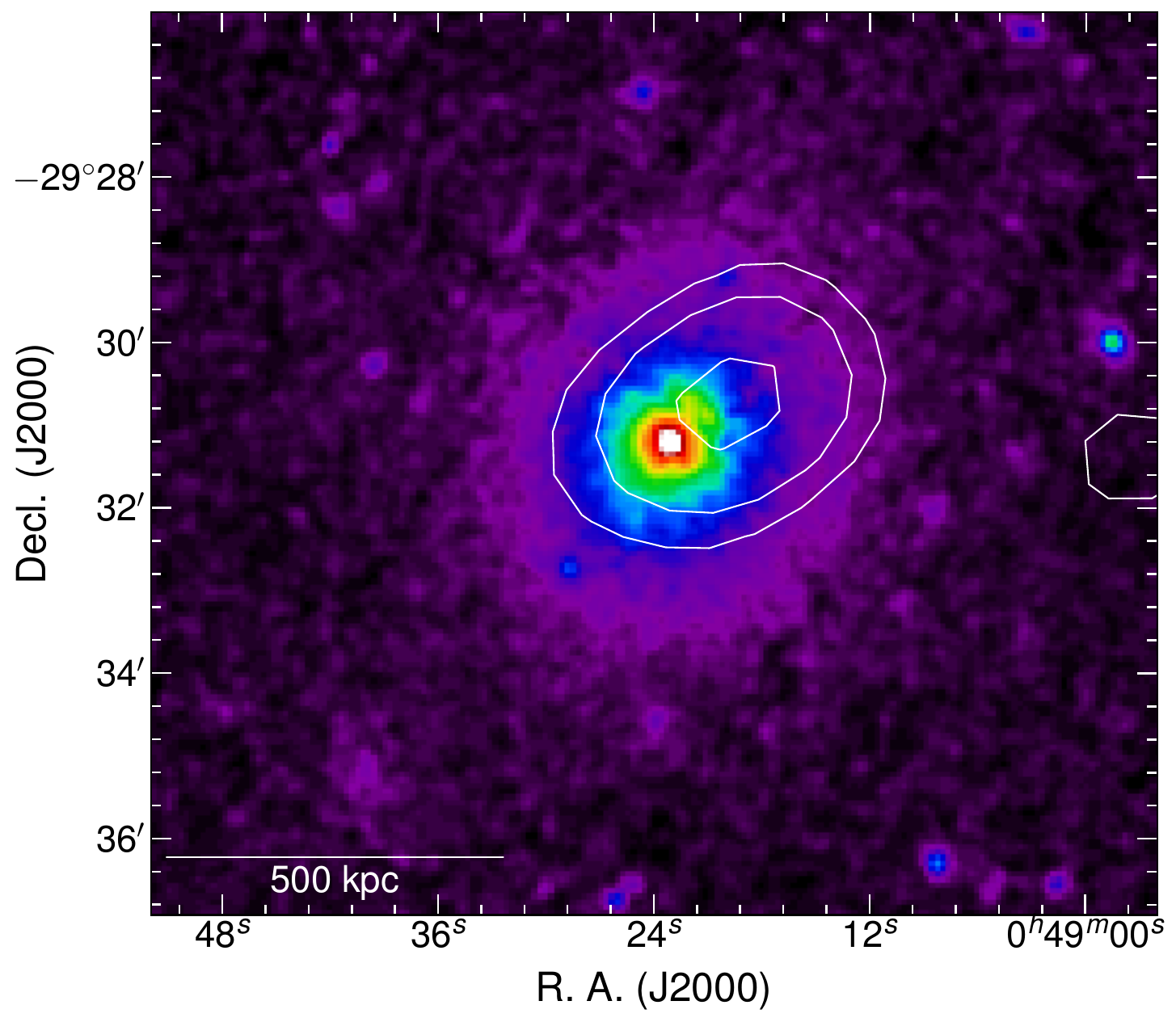}
\caption{The centre of Abell S0084. \emph{Left:} \corrs{DSS2} RGB image with contours overlaid as follows: EoR0 field, white, beginning at 7~\mjybeam; NVSS, \nvsscontour, beginning at 1.5~\mjybeam; TGSS, \tgsscontour, beginning at 12~\mjybeam. \emph{Right:} {Exposure corrected, smoothed XMM-\emph{Newton}} X-ray image from the {\sffamily REXCESS} survey with EoR0 contours overlaid as in the left panel along with X-ray contours.}
\label{fig:as84}
\end{figure*}

{We detect diffuse radio emission at the centre of Abell S0084 (Fig.~\ref{fig:as84}).} The cluster was part of the ARDES sample of \citet{sjp16} though no diffuse emission was detected at the centre of the cluster. We measure {$S_{168} = 32 \pm 5$~mJy} and calculate a limit of $S_{1400} \leq 2.2$~mJy from the rms noise in the ARDES data (Shakouri, priv. comms.). This results in $\alpha_{168}^{1400} \leq -1.3 \pm 0.1$. We measure an LAS of 3.5~arcmin (LLS of 420~kpc). \par
The right panel of Fig.~\ref{fig:as84} shows the {\sffamily REXCESS} X-ray data with MWA contours overlaid. There is no cavity present in the X-ray data to suggest that the emission could be the lobes of an AGN and thus may be associated with the cluster itself. Further, Abell S0084 is not a cool core cluster \citep{pcab09} and so we do not suspect this emission is a mini-halo. Given that the radio emission sits offset from the X-ray peak by $\sim$100~kpc and that the X-ray plasma appears undisturbed, we only tentatively classify this as a candidate radio halo, though note that the emission may be from a centrally located radio galaxy, possibly dying or otherwise of old age.%

\hypertarget{link:as1099}{}\subsubsection{Abell S1099}

\begin{figure}[t!]
\includegraphics[width=0.99\linewidth]{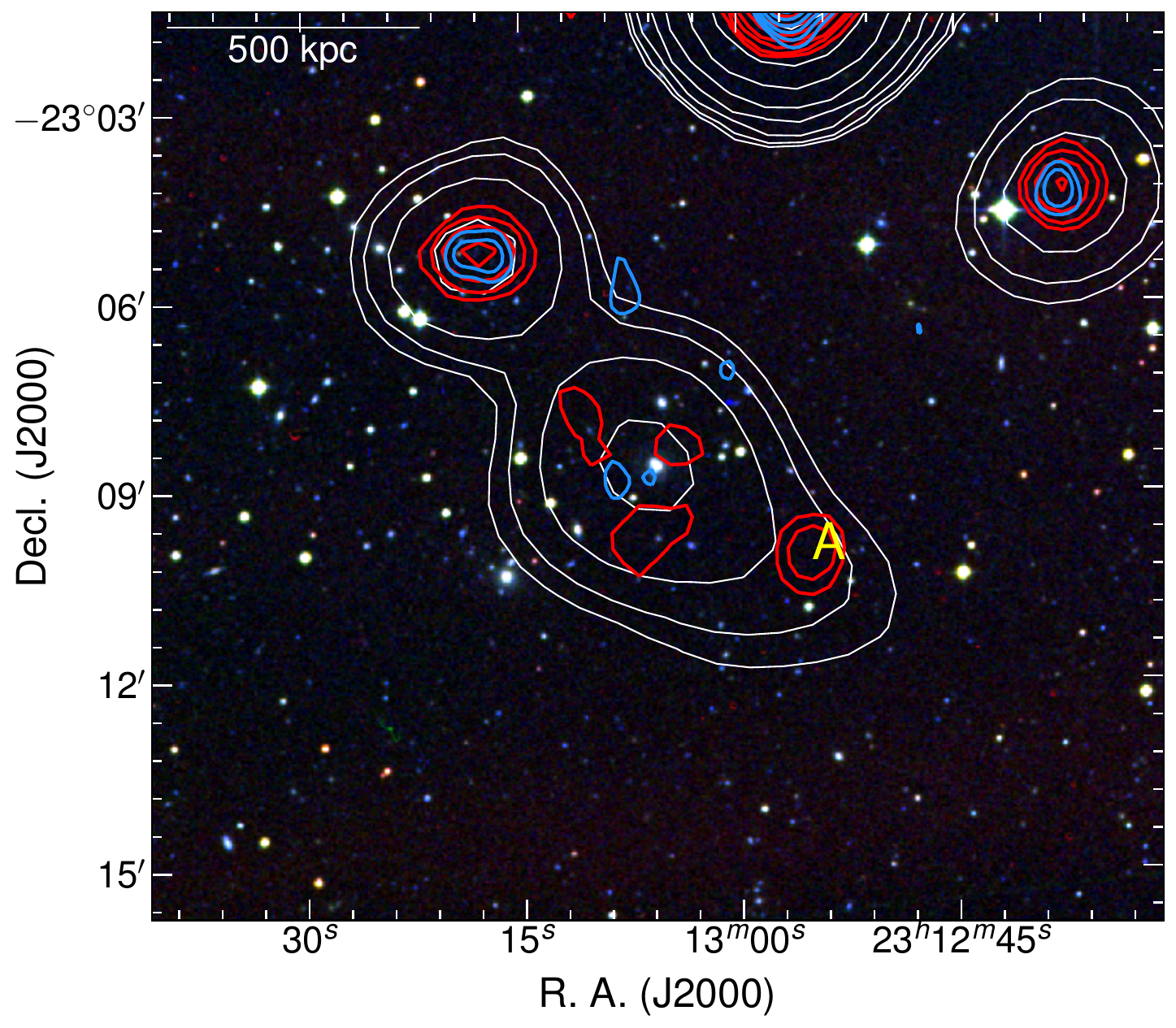}
\caption{Diffuse emission in Abell S1099. \corrs{DSS2} RGB image with contours overlaid as follows: EoR0 field, white, beginning at 10~\mjybeam; NVSS, \nvsscontour, beginning at 1.5~\mjybeam; TGSS, \tgsscontour, beginning at 11.1~\mjybeam. The linear scale is at the redshift of Abell S1099. `A' marks a likely embedded source.}
\label{fig:as1099}
\end{figure}

{Fig.~\ref{fig:as1099} shows Abell S1099 as an RGB image with MWA, NVSS, and TGSS contours overlaid.} The cluster hosts extended, diffuse emission coincident with one of the BCGs, 2MASX~J23130574-2308369 ({$z=0.1086 \pm 0.0002$}; \citealt{cmw04}), which coincides with the peak of the emission at 168~MHz. We measure an LAS of $\sim$9.5~arcmin (LLS of $\sim$1.1~Mpc). We measure $S_{168} = 180 \pm 20$~mJy, where the uncertainty includes a term to account for the slight blending towards the northwestern source. Further, Obj.~A in Fig~\ref{fig:as1099} appears to be an embedded point source, catalogued as NVSS~J231255$-$230959 \citep{ccg+98}, which is not accounted for. We convolve the NVSS image to $108.87$ arcsec $\times$ $108.87$ arcsec and obtain $S_{1400} = 22.3 \pm 6.4$~mJy. This yields $\alpha_{168}^{1400} = -1.0 \pm 0.2$. This is consistent with radio halos or perhaps an intervening radio relic. Without significant X-ray emission detected by \textit{ROSAT}, we suggest it is unlikely to be a radio halo. We cannot confirm the nature of the emission presently. 

\hypertarget{link:as1121}{}\subsubsection{Abell S1121}

\begin{figure*}[t!]
\includegraphics[width=0.478\linewidth]{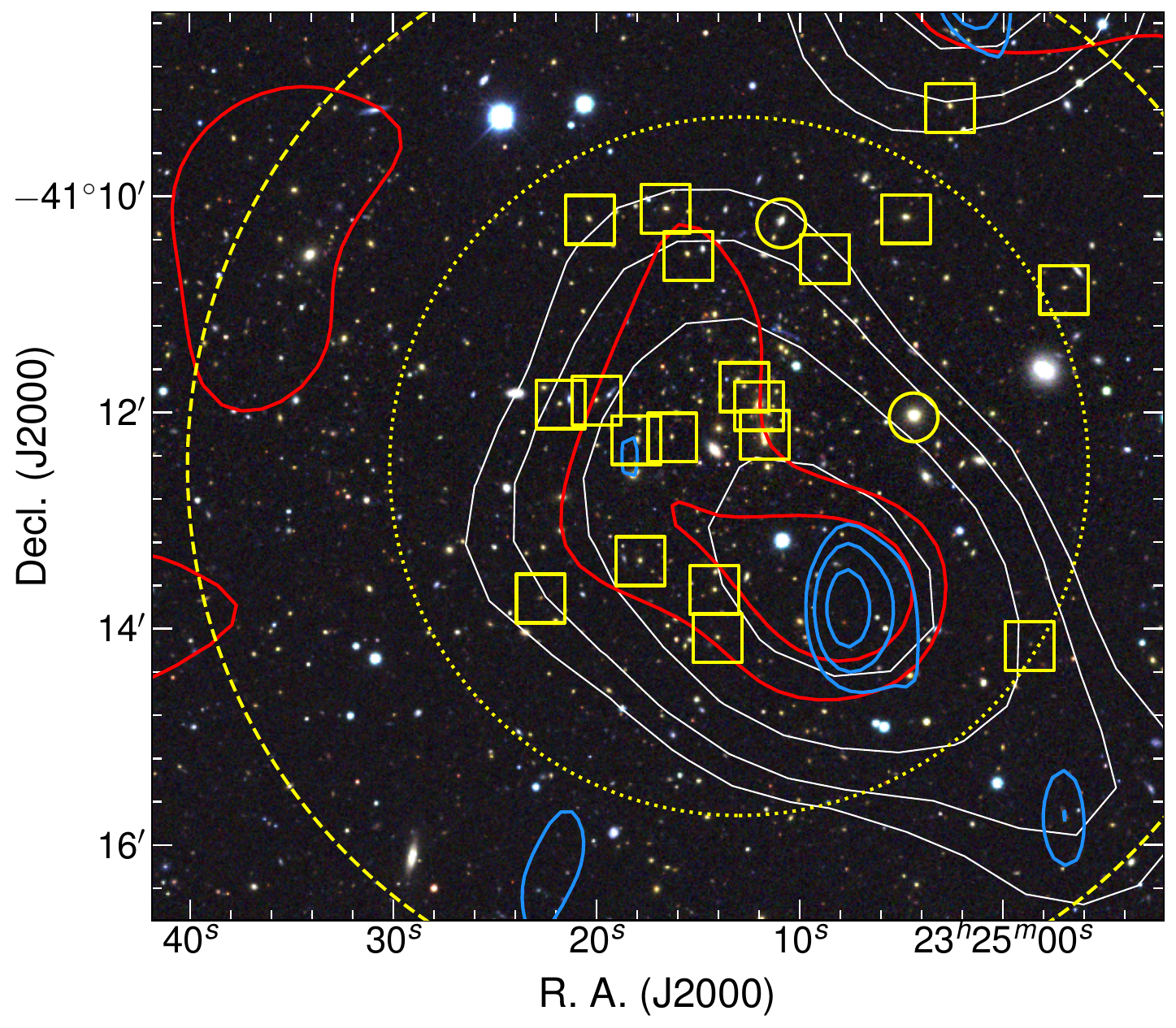}\hfill
\includegraphics[width=0.478\linewidth]{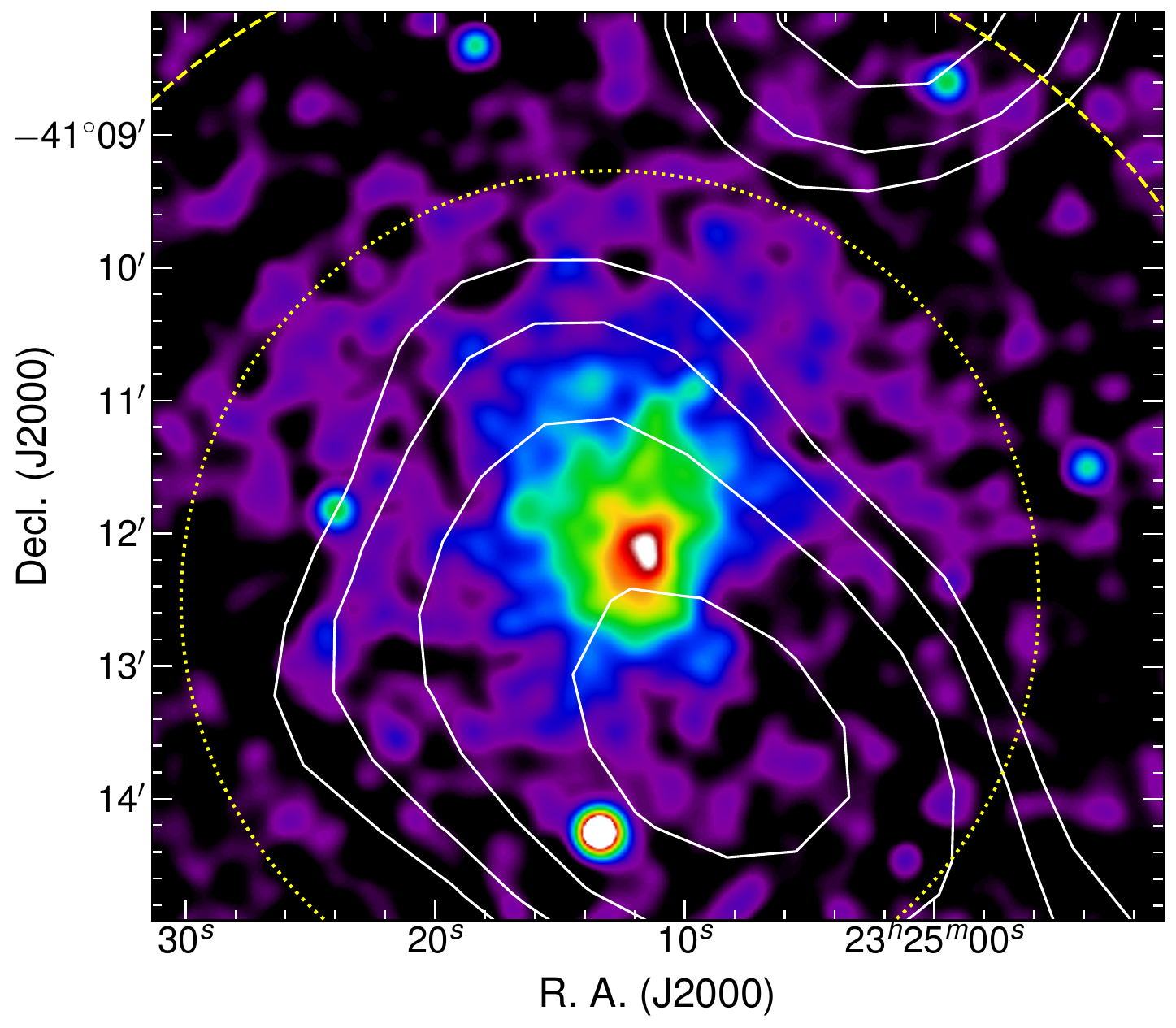}
\caption{Candidate radio halo within Abell S1121. \emph{Left:} \corrs{DES~DR1} RGB image with contours overlaid as follows: EoR0, 15~\mjybeam; smoothed SUMSS, \sumsscontour, beginning at 7 \mjybeam; and TGSS, purple, beginning at 13.8~\mjybeam. The dashed and dotted circles are centred on the cluster with 1~Mpc radii at each of the reported redshifts. The squares and small circles indicate galaxies with available redshifts within {$cz \approx 2000$}~km\,s$^{-1}$ at the two reported redshifts (see text). \emph{Right:} Smoothed, archival \emph{Chandra} data with EoR0 contours as in the left panel. Note that the right panel has a smaller field of view as the \emph{Chandra} image does not cover the entire region shown in the left panel\corrs{---the dotted and dashed circles are identical to those in the left panel}.}
\label{fig:as1121}
\end{figure*}

We detect diffuse emission in Abell~S1121 located near the cluster centre, with \corrs{counterpart emission} in SUMSS (Fig. \ref{fig:as1121}). Note \corrs{the} artefacts in the SUMSS data from a nearby bright source \corrs{(top left contour in Fig. \ref{fig:as1121})}. Abell S1121 is reported by \citet{cac+09} to have a redshift of $z=0.19043$ though \citet{lhd+15} report a redshift of $z=0.3580$ for this system. The left panel of Fig.~\ref{fig:as1121} shows galaxies with available redshifts in the range $cz \approx 2000$~km\,s$^{-1}$} around the reported redshifts, with the small circles associated with {$z=0.19043$} and the small boxes associated with $z=0.3580$. Given the location and numbers of each galaxy distribution, we consider the emission (and the cluster) to be at the redshift reported by \citet{lhd+15}, $z=0.3580$. There is likely a separate, intervening system along the line-of-sight that \citet{cac+09} are measuring. \par
The right panel of Fig.~\ref{fig:as1121} shows archival \emph{Chandra} data (Obs. ID 13405, PI Garmire, exposure time 8.94 ks, 0.1--13.1 keV). This X-ray emitting plasma is situated in the core of the cluster, but shows cone-like morphology suggesting a complex dynamical state. A significant component of the low-frequency emission coincides with the X-ray emission. \corrs{We measure the flux density within an approximate region around the source, obtaining $S_{168} = 80 \pm 13$~mJy.} The SUMSS emission does not coincide with the radial artefacts so is real and we measure $S_{843} = 11 \pm 2$~mJy yielding $\alpha_{168}^{843} = -1.2 \pm 0.2$. We do not detect significant emission in the TGSS data suggesting no significant point source contribution. \par
We estimate an LAS of $\sim$3.6 arcmin (LLS of $\sim$1.1~Mpc at $z=0.3580$). The slight offset from the X-ray centroid and the elongation in SUMSS might suggest a radio relic, perhaps intervening along the line of sight, though the properties are consistent with a radio halo.

\hypertarget{link:as1136}{}\subsubsection{Abell S1136}

\begin{figure}[t!]
\includegraphics[width=0.99\linewidth]{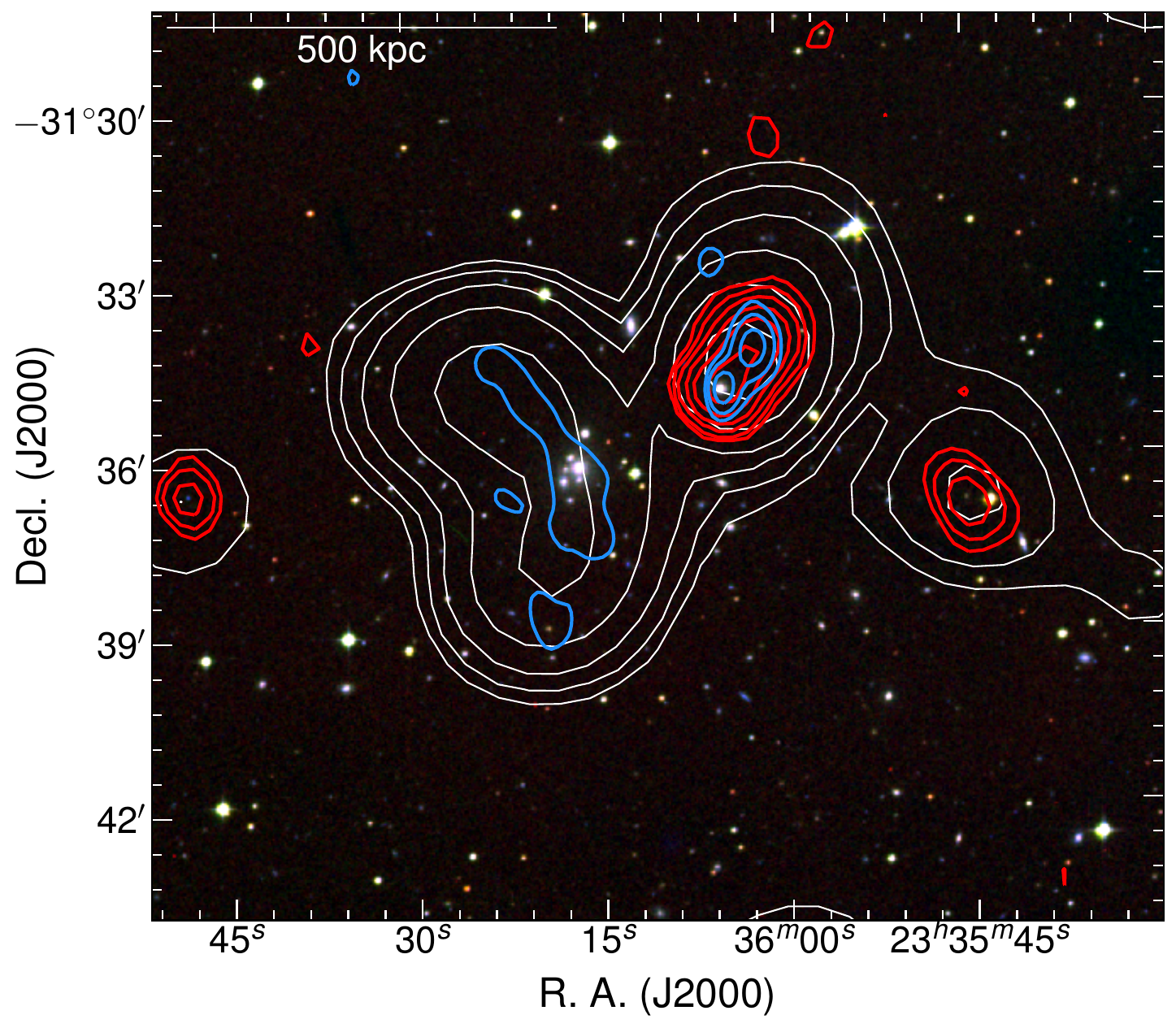}
\caption{Diffuse emission within Abell S1136. \corrs{DSS2} RGB image with contours overlaid as follows: EoR0 field, white, beginning at 10~\mjybeam; NVSS, \nvsscontour, beginning at 1.5~\mjybeam; TGSS, \tgsscontour, beginning at 14.4~\mjybeam. The linear scale is at the cluster redshift.}
\label{fig:as1136}
\end{figure}

{Fig.~\ref{fig:as1136} shows the centre of Abell S1136} with elongated diffuse radio emission appearing strongly at 168~MHz, with a patchy counterpart in the TGSS survey at 147.5~MHz. There is no corresponding 1.4~GHz or 843~MHz emission seen in the NVSS or SUMSS surveys suggesting a steep spectrum. The total flux density of the diffuse source is decomposed using \texttt{aegean} and measured to be $S_{168} = 586 \pm 46$~mJy. \par
The RASS broad-band 0.1--2.4~keV image does not show particularly strong X-ray emission at the centre, and the RGB image (Fig.~\ref{fig:as1136}) shows the optical concentration of galaxies at the centre is offset towards the west of the bulk of the 168~MHz emission. The elongation is north-south, with an almost bent double-lobed structure, and has an {LAS of $\sim$6.8~arcmin (LLS of $\sim$490~kpc)}. While the emission could be classified as a cluster halo, alternate explanations are those of cluster relic intervening along the line-of-sight towards the cluster, or a dead radio galaxy likely having a previous association with the BCG, ESO~470-G020. Without polarisation data and higher resolution imaging we do not classify this emission here. This cluster will be investigated further using ASKAP data (Macgregor et al., in prep.). 

\hypertarget{link:rxc}{}\subsubsection{RXC~J2351.0$-$1954}\label{sec:rxc}

\begin{figure*}[t!]
\includegraphics[width=0.478\linewidth]{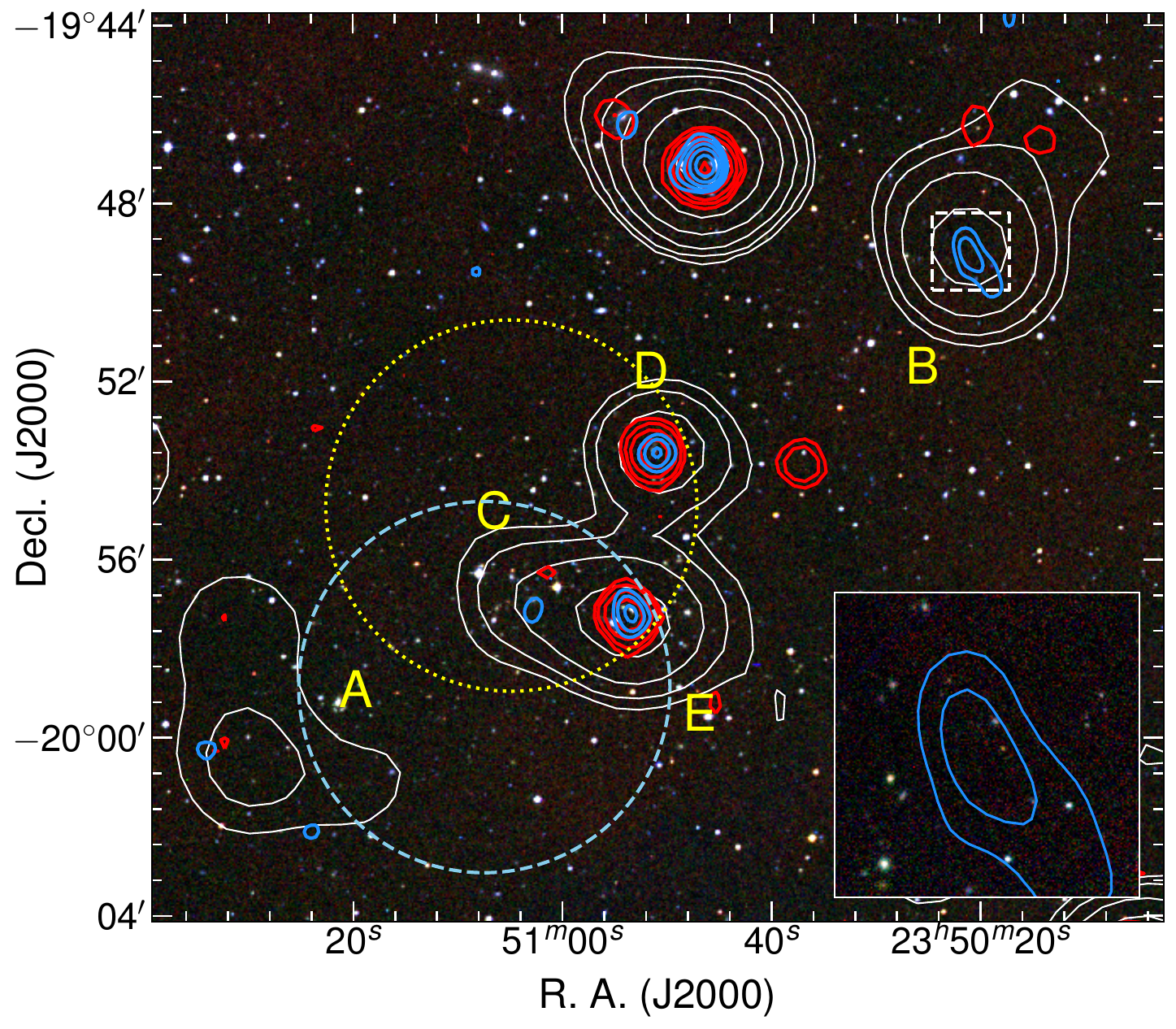}\hfill
\includegraphics[width=0.478\linewidth]{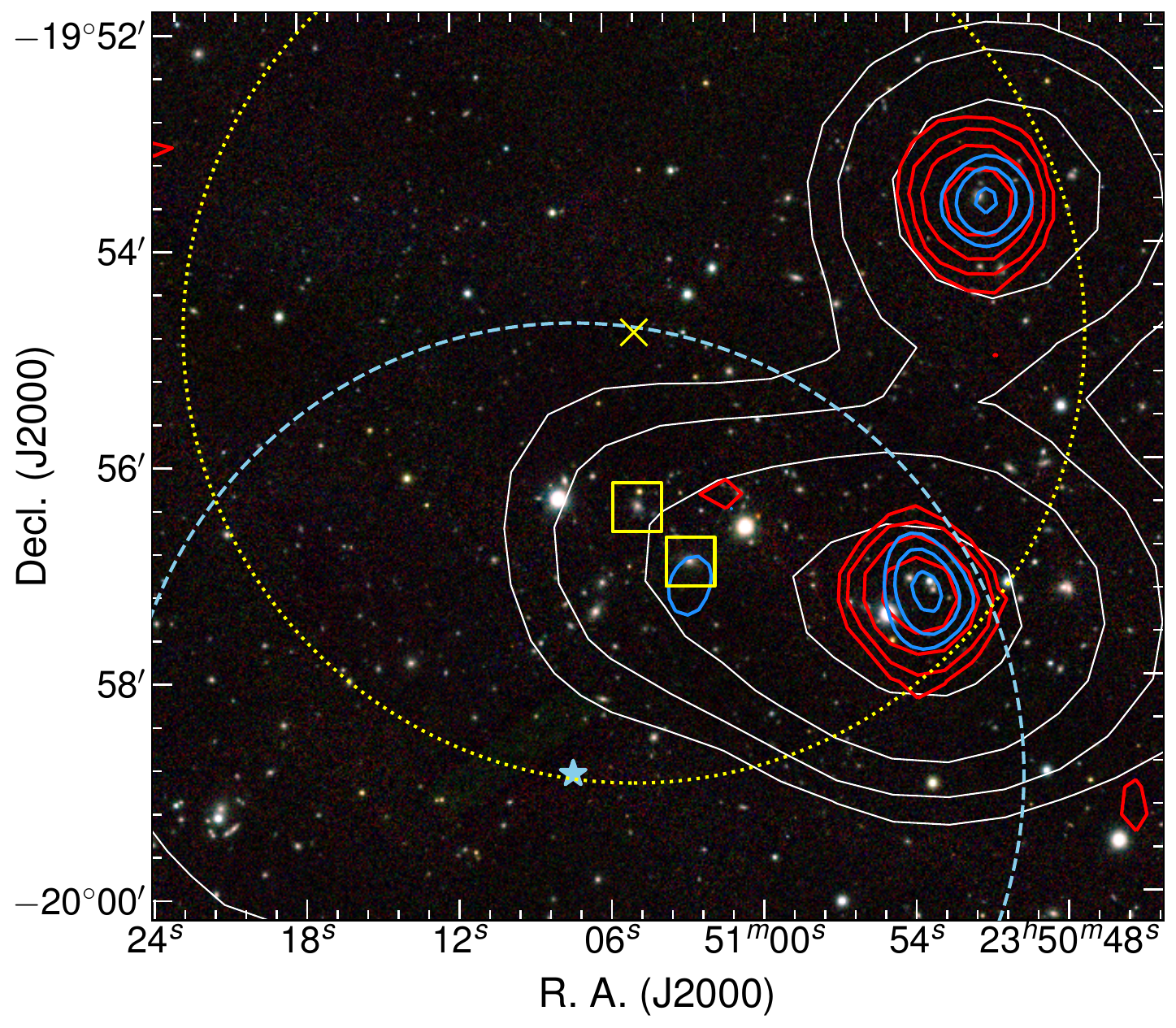}
\caption{RXC~J2351.0$-$1954. \emph{Left:} \corrs{DSS2} RGB image with contours overlaid as follows: EoR0 field, white, beginning at 10~\mjybeam; NVSS, \nvsscontour, beginning at 1.5~\mjybeam; TGSS, \tgsscontour, beginning at 13.8~\mjybeam. The dashed, blue circle is centred on the PSZ1 coordinates, and a dotted, yellow circle is centered on the X-ray coordinates, both with radii of 1~Mpc. `A' and `B' mark candidate relics, `C' a candidate halo, and `D' and `E' are other radio sources mentioned in the text. \corrs{The inset is the PS1 data with its location indicated on the image as a dashed, white box. The TGSS contours are shown on the inset as in the main figure.} \emph{Right:} A smaller field of view of the left panel \corrs{a with PS1 RGB background}, a cross to denote the cluster's coordinates given by \citet{cb12}, a star to denote the coordinates given by \citetalias{planck15}, and squares showing galaxies with spectroscopic redshifts in the region.}
\label{fig:rxc}
\end{figure*}

{The left panel of Fig.~\ref{fig:rxc} shows RXC~J2351.0$-$1954 (PSZ1~G057.09$-$74.45) and the surrounding field,} and the right panel shows the central region of the cluster. The dotted and dashed circles are centered on the RXC and PSZ1 coordinates, respectively. In the left panel of Fig.~\ref{fig:rxc} two steep-spectrum, diffuse sources are located to the southeast (Obj.~A) and northwest (Obj.~B). The right panel shows extended emission coinciding with the optical center, marked as Obj.~C. \citet{cb12} report this cluster as X-ray luminous. \par

Obj.~A (left panel Fig. \ref{fig:rxc}) may be a radio relic, with no optical ID and no counterpart emission in either NVSS or SUMSS. We measure {$S^{\rm{A}}_{168} = 57 \pm 9$~mJy} and $S^{\rm{A}}_{1400} \leq 4.2$}, yielding {$\alpha_{168}^{1400,\rm{A}} \leq -1.2 \pm 0.1$}, consistent with relic sources. A has an {LAS of 5.8~arcmin (LLS of 1.4~Mpc)}. We consider Obj.~A a candidate relic. \par

Obj.~B (left panel Fig. \ref{fig:rxc} \corrs{and inset}) is a candidate for a second relic on the opposite side of the cluster to Obj.~A. B is significantly brighter than A, with partial detection in the TGSS data but no detection at 1.4-GHz with the NVSS. There is no visible optical ID. We measure $S^{\rm{B}}_{168} = 147 \pm 13$~mJy and $S^{\rm{B}}_{1400} \leq 4.7$~mJy, yielding $\alpha_{168}^{1400,\rm{B}} \leq -1.68 \pm 0.04$. We measure an {LAS of $5.4$~arcmin (LLS of 1.3~Mpc)} and it is located 2.8~Mpc away from the cluster center---an extreme distance for a radio relic \citep[the largest relic distance is $\sim 3$~Mpc, reported by][]{Cuciti2018}. We consider this a candidate radio relic. 

Located near the cluster center, between the two reported centers, Obj.~C may be a faint radio halo. We subtracted the flux density contributions of Obj. E and D (left panel of Fig. \ref{fig:rxc}) from the total emission within contours of the Obj.~C$+$D$+$E complex after extrapolation using NVSS and TGSS measurements. This results in $S_{168} = 87 \pm 17$~mJy. We estimate a corresponding limit of $S_{1400} \leq 4.3$~mJy from the NVSS data, yielding $\alpha_{168}^{1400} \leq -1.4 \pm 0.1$, consistent with radio halo sources. We estimate an LAS of $\sim$1.6 arcmin (LLS of $\sim$370~kpc), though note the source blending into Obj. E makes its full extent unclear. Without supplementary archival \emph{Chandra} or XMM-\emph{Newton} data (and note its reported center is based on \textit{ROSAT} data) it is difficult to definitively classify this emission. Nevertheless, given the location between the reported X-ray and SZ centres, we consider this a newly detected candidate radio halo.

\hypertarget{link:macs}{}\subsubsection{MACS~J2243.3$-$0935}
\label{sec:macs}

\begin{figure}[t!]
\includegraphics[width=0.99\linewidth]{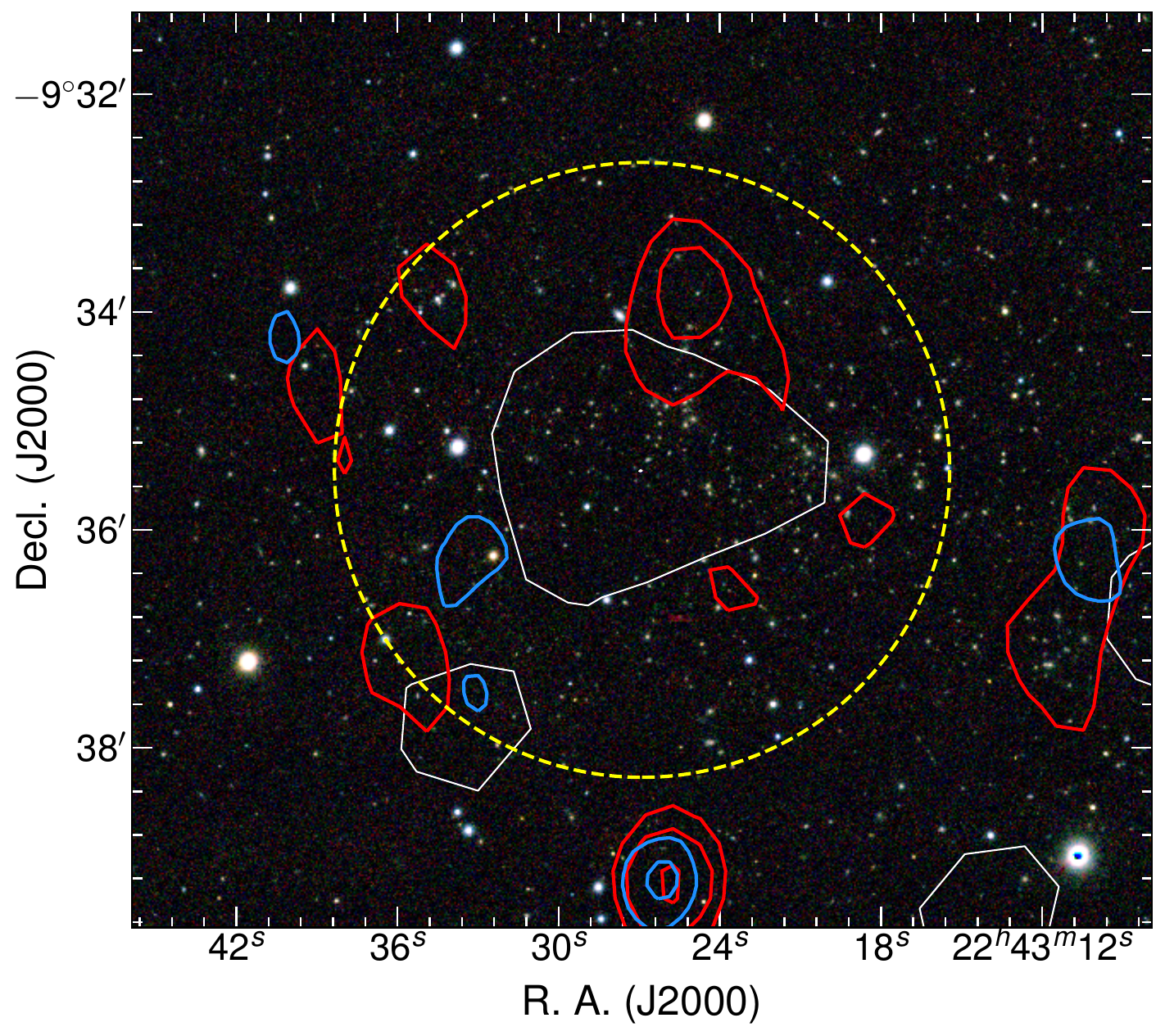}
\caption{MACS~J2243.3$-$0935 with radio halo. \corrs{PS1} RGB image with contours overlaid as follows: EoR0 field, white, beginning at 60~\mjybeam; NVSS, \nvsscontour, beginning at 1.5~\mjybeam; TGSS, \tgsscontour, beginning at 10.2~\mjybeam. The dashed circle is centred on the cluster with a radius of 1~Mpc.}
\label{fig:macs}
\end{figure}

{\citet{cso+16} report the detection of a radio halo in the merging cluster MACS~J2243.3$-$0935 (MCXC~J2243.3$-$0935; PSZ1~G056.94$-$55.06), detected using the Karoo Array Telescope-7 telescope and GMRT.} Fig.~\ref{fig:macs} shows the MWA contours overlaid on the RGB image. MACS~J2243.3$-$0935 is near the edge of the EoR0 field \corrs{where the rms noise is highest.} Because of this, the detection is tentative. Fig.~\ref{fig:macs} shows the cluster with 2$\sigma_{\mathrm{rms}}$ contours to emphasise this. At this level we measure the 168~MHz flux density to be {$S_{168} = 80 \pm 40$~mJy}. With the 610~MHz flux density measured by \citet{cso+16} we obtain a spectral index of $\alpha_{168}^{610} = -1.6 \pm 0.4$. These results should be taken with caution due to the noise in this region of the EoR0 field. In particular, the source size is not sufficient to be considered extended and without the previous detection at 610 and 1826~MHz by \citet{cso+16} of the halo we would not consider this detection sufficient to consider the emission as real and extended.

\hypertarget{link:whl}{}\subsubsection{GMBCG~J357.91841$-$08.97978 and WHL~J235151.0$-$085929}
\label{sec:whlj357}

\begin{figure}[t!]
\includegraphics[width=0.99\linewidth]{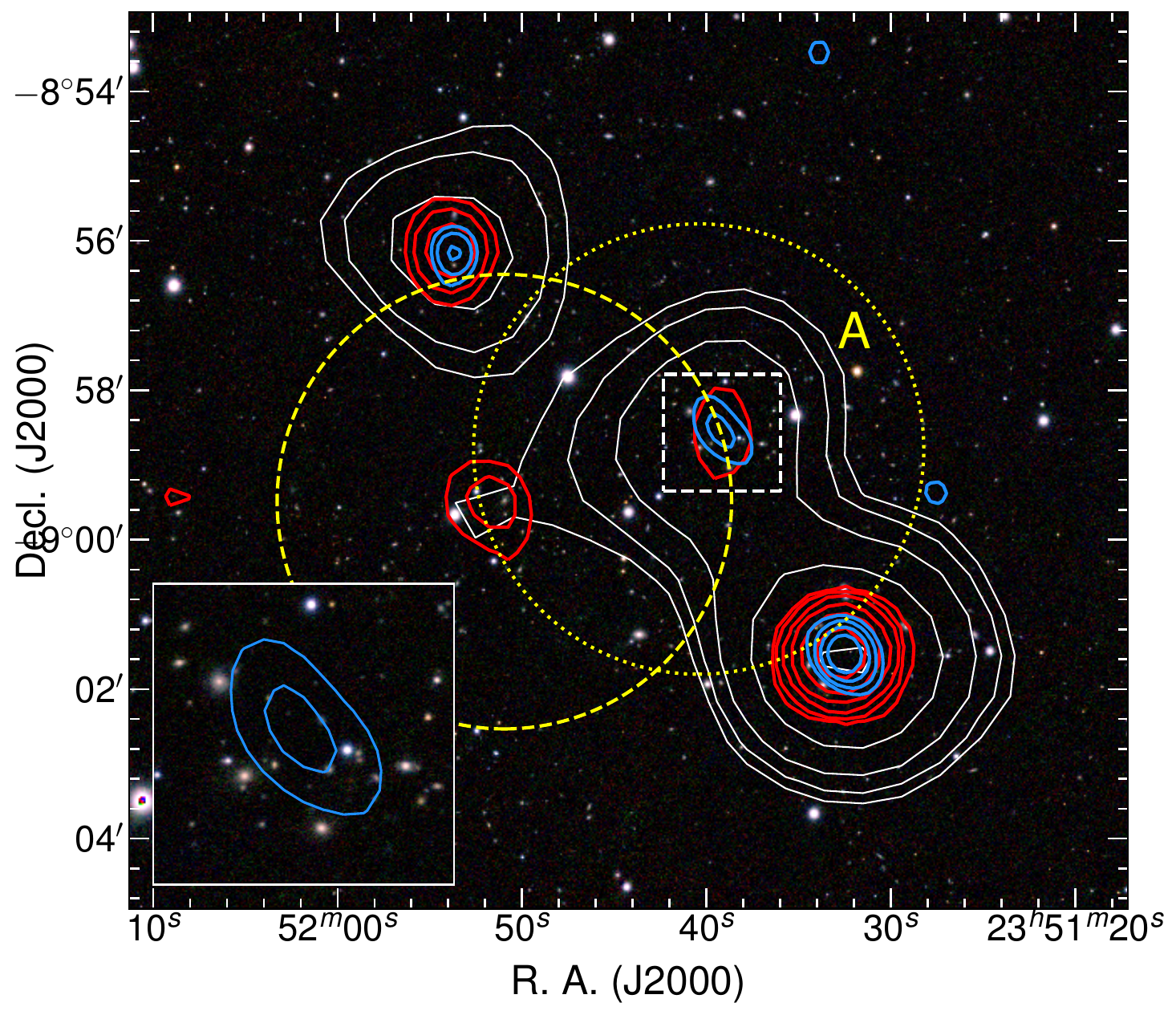}
\caption{Diffuse emission on the periphery of WHL~J235151.0$-$085929 or centre of GMBCG~J357.91841$-$08.97978. \corrs{PS1} RGB image with contours overlaid as follows: EoR0 field, white, beginning at 10~\mjybeam; NVSS, \nvsscontour, beginning at 1.5~\mjybeam; TGSS, \tgsscontour, beginning at 11.7~\mjybeam. The dashed circle is centred on the PSZ1 coordinates of WHL~J235151.0$-$085929 with radius 1~Mpc. The dotted circle is centred on the cluster GMBCG~J357.91841$-$08.97978 with the same 1~Mpc radius. \corrs{The inset is the PS1 data with its location indicated on the image as a dashed, white box. The TGSS contours are shown on the inset as in the main figure.}}
\label{fig:whlj357}
\end{figure}

{Fig.~\ref{fig:whlj357} shows the cluster WHL J235151.0$-$085929 (PSZ1~G082.31$-$67.01, dashed circle)} with an RGB background and radio contours overlaid. Obj.~A is a possible diffuse source on the cluster's periphery. The cluster does not show significant X-ray emission in the RASS broad-band image. The location of the emission relative to the cluster centre and the lack of optical ID (see \corrs{the inset on} Fig.~\ref{fig:whlj357}) are suggestive of a cluster relic. \corrs{Similarly, the NVSS and TGSS data show slightly extended emission, though the 1.4-GHz NVSS detection is at reasonably low significance.} We measure an LAS of 3.2~arcmin (LLS of 1.0~Mpc). The NVSS catalogue flux density is $S_{1400} = 4.1 \pm 0.6$~mJy \citep{ccg+98}. With \texttt{aegean} with find $S_{168} = 128 \pm 20$~mJy, resulting in $\alpha_{168}^{1400} = -1.62 \pm 0.10$. \par
We note that the Gaussian Mixture Brightest Cluster Galaxy \citep[GMBCG;][]{gmbcg} catalogue reports a cluster at the centre of the emission: GMBCG~J357.91841$-$08.97978, with a photometric redshift of $z=0.4$, and the emission may reside within this cluster. If this is the case the steep spectral index and central location would imply a cluster halo. The two clusters, WHL J235151.0$-$085929 and GMBCG~J357.91841$-$08.97978, have centres separated by $\sim$2.7~arcmin which at $z=0.3939$ is $\sim$890~kpc. This separation in both angular distance and redshift would suggest either the clusters may be interacting or that they are the same cluster. With this in mind we suggest that the emission is a candidate cluster halo, at a redshift of $z=0.3939$, associated with the cluster GMBCG~J357.91841$-$08.97978.

\hypertarget{link:as1063}{}\subsubsection{Abell S1063}

\begin{figure*}[t!]
\includegraphics[width=0.478\linewidth]{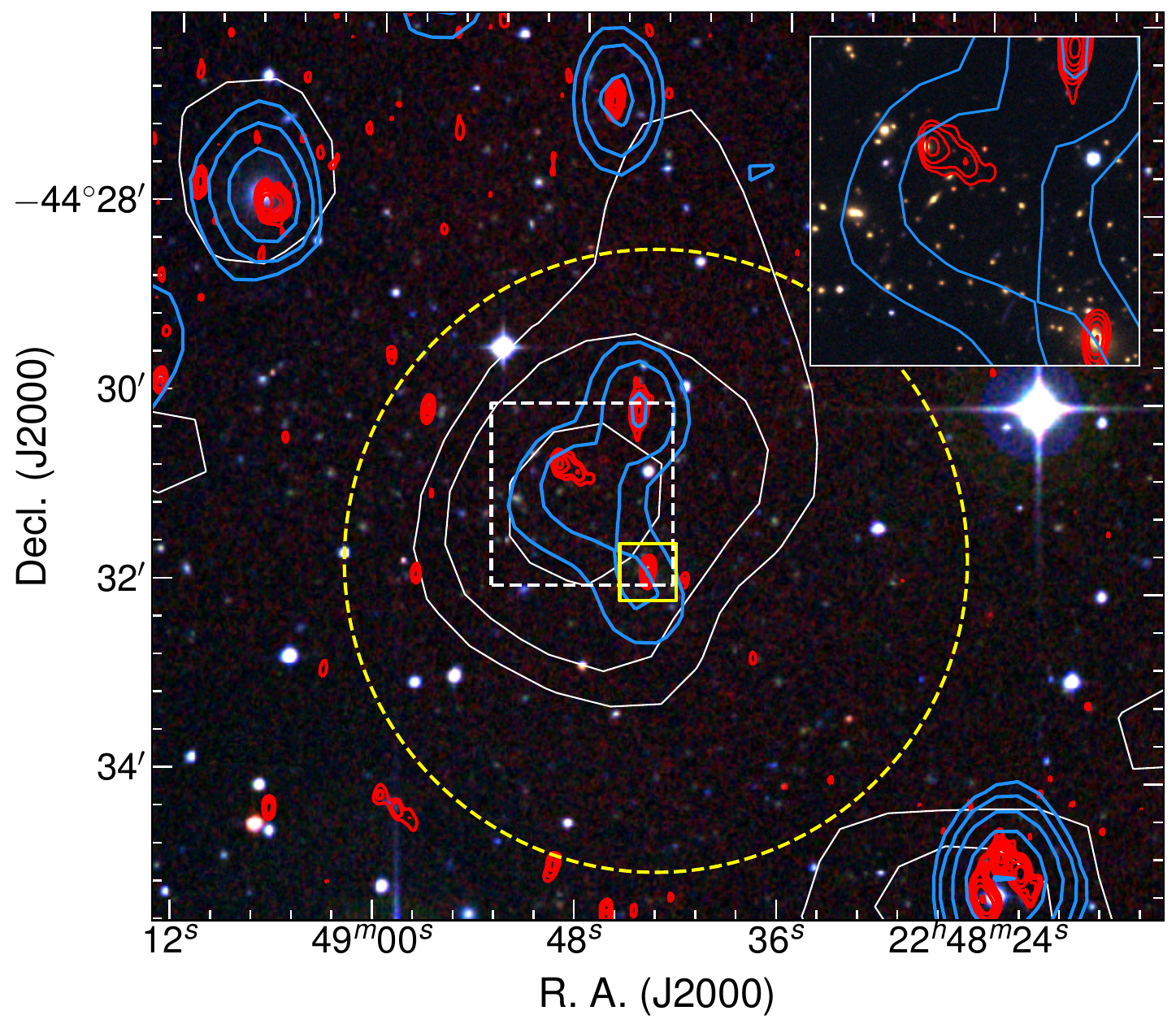}\hfill
\includegraphics[width=0.478\linewidth]{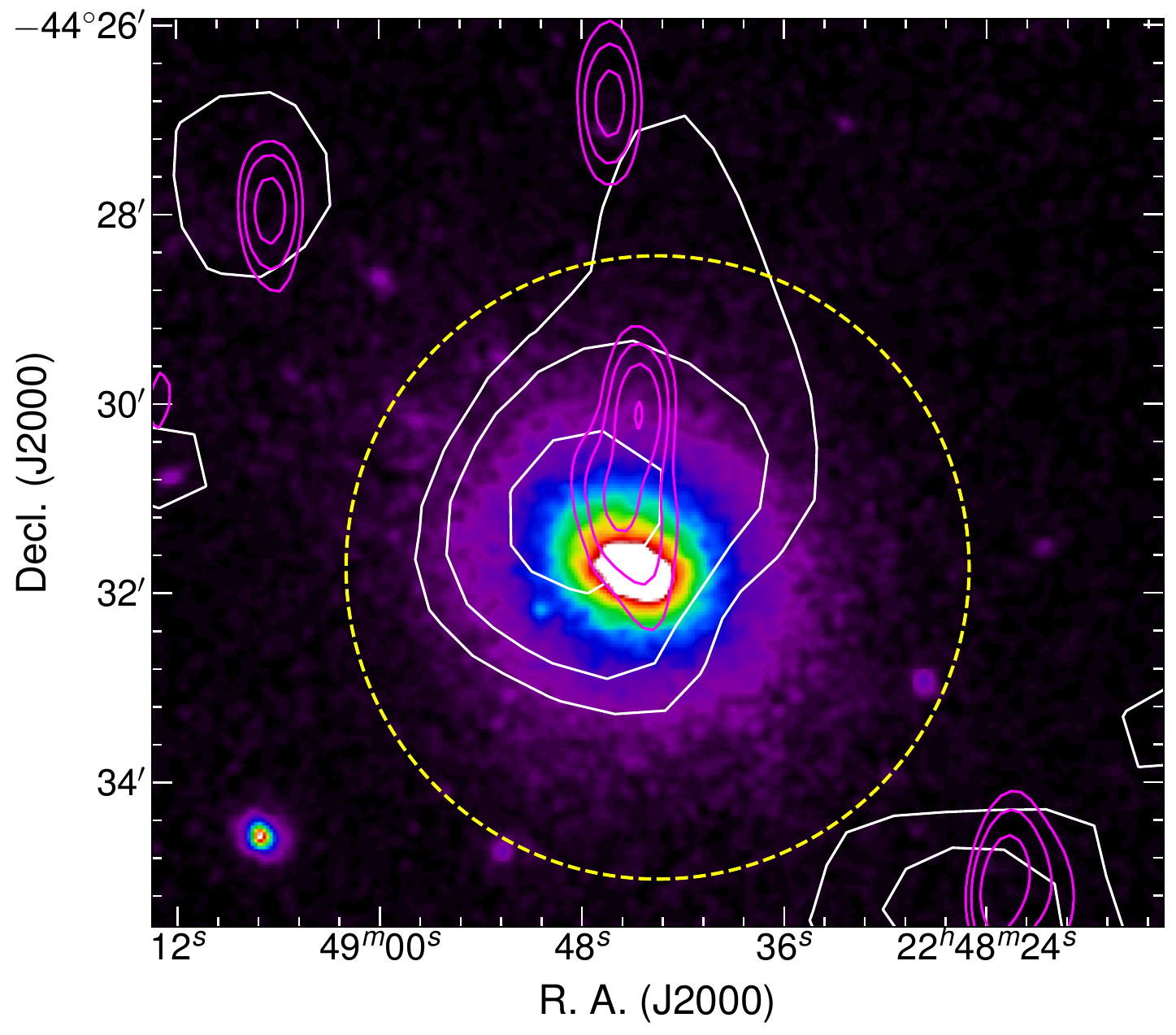}
\caption{Abell S1063. \emph{Left:} \corrs{DSS2} RGB image with contours overlaid as follows: EoR0 field, white, beginning at 50~\mjybeam; ATCA-stacked, \sumsscontour, beginning at 72~$\mu$Jy\,beam$^{-1}$; ATCA-tapered, \tgsscontour, beginning at 1.08~\mjybeam. The small box indicates the BCG of the cluster. \corrs{The inset is the DES~DR1 data with its location indicated by the dashed, white box. ATCA contours are overlaid as in the main image.} \emph{Right:} Exposure corrected, smoothed XMM-\emph{Newton} image with contours EoR0 field contours overlaid as in the left panel, but with TGSS contours, magenta, beginning at 7.8~\mjybeam. The dashed circle in both panels is centred on the cluster and has a radius of 1~Mpc.}
\label{fig:as1063}
\end{figure*}

{Abell S1063 is a \emph{Hubble} Frontier Fields cluster and features heavy gravitational lensing of the distant optical galaxies \citep[see e.g.][]{dbw+16}.} The cluster is near the Southern edge of the EoR0 field and so is more affected by noise. Despite this, above a 50~\mjybeam\ level, a diffuse and elongated piece of emission is seen within the cluster. Exposure corrected, smoothed XMM-\emph{Newton} data (Obs. ID 0504630101, PI Andersson) shows strong X-ray emission coinciding with the 168~MHz radio emission though their respective peaks lie offset from one another, with the X-ray peak situated at the position of the BCG (see Fig.~\ref{fig:as1063}). The X-ray emission can be seen to extend further northeast with the peak of the 168~MHz emission occurring in this same direction. The BCG, LCRS B224549.3-444744, with redshift {$z=0.34711 \pm 0.00025$} \citep{gsb+09} is marked with a square in Fig.~\ref{fig:as1063}. \par
While the EoR0 field 168-MHz data may suggest a radio halo, the higher-resolution ATCA data at 2-GHz reveals the emission to be made up of four individual discrete sources (seen as contours in the left panel of Fig. \ref{fig:as1063}), including the BCG. We find that the BCG has $\alpha_{1332}^{2868} = -1.05 \pm 0.03$, and the apparent spectral steepness of the emission is likely due to contribution from this source. \par
Despite the discrete sources in the cluster, during the preparation of this paper \footnote{During the three year hiatus since the original submission.} \citet{Xie2020} reported the detection of a radio halo from 325--3000~MHz, though the data presented here are unable to provide a clear detection.

\section{Discussion}

\subsection{{Contaminating and blended sources}}

{
The MWA in Phase I has a reasonably low angular resolution when compared to telescopes such as the VLA, GMRT, ATCA, or WRST \footnote{Westerbork Synthesis Radio Telescope}. This is a limitation that arises simply due to the lack of baselines greater than 2873.3~m (in Phase I) and \corrs{its low observing frequency}. Without follow-up observations with higher-resolution instruments it becomes difficult to confirm the nature of emission we find here due to source blending and embedded discrete radio sources or faint point source populations. Two main confusing cases may arise: (i) a source detected at 168~MHz has no counterpart in other survey images, or (ii) \corrs{an apparently} extended source at 168~MHz corresponds to extended sources in other images. \par
(i) A non-detection of a point source with the TGSS ADR1 data with $\sim25$~arcsec resolution provides some confidence in the emission being extended rather than made up of blended point sources. While the sensitivty of the TGSS ADR1 ($\gtrsim 3.5$~mJy\,beam$^{-1}$) does limit the information we get from a non-detection, it does allow us to consider whether a faint source population could reasonably explain the 168-MHz emission. As an example, consider that Abell~2693 would require a faint population of point sources with a total flux density of $S_{150} \sim 56$~mJy. This would require $\sim 5$ discrete sources (unconfused at the TGSS resolution) within an emission area smaller than the MWA beam, which is not possible. Certainly, a portion of the emission may be made up of a faint point source population, but some residual emission must remain. \par 
(ii) This case is more difficult to confirm, however, a deficit in flux between the MWA and TGSS ADR1 data would provide insight into whether there is an extended, low surface-brightness component associated with the emission which may provide support for the existence of diffuse cluster emission. This requires careful measurement of flux densities to have any practical value, and as we near the respectively noise levels in each image this becomes prohibitively more difficult to confirm. Additionally, any difference in flux scale will hinder such an approach \footnote{Noting also a known systematic flux scale discrepancy between TGSS ADR1 and MWA data: \url{http://tgssadr.strw.leidenuniv.nl/doku.php?id=knownproblems}}. \par
In future work with the MWA, a test for the compact nature of emission may be possible through the interplanetary scintillation \citep[IPS; see][]{mme+18,cme+18}. Specifically, \citet{cme+18} show that extended, diffuse sources with steep spectral indices (e.g. the relic in Abell~0085) have low scintillation indices that preclude their emission from being dominated by compact sources. At present, these IPS observations do not reach the required sensitivity to detect most sources in this paper. Making use of these techniques in the future may alleviate some of the issues surrounding low-resolution radio imaging and source-blending.

\subsection{{The scaling relations of cluster radio haloes}}\label{sec:scaling}

\subsubsection{The \texorpdfstring{$P_{1.4}$}--$L_{\mathrm{X}}$ and $P_{1.4}$--$M_{500}$ relations}\label{sec:scaling_relations}

\begin{figure*}[t!]
\includegraphics[width=0.49\linewidth]{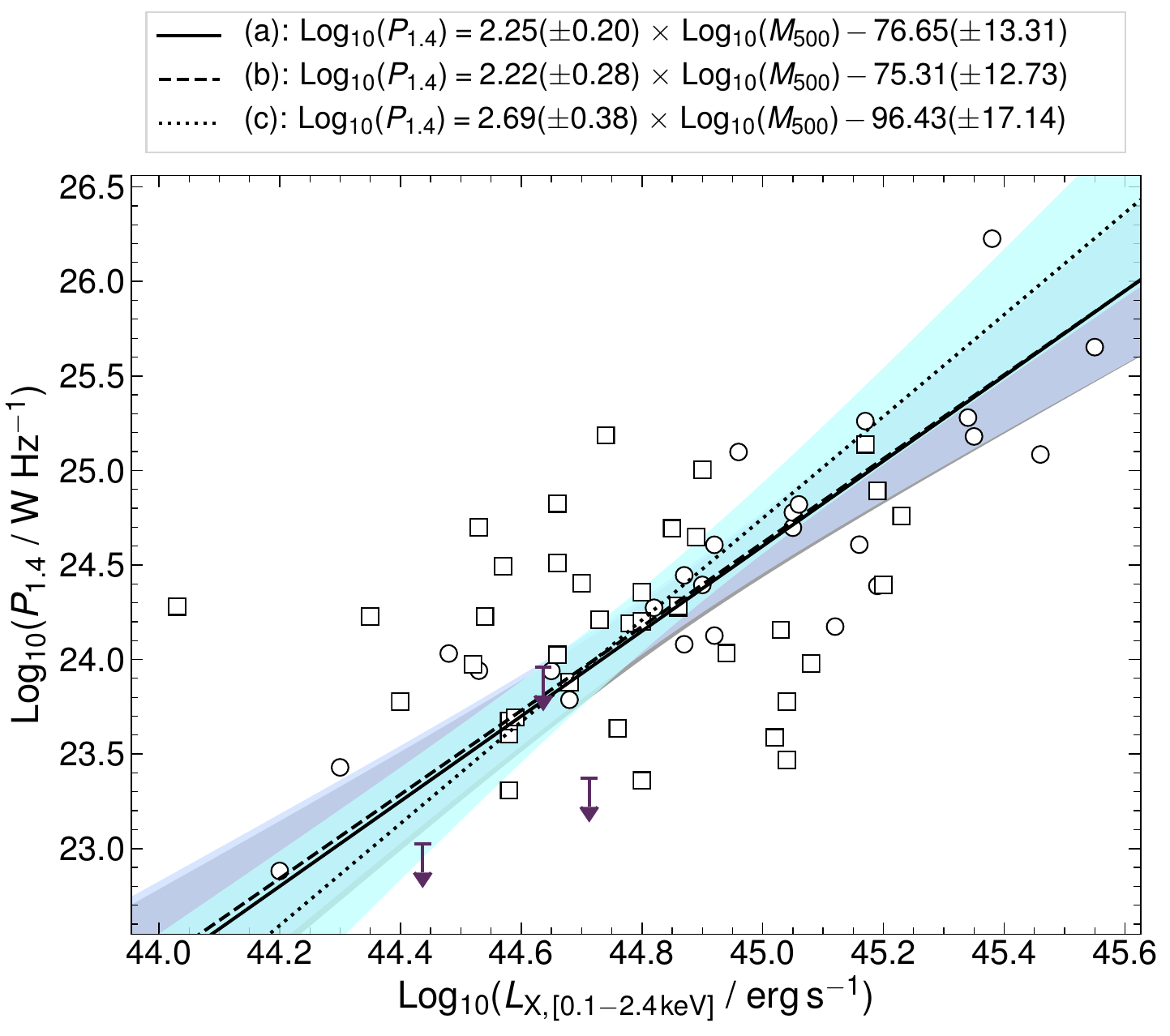} \hfill
\includegraphics[width=0.49\linewidth]{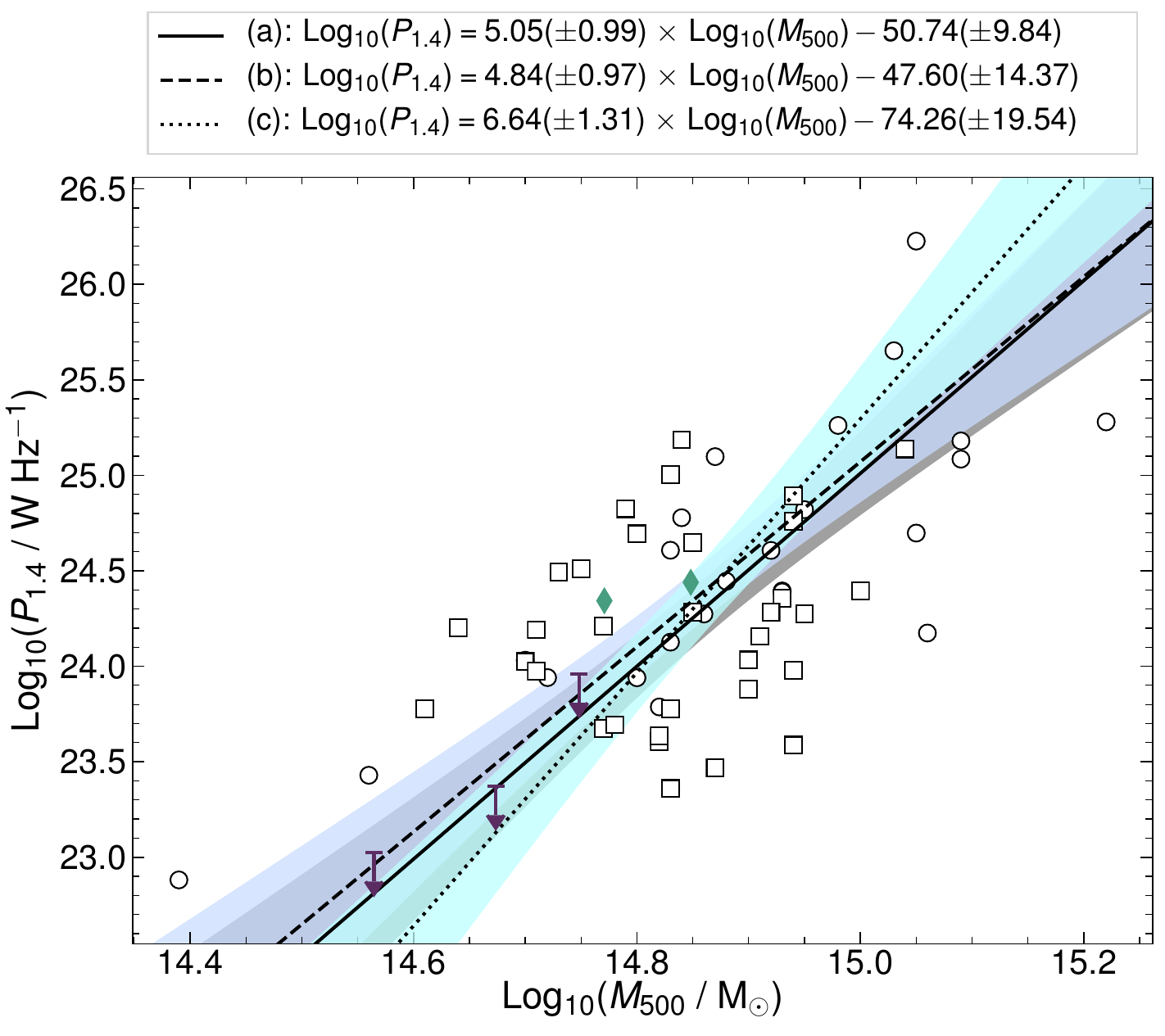}
\caption{{\emph{Left:} The $P_{1.4}$--$L_{\mathrm{X}}$ scaling relation. \emph{Right:} The $P_{1.4}$--$M_{500}$ scaling relation. Error bars have been omitted for the sake of clarity. Unfilled markers are for clusters hosting haloes from the literature. Circles represent those clusters with measured spectral indices, and squares are those assumed to have the average spectral index $\langle \alpha \rangle = -1.47 \pm 0.30$. The dark-purple upper limits are from haloes presented in this paper. The green diamonds are for haloes in GMBCG~J357.91841$-$08.97978  and Abell S1121. The fits presented are from (a) \citet{ceb+13} and (b)/(c) this work---see text for details. The shaded regions represent 95\% confidence intervals~\protect\footnotemark.}}
\label{fig:scaling}
\end{figure*}
\footnotetext{Adapted from \url{https://github.com/rsnemmen/nemmen/blob/master/nemmen/stats.py}}

{An empirical relation exists between the thermal and non-thermal emission of galaxy clusters traced by the synchrotron emission 1.4~GHz power, $P_{1.4}$, and the thermal Bremsstrahlung X-ray luminosity, $L_{\mathrm{X}}$. The $P_{1.4}$--$L_{\mathrm{X}}$ scaling relations have been updated with new surveys halo detections (e.g.~GRHS: I; \citealt{vgb+07} and II; \citealt{vgd+08}, EGRHS \footnote{Extended GMRT Radio Halo Survey}: I; \citealt{kvg+13} and II; \citealt{kvg+15}, KAT-7 observations: \citealt{bvc+16}, ARDES: I; \citealt{sjp16}) to try to understand the link between the thermal X-ray--emitting plasma and radio-halo--emitting electron population and the link to the dynamical state of the cluster \citep[e.g.][]{Cassano2007}. Additional relationships between radio halo power and the cluster's SZ effect with $P_{1.4}$--$Y_{\mathrm{SZ}}$ \citep{basu12} and SZ-dervied mass $P_{1.4}$--$M_{\mathrm{YZ,}500}$ \citep[][but see also \citealt{app+10}]{ceb+13} also exist. 

We compare our newly detected halo/candidate  haloes (that have required mass and X-ray luminosities) with the literature sample of haloes (as of July 2017). We consider the following results in the event the emission within Abell~S1121 is a radio halo. For haloes not measured at 1.4~GHz we extrapolate the flux density measurements to 1.4~GHz using measured spectral indices where available or assuming a spectral index of $\alpha = -1.47 \pm 0.30$ which is the average of the measured indices. We then determine a $k$-corrected $P_{1.4}$ \citep[see][]{hogg99, hbbe02} via \begin{equation}
P_{1.4} = \dfrac{4 \pi {D_{\text{L}}}^2(z)}{\left( 1 + z \right)^{1 + \alpha}} S_{1.4} \quad [\mathrm{W}\,\mathrm{Hz}^{-1}] \, ,
\end{equation}
with the luminosity distance, $D_{\text{L}}(z)$, at the cluster's redshift, and associated error, $\sigma_{P_{1.4}}$, \begin{equation}\label{eq:p_unc}
\sigma_{P_{1.4}} = \dfrac{P_{1.4}}{S_{1.4}} \sqrt{
\left[ S_{1.4} \ln\left( 1 + z \right) \sigma_\alpha \right]^2 +
\left( \sigma_{S_{1.4}} \right)^2
} \quad [\mathrm{W}\,\mathrm{Hz}^{-1}] \, . \end{equation}
Clusters \corrs{used here} from the literature with radio haloes are presented in Appendix~\ref{sec:appendix1}. From here, we consider two radio halo samples: those with measured spectral indices and the full sample.
Fig.~\ref{fig:scaling} shows the $P_{1.4}$--$L_{\mathrm{X}}$ (left panel) and $P_{1.4}$--$M_{500}$ (right panel) relations, with the global sample of clusters as well as clusters from this work with upper limits (dark purple) and exact radio halo powers (green diamonds in the right panel). In Fig.~\ref{fig:scaling}, clusters represented with a circle are those with a measured spectral index and squares are those assuming an average spectral index of $\langle \alpha \rangle = -1.47 \pm 0.30$. Fig.~\ref{fig:scaling} also shows the best-fitting orthogonal BCES \footnote{Bivariate Correlated Errors and intrinsic Scatter, \citep{ab96}} linear regression lines to both relations presented by \citet{ceb+13} (solid line). Additionally, we fit each sample (measured $\alpha$ only and all haloes) via the same method. These fits are shown as dashed and dotted lines, respectively.}\par
{We find that some of the haloes in our newly detected sample lie below the $P_{1.4}$--$L_{\mathrm{x}}$ relation and above the $P_{1.4}$--$M_{500}$ relation. In particular, Abell~2811 and Abell~0141 sit below the $P_{1.4}$--$L_{\mathrm{x}}$ relation, though still fall within the general scatter of the remaining cluster radio halo locations. Though their locations are upper limits, and they may lie even further below if their powers are lower. Similarly, GMBCG~J357.91841$-$08.97978 lies above the $P_{1.4}$--$M_{500}$ relation.}

\subsubsection{{Comparing the scaling relations}}

{We compare the robustness of the $P_{1.4}$--$L_{\mathrm{X}}$ and $P_{1.4}$--$M_{500}$ relations using a measure of the raw scatter, $\sigma_{\mathrm{raw}}$, of the best-fitting BCES regression lines from \citet{ceb+13} and this work. The raw scatter is calculated as the error-weighted orthogonal distances to the best-fitting regression line via \citep[e.g.][]{pcab09, ceb+13} \begin{equation}
\sigma_{\mathrm{raw}}^2 = \dfrac{N}{\left( N-2 \right) \sum_{i=1}^{N} 1/{\sigma_{i}}^2} \displaystyle \sum_{i=1}^{N} \dfrac{1}{{\sigma_i}^2} \left( Y_i - aX_i -b \right) \, ,
\label{eq:raw_scatter}
\end{equation}
where $N$ is the sample size, ${\sigma_i}^2 = \sigma_{y_i}^2 + a^2 \sigma_{x_i}^2$ for uncertainties ${\sigma_y}$, ${\sigma_x}$ in $Y$, $X$, and fitting parameters $a$, $b$.}  \par

\begin{table}[t!]
\centering
\begin{threeparttable}
\caption{Raw scatter between cluster halo samples with best-fitting BCES regression lines to the scaling relations by (a) \citet{ceb+13} and (b) this work.\label{tab:scatter}}
\begin{tabular}{l c c c c}
\toprule
Ref. & $a$ \tnote{a} & $b$ \tnote{a} & Sample & $\sigma_{\mathrm{raw}}$ \\
\midrule
\multicolumn{5}{c}{$P_{1.4}$--$L_\mathrm{X}$} \\
\midrule
\multirow{2}*{(a)} & \multirow{2}*{$2.25 \pm 0.25$} & \multirow{2}*{$-76.65 \pm 13.31$} & Full &0.393 \\
                   &                                &                                   & $\alpha$ &0.305 \\
\multirow{2}*{(b)} &$2.69 \pm 0.38$ & $-96.43 \pm 17.14$ &  Full & 0.423 \\
&$2.22 \pm 0.28$ & $-75.31 \pm 12.73$ &  $\alpha$ & 0.299 \\

\midrule
\multicolumn{5}{c}{$P_{1.4}$--$M_{500}$} \\
\midrule
\multirow{2}*{(a)} & \multirow{2}*{$5.05 \pm 0.99$} & \multirow{2}*{$-50.74 \pm 9.84$} & Full &0.594 \\
                   &                                &                                   & $\alpha$ &0.536 \\
\multirow{2}*{(b)} &$6.64 \pm 1.31$ & $-74.26 \pm 19.54$ &  Full & 0.771 \\
&$4.84 \pm 0.97$ & $-47.60 \pm 14.37$ &  $\alpha$ & 0.550 \\

\bottomrule
\end{tabular}
\begin{tablenotes}
\footnotesize 
\item[a] For $\log_{10}(P_{1.4}) = a \log_{10}(X) + b$ for $X \in \{ L_{\mathrm{X}}, M_{500} \}$.
\end{tablenotes}
\end{threeparttable}
\end{table}

{Table~\ref{tab:scatter} presents the calculated raw scatter in each fit for each of the samples. We see clearly that the measured $\alpha$ sample shows considerably less scatter in all cases, though is also the smaller sample. The sample size from \citet{ceb+13} is 25 haloes, while the full sample used here is 63(59) and the measured $\alpha$ sample is 25(24) for $P_{1.4}$--$L_{\mathrm{X}}$($P_{1.4}$--$M_{500}$).}

\begin{figure}[t!]
\includegraphics[width=0.99\linewidth]{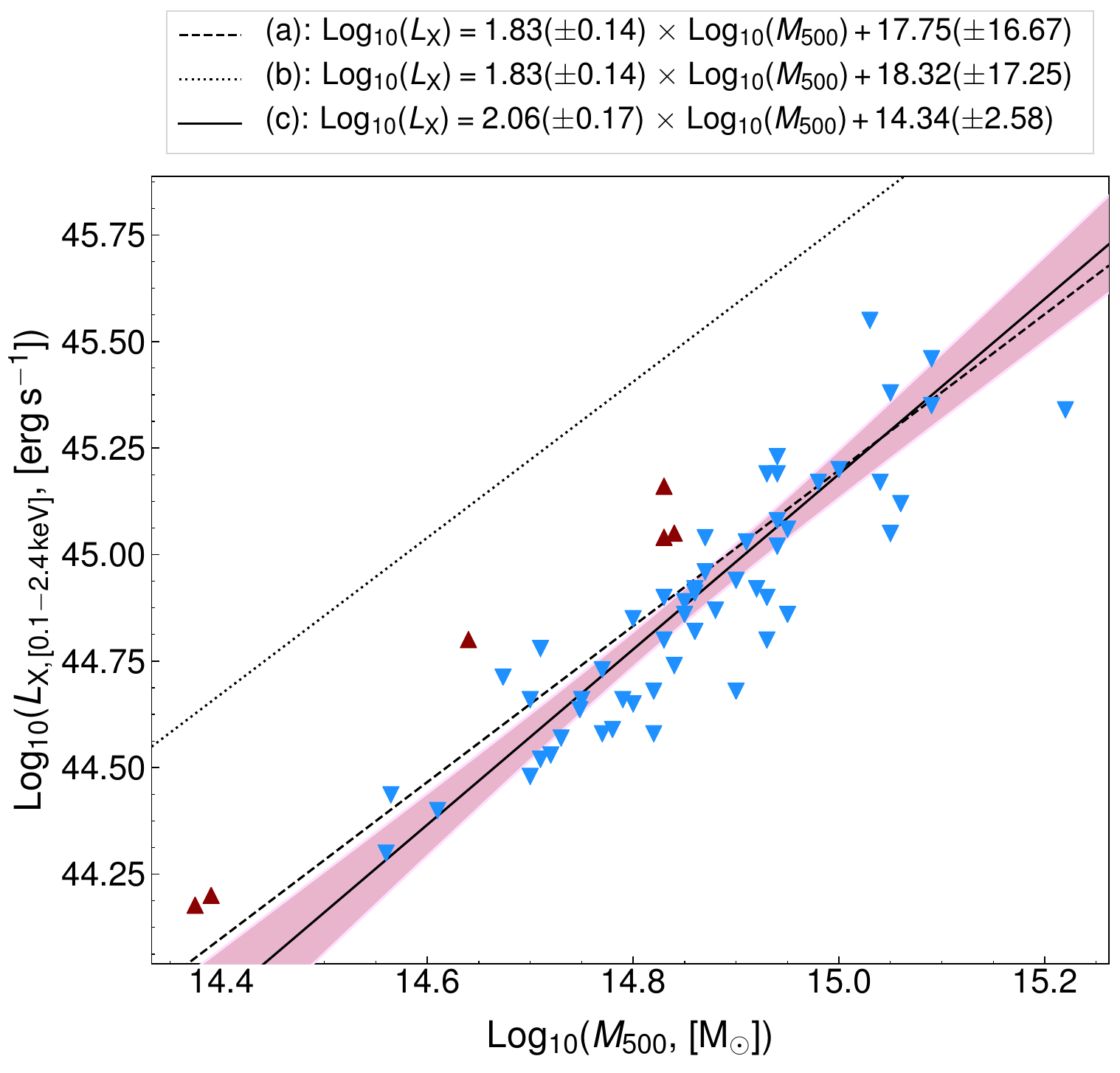}
\caption{{X-ray luminosity against mass for clusters hosting radio haloes. The solid, black fit (c) is made using the BCES orthogonal method for clusters with haloes with the mauve shaded region the 95\% confidence interval as per Fig.~\ref{fig:scaling}. The dashed, black fit (a) is from \citet{pcab09} for clusters within the {\sffamily REXCESS} sample using the same BCES orthogonal fitting method, assuming a redshift of 0. The dotted, black fit (b) is the same fit from \citeauthor{pcab09} but assuming a redshift of 1. The red, up-pointing triangles indicate clusters with predicted $L_\mathrm{X}$ below measured values, and blue, down-pointing triangles are clusters with predicted $L_\mathrm{X}$ above measured values. This sample is the global sample with the addition of the clusters found to host haloes (or candidates) from this work.}}
\label{fig:lum_mass}
\end{figure}

{\citet{pcab09} find, for a representative sample of clusters, that the X-ray luminosities show more scatter when cool core clusters are in the sample. Fig.~\ref{fig:lum_mass} shows the $L_{\mathrm{X}}$--$M_{500}$ relation for clusters hosting radio haloes (literature and this work). The solid, black fit with mauve 95\% confidence region is a BCES orthogonal fit to the cluster hosting haloes, whereas the dotted and dashed, black fits (for $z=1$ and $z=0$) are the equivalent BCES orthogonal fits presented by \citet{pcab09} for the {\sffamily REXCESS} sample of clusters, which comprises both cool and non-cool core clusters. The blue, down-pointing triangles indicate clusters with $L_{\mathrm{X}}$ below their values predicted by the {\sffamily REXCESS} sample and red, up-pointing triangles are with $L_\mathrm{X}$ above predicted values. The six clusters above their predicted $L_\mathrm{X}$ are: CL~1821$+$643, Abell 1914, Abell 0545, Abell 3562, and RXC~J1314.4$-$2512. We note that CL~1821$+$643 hosts a cool core and features a giant radio halo with LLS of $\sim$1.1~Mpc \citep{bib+14}. Abell~3562 hosts a cooling flow \citep{pfe+98} and the lowest-power halo in the global sample \citep{2003A&A...402..913V,gbf+09} and also features the second-lowest mass and lowest luminosity cluster. The remaining three clusters are described as disturbed or dynamically active: Abell~1914 has no cooling flow \citep{whi00} and a morphological analysis performed by \citet{bt96} suggests a dynamical nature; Abell~0545 has an unrelaxed and highly elongated X-ray structure \citep{2003A&A...400..465B}; RXC~J1314.4$-$2512 is dynamically perturbed with bi-modal structure and odd elongation \citep{vmp+02}.}\par
{The sample of clusters used by \citet{pcab09} feature $\sim$32\% cool cores, which in the $L_{\mathrm{X}}$--$T$ relation preferentially lie above the best-fitting parameters. With a significant lack of cool core clusters, we find best-fitting parameters to the $L_{\mathrm{X}}$--$M_{500}$ relation that shows a slightly steeper slope. There is only one cool-core cluster known to host a radio halo (CL~1821$+$643) and this may explain the lower scatter in the $P_{1.4}$--$L_{\mathrm{X}}$ relation compared to that in the $P_{1.4}$--$M_{500}$ relation, where we would otherwise expect a relation derived from homogeneous mass measurements to be more tightly constrained than that of inhomogeneous X-ray measurements. Simulations show that there may be a transient boost to $L_{\mathrm{X}}$ during the course of a cluster merger \citep{ddbc13}, though it is not clear whether this will push the cluster above the $P_{1.4}$--$L_{\mathrm{X}}$ relation or along it with a simultaneous increase to the radio halo power. Less scatter in the $P_{1.4}$--$L_{\mathrm{X}}$ relation may suggest the latter, where transient boosts above the relation would otherwise increase the raw scatter.} \par
A significant contribution to raw scatter in the data (which exists for both $P_{1.4}$--$L_{\mathrm{X}}$ and $P_{1.4}$--$M_{500}$ relations) may arise from inhomogeneous $P_{1.4}$ measurements and determination of $\alpha$. Radio flux densities are often measured on maps made with differing beam sizes and $u$--$v$ plane coverage. In the case of missing $u$--$v$ coverage, not all spatial scales are recovered which results in missing flux, yielding lower limits to integrated flux densities. Additionally, there is no single method used for measuring flux density \citep[see e.g.][]{sb14}. Flux densities are not always measured at 1.4~GHz, and sometimes---as in this work---a lower-frequency integrated flux density is measured for the halo, and a corresponding 1.4~GHz flux density (hence, power) is extrapolated from a calculated or assumed spectral index. The integrated spectral index itself may introduce additional scatter without a well-sampled spectrum. It is typical to assume a spectral index of $-1.3$ \citep[e.g.][]{ceb+13}, though here we use $-1.47$ which is found to be the mean value of the measured spectral indices. Even  haloes measured at 1.4~GHz require an accurate determination of the spectral index as it is important for the \emph{k}-correction in calculating $P_{1.4}$ as $P_{1.4} \propto 1/(1+z)^{1+\alpha}$.  This can result in a $\sim$20\% difference at $z=0.2$ between $\alpha=-1$ and $\alpha=-2$.   \par

\subsection{The incidence of diffuse cluster emission within the EoR0 field}\label{sec:detection_rate}

{This works presents a number of new haloes, relics, and phoenices or candidates of each along with previously objects previously detected. Given the resolution of the EoR0 field (or its approximate beam size) of $\sim$2.3 arcmin, we can estimate the limits in mass and redshift for detecting radio haloes and relics using the $P_{1.4}$--$M_{500}$ scaling relations of \citeauthor{ceb+13} (\citeyear{ceb+13}; for haloes) and \citeauthor{dvb+14} (\citeyear{dvb+14}; for relics). We do not consider the detection limit for phoenices as no scaling relations exist for these objects.} \par
{The two major limiting factors in the detection of such emission are the resolution and sensitivity of the telescope. With the EoR0 field's approximate beam size of $\sim$2.3 arcmin, the viable detection range for distant haloes is $z \leq 0.22$ for $\mathrm{LLS} \leq 500$~kpc or $z \leq 0.67$ for $\mathrm{LLS} \leq 1000$~kpc. Beyond this, haloes become point sources. The second issue is sensitivity; the EoR0 field reaches a sensitivity of approximately 2.3~\mjybeam\ in the central regions of the image. The lowest theoretical sensitivity of the Phase I MWA is approximately 1.7~\mjybeam\ \citep{fjo+16}. With the redshift limits above for the 500 and 1000 kpc radio haloes (or relics, as the argument is the same when approaching the beamsize) the sensitivity is not the major limiting factor. If we assume the smallest-power halo can be detected at $>6.9$~mJy, and if we assume $\alpha = -1.47$ that goes into the \emph{k}-correction, then the limits on detectable radio halo power are $P_{1.4}(z=0.22) \geq 0.7 \times 10^{23}$ and $P_{1.4}(z=0.67) \geq 10 \times 10^{23}$~W\,Hz$^{-1}$. This entire range falls below what is typically seen of cluster haloes \citep[e.g.][]{ceb+13,kvg+15,sjp16}.}  \par

\begin{figure}[t!]
\includegraphics[width=0.99\linewidth]{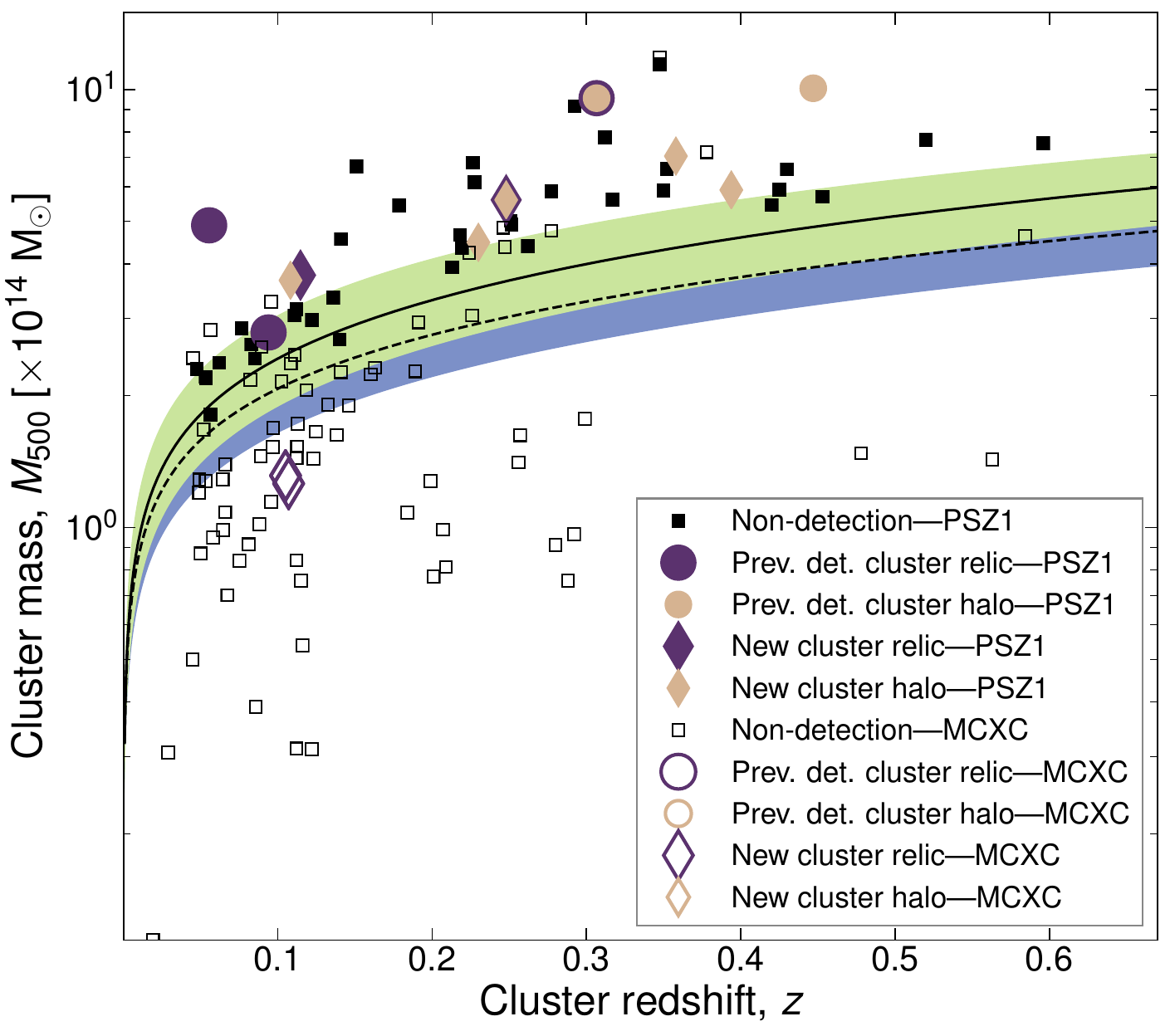}
\caption{\label{fig:redshift} Cluster mass against redshift for clusters within the MCXC and PSZ1 catalogues. The filled points are those using the PSZ1 $M_{\mathrm{YZ,}500}$ measurements and unfilled points are those using the MCXC $M_{500}$ measurements. Where clusters appear in both catalogues we use the PSZ1 $M_{\mathrm{YZ,}500}$ measurements. {The olive-green and blue shaded regions indicate the limits at which 1~Mpc haloes or relics can be detected given the $P_{1.4}$--$M_{500}$ scaling relations found by \citet{ceb+13} and \citet{dvb+14}, respectively. These are determined assuming a $3\sigma_\mathrm{rms}$ detection of $3\times2.3$~\mjybeam, beam size of 2.3 arcmin, and spectral indices between $-0.9 \leq \alpha \leq -2.1$, where a steeper spectral index requires a lower mass cluster. The solid and dashed curves indicate limits for $\alpha=-1.47$ for haloes and relics, respectively.}}

\end{figure}

{Fig.~\ref{fig:redshift} shows clusters within the MCXC and PSZ1 catalogue located within the EoR0 field plotting mass against redshift.  We indicate clusters hosting haloes and/or relics---both previously detected and those detected as part of this work. We use the scaling relations of \citeauthor{ceb+13} (\citeyear{ceb+13}; for haloes) and \citeauthor{dvb+14} (\citeyear{dvb+14}; for relics) to determine the detection limits of these objects within the EoR0 field. For these detection limits, we assume a limiting flux density of 6.9~mJy (the minimum $3\sigma_{\mathrm{rms}}$ for the field) and extrapolate to $P_{1.4}$ for the redshift range $0 \leq z \leq 0.67$ assuming a range of spectral indices $-0.9 \leq \alpha \leq -2.1$. These limiting $P_{1.4}$ values are used with the aforementioned $P_{1.4}$--$M_{500}$ scaling relations to estimate the required mass for a given $P_{1.4}$. These detection limits are plotted as shaded regions with lines drawn at $\alpha=-1.47$. The halo limit region is better defined as the assumption on source size is for a beam-shaped halo, relics typically have shallow spectral indices \citep{fggm12}, \corrs{and because the $P_{1.4}$--$M_{500}$ relation for haloes is better defined with a larger sample and less scatter.} The newly detected relic in Abell~2751 and the candidate in Abell~2798 lie below the $-0.9 \leq \alpha -2.1$ shaded region for relic detections. For the clusters plotted in Fig.~\ref{fig:redshift}, $\sim$37\% lie above the derived 1~Mpc, $\alpha=-1.47$ halo detection limit and $\sim$45\% lie above the corresponding detection limit for relics.}

\section{Summary}

{We have presented diffuse cluster emission detected by the MWA at 168~MHz within the $45\degr \times 45\degr$ EoR0 field including numerous candidates. The field is searched by eye, focusing on clusters part of the Abell, MCXC, PSZ1 galaxy cluster catalogues with reported redshifts. We report the following cluster-based sources:
\begin{enumerate}
    \item 9 halos and candidate halos, including mini-halo sources, of which 2 are known.
    \item 7 relics and candidate relics, of which 2 are known.
    \item 4 known phoenices and 1 candidate phoenix.
    \item 9 sources in clusters with similar features to the above which we cannot classify easily within the current taxonomy.
\end{enumerate}
Where possible, we measure 168~MHz flux densities, estimate angular and linear sizes, and estimate spectral indices or spectral index limits based on non-detections at other frequencies. In particular, we detect a halo associated with the cluster Abell~0141 which is undergoing a merger as suggested by the bi-modality of \corrs{the galaxy distribution} and X-ray--emitting plasma. This halo appears to have an ultra-steep spectrum with $\alpha_{168}^{610} \leq -2.1 \pm 0.1$. Such ultra-steep--spectrum haloes are predicted to be found in low-frequency surveys \citep{cbn+12} and their detection is suggestive of the validity of current halo acceleration models.} \par
{We consider the impact of the MWA's resolution on its ability to properly measure the flux density of sources and its ability to unambiguously confirm the nature of seemingly extended emission, concluding that for the EoR0, the TGSS works well to check for the worst point source contamination, but higher-resolution followups would be needed for cases where source blending may be at play. We examine the newly detected haloes within the context of the established $P_{1.4}$--$L_{\mathrm{X}}$ and $P_{1.4}$--$M_{500}$ scaling relations, finding their locations fairly consistent with other cluster-hosted haloes. We examine previously-found best-fitting relations and derive new fits based on \corrs{a literature sample} and new halo detections.} \par
{We find that radio haloes are predominantly hosted by clusters below the established $L_{\mathrm{X}}$--$M_{500}$ relation for clusters from the {\sffamily REXCESS} sample, with only four examples above the relation not hosting cool cores. Finally, with these new halo and relic detections we examine the incidence of such emission, finding that the MWA is beginning to see emission with little bias beyond what is present in the catalogues the clusters are drawn from.}

\begin{acknowledgements}
The authors would like to thank Emil Lenc, Thomas Reiprich, Tiziana Venturi, Francesco de Gasperin, and Susannah R.~Keel for their help in preparation of this paper. We would also like to thank two anonymous referees for their valuable comments that have improved this paper. SWD, MJ-H, and QZ acknowledge the Marsden Fund administered by the Royal Society of New Zealand. SWD also acknowledges a Doctoral Scholarship from Victoria University of Wellington and and Australian Government Research Training Programme scholarship administered through Curtin University. This scientific work makes use of the Murchison Radio-astronomy Observatory, operated by CSIRO. We acknowledge the Wajarri Yamatji people as the traditional owners of the Observatory site. Support for the operation of the MWA is provided by the Australian Government (NCRIS), under a contract to Curtin University administered by Astronomy Australia Limited. We acknowledge the Pawsey Supercomputing Centre which is supported by the Western Australian and Australian Governments. \par
{This research made use of \texttt{astropy} \citep{astropy}, \texttt{aplpy} (\url{http://aplpy.github.com}), \texttt{NumPy} \citep{numpy}, \texttt{matplotlib} \citep{mpl}, \texttt{iPython} \citep{ipython} and the \texttt{SciPy} library (\url{https://www.scipy.org/}). This research has made use of the VizieR catalogue access tool, CDS, Strasbourg, France.  The original description of the VizieR service was described in \citet{vizier}. }This research also made use of the NASA/IPAC Extragalactic Database (NED) which is operated by the Jet Propulsion Laboratory, California Institute of Technology, under contract with the National Aeronautics and Space Administration. The Digitized Sky Surveys were produced at the Space Telescope Science Institute under U.S. Government grant NAG W-2166. The images of these surveys are based on photographic data obtained using the Oschin Schmidt Telescope on Palomar Mountain and the UK Schmidt Telescope. The plates were processed into the present compressed digital form with the permission of these institutions.
\end{acknowledgements}

{\footnotesize
\bibliographystyle{pasa-mnras}
\bibliography{bib_file}
}

\begin{appendix}

\section{Clusters with radio haloes}\label{sec:appendix1}
\begin{table*}[b]
\centering
\begin{threeparttable}
\caption{Clusters known to host radio haloes as of July 2017 used in Section~\ref{sec:scaling_relations}.}
\label{table:known_cluster_sources}
\begin{tabular}{l c c c c c c}
\toprule
Cluster & $z$ & $S_\nu$ \tnote{a} & $\nu$ & $\alpha$ & log($P_{1.4} / \text{W\,Hz}^{-1}$) & References \tnote{b} \\
        &     & (mJy)   & (MHz) &          &                &            \\
\midrule
% CL~0217+70 & 0.065 & $58.6 \pm 0.9$ & 1400 & $-1.34 \pm 0.19$ & $23.79 \pm 0.01$ & 0/0 \\
Abell~0523 & 0.1 & $72.0 \pm 3.0$ & 1400 & $-1.47 \pm 0.30$ & $24.28 \pm 0.02$ & 1/- \\
Abell~3562 & 0.04 & $20.0 \pm 2.0$ & 1400 & $-1.5 \pm 0.1$ & $22.88 \pm 0.04$ & 2/3 \\
Abell~2061 & 0.078 & $16.9 \pm 4.2$ & 1400 & $-1.8 \pm 0.3$ & $23.43 \pm 0.11$ & 4/4 \\
Abell~2065 & 0.084 & $32.9 \pm 11.0$ & 1400 & $-1.47 \pm 0.30$ & $23.78 \pm 0.15$ & 4/- \\
CL~1446+26 & 0.37 & $9.2 \pm 0.5$ & 1400 & $-1.47 \pm 0.30$ & $24.70 \pm 0.05$ & 5/- \\
Abell~0746 & 0.232 & $18.0 \pm 4.0$ & 1382 & $-1.47 \pm 0.30$ & $24.49 \pm 0.10$ & 6/- \\
Abell~0399 & 0.071 & $16.0 \pm 2.0$ & 1400 & $-1.47 \pm 0.30$ & $23.31 \pm 0.06$ & 7/- \\
Abell~3411 & 0.168 & $4.8 \pm 0.5$ & 1400 & $-1.47 \pm 0.30$ & $23.61 \pm 0.05$ & 8/- \\
Abell~2034 & 0.113 & $13.6 \pm 1.0$ & 1400 & $-1.47 \pm 0.30$ & $23.67 \pm 0.03$ & 5/- \\
Abell~2294 & 0.169 & $5.8 \pm 0.5$ & 1400 & $-1.47 \pm 0.30$ & $23.69 \pm 0.04$ & 5/- \\
Abell~0781 & 0.3 & $20.5 \pm 5.0$ & 1400 & $-1.47 \pm 0.30$ & $24.83 \pm 0.11$ & 9/- \\
Abell~2069 & 0.116 & $28.8 \pm 7.2$ & 1400 & $-1.47 \pm 0.30$ & $24.02 \pm 0.11$ & 4/- \\
Abell~2254 & 0.178 & $33.7 \pm 1.3$ & 1400 & $-1.47 \pm 0.30$ & $24.51 \pm 0.03$ & a/- \\
Abell~0851 & 0.406 & $3.7 \pm 0.3$ & 1400 & $-1.47 \pm 0.30$ & $24.40 \pm 0.06$ & 5/- \\
RXC~J0107.8+5408 & 0.106 & $55.0 \pm 5.0$ & 1382 & $-1.47 \pm 0.30$ & $24.21 \pm 0.04$ & 6/- \\
Abell~1351 & 0.322 & $39.6 \pm 3.5$ & 1400 & $-1.47 \pm 0.30$ & $25.19 \pm 0.05$ & 5/- \\
Abell~2218 & 0.175 & $4.7 \pm 0.47$ & 1400 & $-1.47 \pm 0.30$ & $23.64 \pm 0.05$ & b/- \\
Abell~0209 & 0.206 & $16.9 \pm 1.0$ & 1400 & $-1.47 \pm 0.30$ & $24.36 \pm 0.04$ & 5/c \\
Abell~0401 & 0.073 & $17.0 \pm 1.0$ & 1400 & $-1.47 \pm 0.30$ & $23.36 \pm 0.03$ & d/- \\
RXC~J1514.9-1523 & 0.22 & $102.0 \pm 9.0$ & 327 & $-1.47 \pm 0.30$ & $24.27 \pm 0.20$ & e/- \\
PSZ1~G108.18-11.53 & 0.335 & $6.8 \pm 0.2$ & 1380 & $-1.4 \pm 0.07$ & $24.45 \pm 0.02$ & f/f \\
MACS~J1752.0+4440 & 0.366 & $164.0 \pm 13.0$ & 323 & $-1.47 \pm 0.30$ & $25.00 \pm 0.20$ & g/- \\
Abell~2142 & 0.09 & $18.3 \pm 1.83$ & 1400 & $-1.47 \pm 0.30$ & $23.59 \pm 0.04$ & b/- \\
PSZ1~G216.60+47.00 & 0.382 & $21.0 \pm 2.2$ & 323 & $-1.47 \pm 0.30$ & $24.16 \pm 0.20$ & h/- \\
RXC~J1314.4-2515 & 0.243 & $10.3 \pm 0.3$ & 610 & $-1.47 \pm 0.30$ & $23.78 \pm 0.11$ & c/- \\
PSZ1~G086.47+15.31 & 0.26 & $11.0 \pm 1.2$ & 323 & $-1.47 \pm 0.30$ & $23.47 \pm 0.20$ & h/- \\
Abell~1689 & 0.183 & $11.7 \pm 3.4$ & 1200 & $-1.47 \pm 0.30$ & $23.98 \pm 0.13$ & i/- \\
MACS~J0553.4-3342 & 0.431 & $62.0 \pm 5.0$ & 323 & $-1.47 \pm 0.30$ & $24.76 \pm 0.20$ & g/- \\
MACS~J0417.5-1154 & 0.443 & $10.6 \pm 1.0$ & 1575 & $-1.72 \pm 0.15$ & $25.08 \pm 0.05$ & j/j \\
El~Gordo & 0.87 & $29.0 \pm 3.0$ & 610 & $-1.2 \pm 0.1$ & $25.65 \pm 0.06$ & k/k \\
Abell~0800 & 0.222 & $10.6 \pm 0.8$ & 1400 & $-1.47 \pm 0.30$ & $24.23 \pm 0.04$ & l/- \\
Abell~2255 & 0.081 & $56.0 \pm 3.0$ & 1400 & $-1.47 \pm 0.30$ & $23.97 \pm 0.03$ & m/- \\
Abell~1656~(Coma) & 0.023 & $720.0 \pm 130.0$ & 1400 & $-1.34 \pm 0.06$ & $23.94 \pm 0.08$ & n/n \\
Abell~1550 & 0.254 & $7.7 \pm 1.6$ & 1400 & $-1.47 \pm 0.30$ & $24.23 \pm 0.09$ & l/- \\
Abell~2256 & 0.058 & $103.4 \pm 1.1$ & 1400 & $-1.8 \pm 0.3$ & $23.94 \pm 0.01$ & o/p \\
Abell~0754 & 0.054 & $86.0 \pm 4.0$ & 1400 & $-1.5 \pm 0.3$ & $23.79 \pm 0.02$ & d/d \\
Abell~1995 & 0.319 & $4.1 \pm 0.7$ & 1400 & $-1.47 \pm 0.30$ & $24.19 \pm 0.08$ & 5/- \\

\bottomrule
\end{tabular}
\begin{tablenotes}[flushleft]
\footnotesize
\item[a] Flux densities are converted to a radio power at 1.4~GHz using a measured $\alpha$ where available otherwise assuming an average spectral index of $-1.47\pm0.30$. For measured flux densities without an associated uncertainty, we assume 10\%.
\item[b] References are provided as `reference for $S_\nu$/reference for $\alpha$'.
\end{tablenotes}
\end{threeparttable}
\end{table*}

% continued table environment? 
% \setcounter{table}{8}

\begin{table*}
\centering
\renewcommand\thetable{8}
\begin{threeparttable}
\caption{\emph{continued.}}
\begin{tabular}{l c c c c c c}
\toprule
Cluster & $z$ & $S_\nu$ & $\nu$ & $\alpha$ & log($P_{1.4}$) & References \\
        &     & (mJy)   & (MHz) &          &                &            \\\midrule
        Abell~0545 & 0.154 & $23.0 \pm 1.0$ & 1400 & $-1.47 \pm 0.30$ & $24.20 \pm 0.03$ & d/- \\
Abell~3888 & 0.153 & $27.6 \pm 3.1$ & 1400 & $-1.48 \pm 0.14$ & $24.27 \pm 0.05$ & q/q \\
PLCK~G147.3-16.6 & 0.65 & $7.3 \pm 1.1$ & 610 & $-1.47 \pm 0.30$ & $24.69 \pm 0.14$ & r/- \\
Abell~0773 & 0.217 & $12.7 \pm 1.3$ & 1400 & $-1.47 \pm 0.30$ & $24.28 \pm 0.05$ & a/- \\
MACS~J0416.1-2403 & 0.396 & $1.6 \pm 0.14$ & 1500 & $-1.6 \pm 0.5$ & $24.08 \pm 0.08$ & s/s \\
Abell~0520 & 0.203 & $34.4 \pm 1.5$ & 1400 & $-1.47 \pm 0.30$ & $24.65 \pm 0.03$ & a/- \\
Abell~2319 & 0.056 & $328.0 \pm 28.0$ & 1400 & $-1.2 \pm 0.3$ & $24.39 \pm 0.04$ & 4/4 \\
Abell~0521 & 0.248 & $5.9 \pm 0.5$ & 1400 & $-1.88 \pm 0.07$ & $24.13 \pm 0.04$ & 5/t \\
Abell~0665 & 0.182 & $43.1 \pm 2.2$ & 1400 & $-1.04 \pm 0.02$ & $24.61 \pm 0.02$ & b/u \\
Abell~1758 & 0.28 & $3.9 \pm 0.4$ & 1400 & $-1.47 \pm 0.30$ & $24.03 \pm 0.05$ & 5/- \\
RXC~J2003.5-2323 & 0.317 & $35.0 \pm 2.0$ & 1400 & $-1.32 \pm 0.06$ & $25.10 \pm 0.03$ & v/v \\
PLCK~G171.9-40.7 & 0.27 & $18.0 \pm 2.0$ & 1400 & $-1.84 \pm 0.14$ & $24.70 \pm 0.05$ & w/w \\
Abell~1914 & 0.171 & $64.0 \pm 3.0$ & 1400 & $-1.91 \pm 0.03$ & $24.78 \pm 0.02$ & d/x \\
Abell~1300 & 0.308 & $20.0 \pm 2.0$ & 1400 & $-1.28 \pm 0.09$ & $24.82 \pm 0.04$ & y/z \\
Abell~0697 & 0.282 & $5.2 \pm 0.5$ & 1382 & $-1.64 \pm 0.06$ & $24.17 \pm 0.04$ & 6/6 \\
CL~1821-643 & 0.299 & $11.9 \pm 0.5$ & 1665 & $-1.0 \pm 0.1$ & $24.61 \pm 0.02$ & A/B \\
Abell~2219 & 0.228 & $81.0 \pm 4.0$ & 1400 & $-1.47 \pm 0.30$ & $25.14 \pm 0.03$ & d/- \\
Abell~2744 & 0.307 & $57.0 \pm 3.0$ & 1400 & $-1.19 \pm 0.11$ & $25.26 \pm 0.03$ & z/z \\
CL~0016+16 & 0.541 & $5.5 \pm 0.55$ & 1400 & $-1.47 \pm 0.30$ & $24.89 \pm 0.07$ & b/- \\
MACS~J1149.5+2223 & 0.544 & $1.2 \pm 0.5$ & 1450 & $-2.1 \pm 0.3$ & $24.39 \pm 0.19$ & g/g \\
Abell~2163 & 0.203 & $155.0 \pm 2.0$ & 1400 & $-1.18 \pm 0.04$ & $25.28 \pm 0.01$ & C/u \\
1E~0657~(Bullet) & 0.296 & $24.7 \pm 1.5$ & 2100 & $-1.57 \pm 0.05$ & $25.18 \pm 0.03$ & D/D \\
MACS~J0717.5+3745 & 0.548 & $118.0 \pm 5.0$ & 1465 & $-1.27 \pm 0.02$ & $26.23 \pm 0.02$ & E/E \\
PSZ1~G285.0-23.7 & 0.39 & $2.02 \pm 0.25$ & 1867 & $-1.47 \pm 0.30$ & $24.28 \pm 0.08$ & F/- \\
ACT-CL~J0256.5+0006 & 0.36 & $5.6 \pm 1.4$ & 610 & $-1.0 \pm 0.9$ & $24.03 \pm 0.36$ & G/G \\
MACS~J2243.3$-$0935 & 0.44 & $10.0 \pm 2.0$ & 610 & $-1.47 \pm 0.30$ & $24.39 \pm 0.15$ & H/- \\
Triangulum~Australis & 0.051 & $130.0 \pm 4.0$ & 1328 & $-1.47 \pm 0.30$ & $23.88 \pm 0.02$ & I/- \\
\bottomrule
\end{tabular}
\begin{tablenotes}[flushleft]
\footnotesize \item \textit{References}:
(0)~\citet{2011ApJ...727L..25B};
(1)~\citet{2016MNRAS.456.2829G};
(2)~\citet{2003A&A...402..913V};
(3)~\citet{2005A&A...440..867G};
(4)~\citet{2013ApJ...779..189F};
(5)~\citet{2009A&A...507.1257G};
(6)~\citet{2011A&A...533A..35V};
(7)~\citet{2010A&A...509A..86M};
(8)~\citet{2013ApJ...769..101V};
(9)~\citet{2011A&A...529A..69G};
(a)~\citet{gfg+01};
(b)~\citet{gf00};
(c)~\citet{vgb+07};
(d)~\citet{2003A&A...400..465B};
(e)~\citet{2011A&A...534A..57G};
(f)~\citet{div+15};
(g)~\citet{2012MNRAS.426...40B};
(h)~\citet{2015MNRAS.454.3391B};
(i)~\citet{2011A&A...535A..82V};
(j)~\citet{2017MNRAS.464.2752P};
(k)~\citet{2014ApJ...786...49L};
(l)~\citet{2012A&A...545A..74G};
(m)~\citet{2005A&A...430L...5G};
(n)~\citet{1990ApJ...355...29K};
(o)~\citet{2006AJ....131.2900C};
(p)~\citet{1979A&A....80..201B};
(q)~\citet{sjp16};
(r)~\citet{2014ApJ...781L..32V};
(s)~\citet{2015ApJ...812..153O};
(t)~\citet{bgc+08};
(u)~\citet{2004A&A...423..111F};
(v)~\citet{2009A&A...505...45G};
(w)~\citet{2013ApJ...766...18G};
(x)~\citet{1994A&A...285...27K};
(y)~\citet{1999MNRAS.302..571R};
(z)~\citet{vgd+13};
(A)~\citet{bib+14};
(B)~\citet{2016MNRAS.459.2940K};
(C)~\citet{2001A&A...373..106F};
(D)~\citet{sbf+14};
(E)~\citet{2009A&A...503..707B};
(F)~\citet{2016A&A...595A.116M};
(G)~\citet{2016MNRAS.459.4240K};
(H)~\citet{cso+16};
(I)~\citet{2015MNRAS.451.4021S}.
\end{tablenotes}
\end{threeparttable}
\end{table*}
\setcounter{ft}{0}

\end{appendix}

\end{document}